\documentclass[12pt, a4paper]{article}

\usepackage[left=2.54cm,right=2.54cm,top=2.54cm,bottom=2.54cm]{geometry}
\usepackage{graphicx}
\usepackage{pdfpages}
\usepackage{eso-pic}
\usepackage{tabularx}
\usepackage{dcolumn}
\usepackage{booktabs}
\usepackage[authoryear, round]{natbib}
\usepackage[ansinew]{inputenc}
\usepackage{graphicx}
\usepackage{sidecap}
\usepackage{amsmath} %
\usepackage[export]{adjustbox}
\usepackage[skins,theorems]{tcolorbox}
\usepackage{textcomp}
\usepackage[onehalfspacing]{setspace}
\usepackage[nottoc,numbib]{tocbibind}
\usepackage[table]{colortbl}
\usepackage{makecell}
\usepackage{boldline}
\usepackage{rotating}
\usepackage{float}
\usepackage{multirow}
\usepackage[bottom,hang,flushmargin]{footmisc}
\usepackage{lipsum,color,float}
\usepackage{caption}
\usepackage{diagbox}
\usepackage{slashbox,pict2e}
\usepackage{flafter}
\usepackage{placeins}
\setcitestyle{notesep={: }}
\usepackage[hidelinks]{hyperref}
\usepackage{pdflscape}
\usepackage{subcaption}
\usepackage{changepage}
\usepackage{amssymb}
\usepackage{caption}

\usepackage{afterpage}

\usepackage{url}
\expandafter\def\expandafter\UrlBreaks\expandafter{\UrlBreaks%
  \do\a\do\b\do\c\do\d\do\e\do\f\do\g\do\h\do\i\do\j%
  \do\k\do\l\do\m\do\n\do\o\do\p\do\q\do\r\do\s\do\t%
  \do\u\do\v\do\w\do\x\do\y\do\z\do\A\do\B\do\C\do\D%
  \do\E\do\F\do\G\do\H\do\I\do\J\do\K\do\L\do\M\do\N%
  \do\O\do\P\do\Q\do\R\do\S\do\T\do\U\do\V\do\W\do\X%
  \do\Y\do\Z}
\usepackage{fancyhdr}
\usepackage{threeparttable}

\usepackage{arydshln}

\usepackage{lmodern}

\graphicspath{{figures/}}

\newcounter{daggerfootnote}
\newcommand*{\daggerfootnote}[1]{%
    \setcounter{daggerfootnote}{\value{footnote}}%
    \renewcommand*{\thefootnote}{\fnsymbol{footnote}}%
    \footnote[2]{#1}%
    \setcounter{footnote}{\value{daggerfootnote}}%
    \renewcommand*{\thefootnote}{\arabic{footnote}}%
    }

\newcommand*{\asterixfootnote}[1]{%
    \setcounter{daggerfootnote}{\value{footnote}}%
    \renewcommand*{\thefootnote}{\fnsymbol{footnote}}%
    \footnote[1]{#1}%
    \setcounter{footnote}{\value{daggerfootnote}}%
    \renewcommand*{\thefootnote}{\arabic{footnote}}%
    }

\renewenvironment{abstract}
 {\small
  \begin{center}
  \bfseries \abstractname\vspace{-.5em}\vspace{0pt}
  \end{center}
  \list{}{
    \setlength{\leftmargin}{0cm}%
    \setlength{\rightmargin}{\leftmargin}%
  }%
  \item\relax}
 {\endlist}
 
 \usepackage{datetime}

\newdateformat{monthyeardate}{%
  \monthname[\THEMONTH], \THEYEAR}

\usepackage{type1cm}
\usepackage{lettrine}

\usepackage[center]{titlesec}

\usepackage{tikz}
\usetikzlibrary{decorations.text}
\usepackage{framed}
\usepackage[export]{adjustbox}

\begin{document}
 \parindent0mm

\thispagestyle{empty}

 \begin{center}
				\large{\bf{The Local Economic Impact of Mineral Mining in Africa:}} \\
				\large{\bf{Evidence from Four Decades of Satellite Imagery}}\\[1em]
			\vspace{1cm}
			
\large{By Sandro Provenzano\asterixfootnote{\footnotesize Corresponding author, London School of Economics, Department of Geography and Environment, Houghton Street, London WC2A 2AE, UK. E-mail: \href{mailto:s.provenzano@lse.ac.uk}{s.provenzano@lse.ac.uk}.} and Hannah Bull\daggerfootnote{\footnotesize Universit\'e Paris-Saclay, Department of Computer Science, 3 Rue Joliot Curie, 91190 Gif-sur-Yvette. E-mail: \href{mailto:hannah.bull@universite-paris-saclay.fr}{hannah.bull@universite-paris-saclay.fr}.}}

\end{center}

\begin{center}
\small{This Version: \monthyeardate\today}	\\
\end{center}

\vspace{0.25cm}

\par\noindent\rule{\textwidth}{0.1pt}

\vspace{0.15cm}

\begin{abstract}

\noindent \normalsize{Using state-of-the-art techniques in computer vision, we analyze one million satellite images covering 12\% of the African continent between 1984 and 2019 to track local development around 1,658 mineral deposits. We use stacked event studies and difference-in-difference models to estimate the impact of mine openings and closings. The magnitude of the effect of mine openings is considerable - after 15 years, urban areas within 20km of an opening mine almost double in size. We find strong evidence of a political resource curse at the local level. Although mining boosts the local economy in democratic countries, these gains are meager in autocracies and come at the expense of tripling the likelihood of conflict relative to prior the onset of mining. Furthermore, our results suggest that the growth acceleration in mining areas is only temporary and diminishes with the closure of the mine. }

\end{abstract}

\vspace{1.6cm}

\begin{tabular}{ll}
Keywords: & Africa, Local Development, Natural Resources, Remote Sensing \\
JEL:      & C81, L72, O13, Q32, R11
\end{tabular}

\pagebreak

\section{Introduction}

Africa's great wealth in natural resources attracts tens of billions of dollars of investments by international mining companies every year. Since the early 2000s, increased commodity prices have led to a surge in investor interest and mine openings on the continent. 
Mining projects often gain support from communities through promises of benefits to the local economy, though the empirical evidence is mixed: mineral assets are advantageous in some circumstances but lead to corruption and violence in others. To shed light on this apparent discrepancy, we significantly extend the geographic and temporal coverage of previous work in this area by gathering satellite data that spans several decades and encompasses several institutional environments. \\

Previous research on the local impact of mineral mining generally focuses on a few mines or countries. However, due to macro-level factors, such as institutional context, conclusions differ amongst these small scale studies. Some studies find that mine openings boost the local economy \citep{Aragon2013,Mamo2019}, while other studies find no significant effects \citep{Pokorny2019,Bazillier2020}\footnote{Note that these studies only found positive income and employment effects for small artisanal mines, not for industrial mines which are the focus of this study.} or increases in corruption and conflict \citep{Vicente2010,Berman2017,Knutsen2017}. Research at the macro level suggests only countries with good quality institutions benefit from natural riches
\citep{Mehlum2006,Bhattacharyya2010}, suggesting that institutions may play a similar role for the local impact of mining. Using a field experiment, \cite{Armand2020} provide an example of how community activism can prevent civil conflict. Our large-scale long-term study allows us to identify heterogeneities in the local effects of mining across different institutional contexts. Moreover, our work is the first large-scale attempt at measuring the long-term effects of mine closures in local communities. \\

\begin{figure}[!h]
\centering
\caption{Automatic Annotation of Two Landsat Images over Time} \label{satim:pre_post_mine_and_zoom}
\begin{subfigure}{0.5\hsize}\centering
    \caption{\scriptsize \centering 1991: Prior to the onset of mining, with a small urban area (magenta)}
    \includegraphics[width=0.8\textwidth]{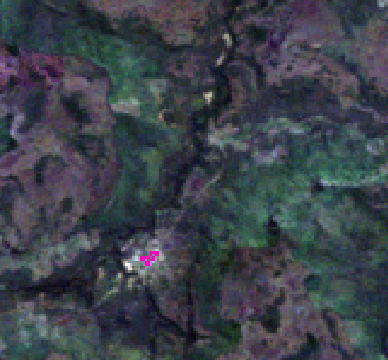}
\end{subfigure}
\begin{subfigure}{0.5\hsize}\centering
    \caption{\scriptsize \centering 2018: An urban area (magenta) has developed next to the mine (cyan)}
    \includegraphics[width=0.8\textwidth]{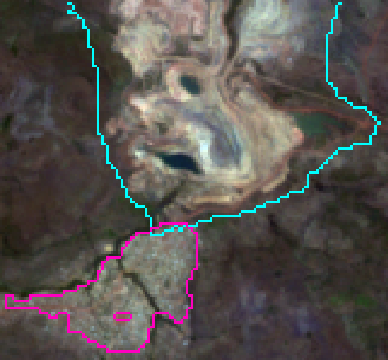}
\end{subfigure}
\begin{tablenotes}[flushleft]
\scriptsize \item \textit{Note:}
Automatic annotation of satellite images allows for large-scale and long-term observation of urban growth, land use and material wealth in mining areas. See Figure~\ref{satim:pre_post_mine_zoom_google_images} for recent high-resolution images of this location.
\end{tablenotes}
\end{figure}

Our novel approach uses one million satellite images to create an extensive panel covering 12\% of the African landmass, tiling the area within 40km$^2$ of 1,658 mines across 47 African countries, over a period of 36 years from 1984 to 2019. The images are acquired from different stages in the mine life-cycle, including prior to the mineral discovery and after the mine's closure, 
as well as at various distances to the mine. This allows us to observe the local effects of mining over long periods of time and under strong and weak institutional contexts. We use state-of-the-art techniques in machine learning and computer vision to translate this vast collection of images into economically meaningful indicators including urban and agricultural land use, as well as material wealth.\footnote{Since wealth predictions are based on satellite images, `wealth' does not explicitly refer to usual measures of wealth such as assets or savings. While the underlying models are trained using household level asset wealth indices as input, the wealth index ultimately reflects correlates and material manifestations of wealth in a local area, such as urban shape and density, infrastructure, or roof reflectivity.} Figure~\ref{satim:pre_post_mine_and_zoom} illustrates how we can effectively detect land cover changes before and after the mine opening from satellite images using computer vision techniques. Using stacked event study designs and difference-in-difference (DiD) models, we estimate the impact of mine openings and closings on our satellite-derived indicators. We exploit spatial variation and temporal differences between mineral discovery and active mining to derive counterfactual groups. 
 
We provide three key results on the local impact of mining. Firstly, mine openings -  particularly large mine openings - considerably increase urbanization, our main measure of economic development. Using an event-study, we show that after 15 years, areas within 20km from opening mines gain on average around 80\% in urban extent relative to not-yet-mined mine areas. In addition, we show that mining primarily impacts the area within 20km of the mine, and that large mines tend to have a stronger effect than small mines. We find that part of the economic gains from mining are due to increased agricultural activities. \\

Secondly, we find strong support for the presence of a political resource curse at the local level. Although mining boosts the local economy in democratic countries, these gains are meager in autocracies and come at the expense of a nearly 5-fold increase in localized conflict. \\

Thirdly, our results suggest that the growth boost in mining areas is temporary. After a mine's closure, there is a drop in the economic growth rate around closing mines. While former mining areas maintain their increased urban development with respect to non-mining areas, gains from mining diminish after the mine's closure. \\

Furthermore, we contribute code, model parameters and implementation details of two models trained on Landsat data for future research. The first model segments land use classes to annotate water, urban areas and cropland in satellite images. The second model segments mine areas.

\subsection*{Related Work}

Our large-scale and long-term approach consolidates macro-economic findings on the potential negative impacts of natural resources in countries with poor institutions, and micro-economic research on the local impact of mines with conflicting conclusions depending on the particular context of the area of study. 
Additionally, the temporal extent of our study allows us to observe mines over the decades both prior and post opening, to cover a blind-spot in the literature on the duration of the effects of mining.
Unlike studies using low-resolution nightlight data, we are able to mask out the mine itself from daytime satellite imagery in order to distinguish between the development of the mine and the development of the neighbouring urban areas. Finally, we complement existing models on estimating poverty from satellite data, by providing trained models to segment urban areas, water, cropland and mines in Landsat imagery. \\

There is extensive macroeconomic research on the economic implications of the extractive resource sector in developing countries, in particular on potential underlying mechanisms of the `resource curse' - the counter-intuitive finding that countries with a larger natural resource sector tend to be less developed \citep{Auty1993}. The literature from the 1980s and 90s points towards macroeconomic mechanisms such as the `dutch disease', i.e. a rise in price level that makes other exporting sectors less competitive \citep{Corden1982,Wijnbergen1984,Krugman1987,Sachs2001}. Other explanations include the high volatility of commodity prices that makes resource rich countries more vulnerable to macroeconomic shocks \citep{Deaton1999}. Subsequently, studies such as \cite{Lane1996}, \cite{Tornell1999}, \cite{Torvik2002}, \cite{Hodler2006} and \cite{Caselli2013} identify the cause of resource misfortune in rent-seeking behaviour, preventing the redistribution of resource windfalls amongst the population and also shifting the economy towards less productive activities. \citet{Caselli2013} find that only a small part of the increased oil revenue of Brazilian municipalities actually benefits the population, probably due to embezzlement. More recent research emphasizes the political dimension of the resource curse, arguing that natural resources are only advantageous in places with good institutions \citep{Mehlum2006, Robinson2006, Collier2009, Bhattacharyya2010}. In places with poor institutions, resource windfalls deteriorate institutions even further by increasing corruption and undermining the political process, resulting in a net negative impact of resources on growth.\footnote{See \cite{Ploeg2011} and \cite{Venables2016} for an extensive literature review and a discussion about the challenges of resource governance in developing countries.}\\

The literature on the local economic impact of mining is relatively recent. In their seminal paper, \cite{Aragon2013} investigate a production expansion of a Peruvian Gold mine over ten years and find strong evidence for increases in real income in the mining city and neighboring areas, albeit with a decreasing positive impact with distance from the mining city. Moreover, service and agricultural workers also benefit from the mine expansion, and the authors argue that this is due to backward linkages. Similarly, \cite{Lippert2014} studies the copper mining boom in Zambia and documents positive effects of mining beyond the mining sector and immediate mine area. \cite{Mamo2019} find positive effects of mining at the district level in Africa at the intensive margin (increased mining production), and especially at the extensive margin (new mineral discoveries and mine openings). However, the authors find little evidence for spillovers to neighboring districts. With regards to the developed world, \cite{Allcott2018} find positive effects of oil and gas production on real wages at the US county level, and no negative effects on the productivity of the tradeable manufacturing sector.\\

On the other hand, \cite{Hirschman1958} and \cite{McMillan2014} hold a critical view of mining, arguing that mining ventures cannot induce structural change, as they are capital intensive `enclaves' with little linkage to the local economy and few local employment opportunities. \cite{Bazillier2020} investigate the gold boom in Burkina Faso and find that artisanal mines have a significantly positive impact on the local economy, in constrast to industrial mines - an effect also noted in \cite{Pokorny2019}. There are strong heterogeneities on the local economic impact of mining between capital-intensive large industrial mines with relatively few unskilled job opportunities and artisanal mines providing income to local workers. Mining-related pollution also induces adverse health outcomes in the local population, such as lead toxicity \citep{Goltz2019}. \citep{Aragon2016} find that mining has considerable negative effects on local agricultural productivity that might even outweigh the gains from mining \citep{Aragon2016}, either due to adverse effects on workers' health and labor productivity or due to soil and crop deterioration. However, \cite{Benshaul-Tolonen2019} finds that gold mine openings reduce child mortality by 50\% arguing that the indirect positive employment and income effects outweigh direct negative health effects. \\

Recent studies point to the relevance of political economy mechanisms in local distribution of resource wealth. \cite{Vicente2010} and \cite{Knutsen2017} show that mining deteriorates local institutions by increasing corruption. Furthermore, \cite{Dube2013} and \cite{Berman2017} demonstrate that resource windfalls lead to political instability through increasing conflict.\footnote{\cite{Berman2017} show that one important channel through which mining increases conflict is by helping rebel groups in control of mine sites to finance their military capacity. The authors also point out that there might be other important mechanisms that link mining and conflict, such as increasing potential for rent-seeking.} Out of these studies, only \cite{Berman2017} examines whether the effects are stronger in countries with poor institutions, as suggested by the `resource curse' hypothesis, but does not detect significant heterogenous effects. However, the authors do find stringent mining-specific anti-corruption measures and transparency initiatives reduce conflict. \cite{Armand2020} conduct a field experiment in areas of recent natural gas discoveries in Northern Mozambique, and provide tangible evidence in support of the local political `resource curse' hypothesis. The authors show that information campaigns encouraging communities to participate in decision-making leads to a decrease in conflict. In contrast, when information is only provided to the local leader, there is no reduction in conflict, but rather increased elite capture and rent seeking. \\

Two key shortcomings in the existing literature on the local effects of mining are small samples sizes of only a handful of mines or countries and lack of observation of long-term trends. If the political `resource curse' hypothesis were true, we would expect there to be important heterogeneities between countries depending upon quality of institutions. Consequently, relying on a small sample undermines external validity, and explains why studies reach conflicting conclusions about the local impact of mining. Long-term studies allow for observation of common trends in non-mining and mining areas prior to production in order to isolate the effects of the mine. A few studies, such as \cite{Chuhan-Pole2016} and \cite{Mamo2019}, have a panel of 20 years covering a wide sample of mines and countries. However, these studies are limited by reliance on nightlights as a measure of economic activity. The use of nightlights is likely to result in distorted estimates due to the impossibility of distinguishing between changes in nightlights that reflects the mere activity of the mine infrastructure, and the change related to wealth and income changes in the local area. Daytime satellite images are at a substantially higher resolution ($<$30m) than available nightlight data ($\sim$750m), and allow for masking out the mine to isolate observation of the local area. \\

Daytime satellite images have been used in the development economics literature to estimate various indicators, including income levels in African countries \citep{yeh2020using} and sustainable development goals e.g. agriculture, health, education, water and sanitation \citep{yeh2021sustainbench}. To obtain our indicators of interest, we train models to segment objects in image data - a highly popular field in computer vision literature - and we recommend \cite{minaee2021image} for a detailed summary of deep learning methods for such tasks. In particular, the U-Net architecture \citep{ronneberger2015u} and the ResNet architecture \citep{He2016} are popular choices for image segmentation and object detection, and \cite{diakogiannis2020resunet} use a combination of these two architectures for segmenting objects in satellite imagery. \\

The remainder of this paper is organized into five sections. Section \ref{satim:data} introduces our mine categorization framework for comparing different mining areas across time. Section \ref{satim:sect_remotesens} describes the machine learning methods that we use for processing our remote sensing data. Section \ref{satim:sec_empirical_strategy} discusses our identification strategy for estimating the impact of mine openings and closures. In Section \ref{satim:sect_results}, we present our results on the impact of mines, investigate heterogeneities with respect to mine size and institutional context, and examine conflict as a mechanism counteracting the potential economic gains from mines. Finally, Section \ref{satim:sect_conclusion} concludes our findings.

\section{Categorization of Mineral Deposits and Surrounding Areas} \label{satim:data}

To the best of our knowledge, our dataset is the most extensive panel used to study the effects of mining in Africa, covering 84k areas of 43km$^2$ within 40km of a mineral deposit - corresponding to 12\% of the continent - over a period of 36 years.\footnote{See App.~\ref{satim:app:data_appendix} for a list of our data sources.} This section describes the spatial, temporal and typological categories of this panel data. We classify areas surrounding mines by proximity to mineral deposits, as well as the size and activity status of neighboring mines, and track these categories across 12 three-year time periods. This categorization of mine areas is key to our identification strategy and is used to define treated and untreated units,\footnote{See Section \ref{satim:sec_empirical_strategy}.\ref{satim:sect_counterfact} for a detailed discussion on treatment and control groups.} and we refer to these categories throughout the remainder of the article. \\

Our mining dataset includes information on 1,658 mineral deposits and industrial mines in 47 African countries, including the size and type of mine, the date of discovery, the dates of activity and the geographic location. Mines are classified by experts in the mining sector as \emph{Small} or \emph{Large} based on a variety of indicators including the pre-mined resource, ore value, by-products and the type of mineral.\footnote{Mines in our dataset are classified as Minor, Moderate, Major, Giant and Super Giant, and we refer to Minor and Moderate mines as \emph{Small} and remaining categories as \emph{Large} in our typology.} We observe these mineral deposits from 1984 to 2019, and we divide this 36 year interval into 12 equal-length periods of 3 years, i.e. Period 1 (1984-1986), $\dots$, Period 12 (2017-2019). \\

\begin{figure}[htb]
\centering
\caption{Mine Status Overview in the Period of 1984-2019} \label{satim:mine_groups_overview}
\begin{adjustbox}{width=.7\textwidth,center}
\begin{minipage}{.98\textwidth}
\begin{framed}
\begin{tikzpicture}[every node/.style = {shape=rectangle, rounded corners,draw, align=center},
    level distance=2.5cm,
  	level 1/.style={sibling distance=8cm},
  	level 2/.style={sibling distance=4cm},
 	level 3/.style={sibling distance=3.5cm}]
  \node[top color=red!30, bottom color=red!30] (a) {\LARGE{1,658 Mineral Deposits}}
      child { node[top color=blue!30, bottom color=blue!30] (b) {\Large{Active Mine}} 
      	child { node[top color=blue!20, bottom color=blue!20] (b1) {Continuous} } 
      	child { node[top color=blue!20, bottom color=blue!20] (b2) {Partial}   
      		child { node[top color=blue!10, bottom color=blue!10] (b21) {Opening} } 
      		child { node[top color=blue!10, bottom color=blue!10] (b22) {Closing} } 
      		child { node[top color=blue!10, bottom color=blue!10] (b23) {Opening \& Closing} } } }
      child { node[top color=blue!30, bottom color=blue!30] (c) {\Large{Inactive Mine}} 
      	child { node[top color=blue!20, bottom color=blue!20] (c1) {No Longer Active} }
      	child { node[top color=blue!20, bottom color=blue!20] (c2) {Not Yet Opened} } };
      \draw [postaction={decorate,decoration={raise=0.5ex,text along path,text align=center,text={|\sffamily|671}}}] (b) to (a);
      \draw [postaction={decorate,decoration={raise=0.5ex,text along path,text align=center,text={|\sffamily|987}}}] (a) to (c);
      \draw [postaction={decorate,decoration={raise=0.5ex,text along path,text align=center,text={|\sffamily|194}}}] (c1) to (c);
      \draw [postaction={decorate,decoration={raise=0.5ex,text along path,text align=center,text={|\sffamily|793}}}] (c) to (c2);
      \draw [postaction={decorate,decoration={raise=0.5ex,text along path,text align=center,text={|\sffamily|127}}}] (b1) to (b);
      \draw [postaction={decorate,decoration={raise=0.5ex,text along path,text align=center,text={|\sffamily|544}}}] (b) to (b2);
      \draw [postaction={decorate,decoration={raise=0.5ex,text along path,text align=center,text={|\sffamily|348}}}] (b21) to (b2);
      \draw [postaction={decorate,decoration={raise=0.5ex,text along path,text align=center,text={|\sffamily|77}}}] (b2) to (b22);
      \draw [postaction={decorate,decoration={raise=0.5ex,text along path,text align=center,text={|\sffamily|119}}}] (b2) to (b23);
\end{tikzpicture}
\end{framed}
\end{minipage}
\end{adjustbox}

\begin{tablenotes}[flushleft]
\scriptsize \item \textit{Note:} \emph{Active} mines are active during at least one period. \emph{Continuously Active} mines are active from Period 1 until Period 12 while \emph{Partially Active} mines do not continuously operate between these two periods. \emph{Partially Active} mines may open, close or open and close, and we call these subcategories \emph{Opening}, \emph{Closing} and \emph{Opening \& Closing}. \emph{Inactive} mines either closed prior to our period of interest (\emph{No Longer Active}) or had not opened by the end of our period of interest (\emph{Not Yet Opened}). Our identification strategy exploits differences between these categories in order to measure the effect of active mining. %
\end{tablenotes}
\end{figure}
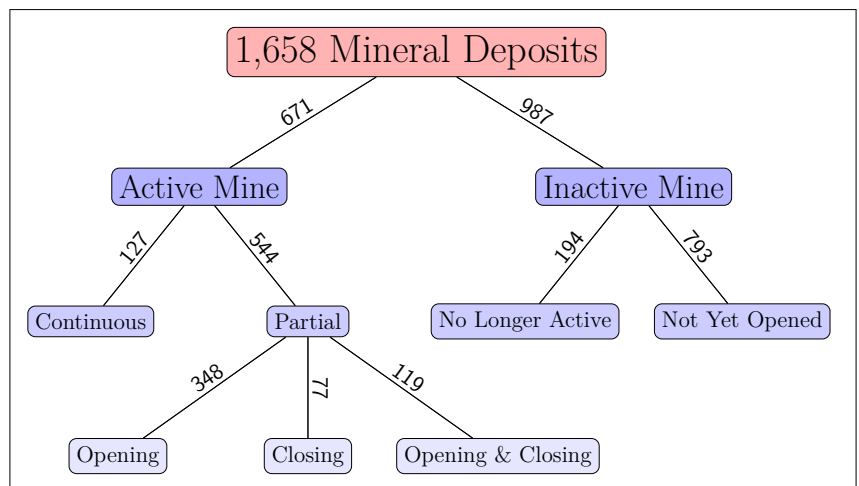

Figure~\ref{satim:mine_groups_overview} contains a tree diagram classifying the activity status of these mineral deposits across our 12 periods. At the first level of grouping, we distinguish mines that were actively operating at some point between 1984 and 2019 (\emph{Active}) from mines that were never operating during our period of interest (\emph{Inactive}). At the second level of grouping, we distinguish \emph{Active} mines between those mines that were active in Period 1 (1984-1986) and continuously operated until at least Period 12 (2017-2019) and other mines that were only partially active (\emph{Continuous} vs. \emph{Partial}). We also group \emph{Inactive} mines into mines that had ceased operation at some point prior to 1984 (\emph{No Longer Active}) and mineral discoveries with no active mine as of 2019 (\emph{Not Yet Opened}). At the third level of grouping, we categorize mines that were partially active in our study period into three groups: mines which began operation at some point between Periods 2 and 12 and continued operation until Period 12 (\emph{Opening}), mines that were operating in Period 1 but that closed and remained inactive until Period 12 (\emph{Closing}), and mines that opened and closed or closed and reopened during our study period (\emph{Opening \& Closing}). 

\begin{figure}[ht]
\centering
\caption{Locations of Mineral Deposits and Tile Grid} \label{satim:fig:africa_locations}
\begin{subfigure}{0.4\hsize}\centering
    \caption{\centering \scriptsize 1,658 mineral deposits} 
    \includegraphics[height=5.5cm]{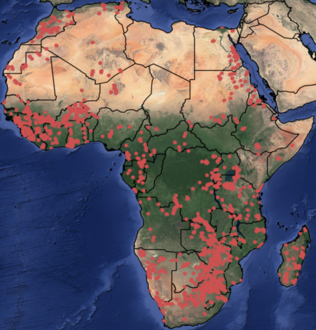}
\end{subfigure}
\begin{subfigure}{0.6\hsize}\centering
    \caption{\centering \scriptsize 40km radius around each mineral deposit}
    \includegraphics[height=5.5cm]{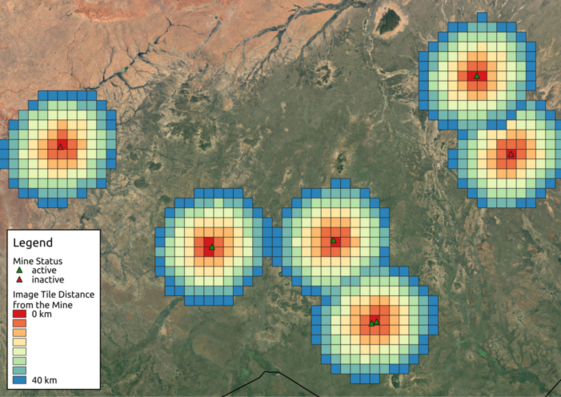}
\end{subfigure}
\begin{tablenotes}[flushleft]
\scriptsize \item \textit{Note:} The area within 40km surrounding a mineral deposit is divided into tiles of $6.5\times 6.5\mathrm{km}\approx43\mathrm{km}^2$, providing a total of 84,207 images covering 12\% of African landmass. Each tile corresponds to a Landsat image with $224\times224$ pixels at 30m resolution, with 7 multi-spectral bands. We acquire one image (3-year median) at each of the 12 three-year periods, totaling around one million tiles.
\end{tablenotes}
\end{figure}

Using the geolocalization of each mineral deposit, we assemble medium-resolution multi-spectral satellite images from a radius of 40km and from each of our 12 three-year periods between 1984 and 2019. We use 3-year periods to obtain multiple cloud-free satellite images from each period, and then take the median pixel value to remove extreme values due to image defects and to smooth seasonal weather trends. This amounts to a total study area of 3.6 million km$^{2}$, representing 12\% of the total African landmass or 15 times the landmass of the United Kingdom. We divide this area into 84,207 tiles of $6.5\times 6.5\mathrm{km}\approx43\mathrm{km}^2$, as depicted in Figure~\ref{satim:fig:africa_locations}. We track the evolution of each tile over our 12 periods and associate this observed evolution to neighboring mine activity.

To assign tiles to mines, we compute the distance from each tile centroid to each active mine in each period.\footnote{See \url{http://spatialreference.org/ref/esri/africa-sinusoidal} for the projection used to calculate distance.} We assign each tile to the closest \emph{Active} mine if there is an \emph{Active} mine within a 40km radius, else to the closest \emph{Inactive} deposit. Each tile is then classified according to the category of the associated mine, as shown in the schema in Figure~\ref{satim:mine_groups_overview}. In order to avoid confounding effects, we remove a relatively small number of tiles in proximity to multiple mines in different \emph{Active} mine categories (e.g. \emph{Continuous} and \emph{Partial}). We call tiles within 20km of the associated mineral deposit \emph{Near} to a mineral deposit and tiles between 20 and 40km \emph{Far} from a mineral deposit. \\ 

In the next section, we describe how we extract quantitatively meaningful indicators from our million satellite image tiles, in particular urban land areas, agricultural areas, the area of the mine itself and an estimation of material wealth.

\section{Using Machine Learning to Extract Economic Indicators from Satellite Data} \label{satim:sect_remotesens}

We use state-of-the-art techniques in computer vision to translate our rich satellite imagery into interpretable variables on land use and material wealth, in order to track changes in these economic indicators with respect to the proximity, size and activity status of neighboring mines. Moreover, we are able to automatically segment the area of the mine from each image, in order to exclude the mine itself when analyzing the impact on the area surrounding the mine. In Section~\ref{satim:subsec_landusemine}, we describe our methodology to train neural networks to predict land use and mine areas from Landsat images, and in Section~\ref{satim:subsec_materialwealthindex} we present an existing model that we use to predict local material wealth \citep{yeh2020using}. \\

For all our models, the input satellite image tiles consist of the Blue, Green, Red, Near Infrared, Short-wave Infrared 1, Short-wave Infrared 2 and Thermal bands from the atmospherically corrected surface reflectance sensors of Landsat-5, 7 and 8. The Thermal band is at 120m resolution and is resampled to 30m pixels, and the remaining bands are at 30m resolution. We take the median of all Landsat images over 3-year periods, excluding any pixels with clouds, cloud shadows, or snow. We divide our area of interest into square tiles of $224\times224$ pixels, or approximately $6.5\times6.5\textrm{km}^2$, where each pixel has 7 channels corresponding to the Landsat bands. We download this data using Google Earth Engine. 

\subsection{Land Use and Mine Segmentation}\label{satim:subsec_landusemine}

We train a convolutional neural network (CNN) model to segment our image tiles into 4 mutually exclusive land use classes: Urban Areas, Croplands, Water and Other. We also train a mine segmentation model to identify the geographic extent of the mine, as our mining dataset only provides a point location rather than the area of the mine itself. Both models use the same architecture and differ only in terms of number of outputs (4 classes for land use and binary outputs for mine segmentation), and initialization of the model parameters. 
The data used for training the land use and the mine segmentation models is described in Table~\ref{satim:tab:data_training}. We combine multiple annotation sources to train the land use segmentation model, and annotations for the mine segmentation model come from \cite{Maus2020}. 

\begin{table}[htb]
\caption{Data for Training and Evaluating the Land Use and Mine Segmentation Models}
\vspace{-0.2cm}
\label{satim:tab:data_training}
\small
\begin{center}
\begin{adjustbox}{width=.8\textwidth,center}
\begin{tabular}{|l|l|l|}
\hline
\textbf{Task}           & \textbf{Land Use Segmentation}                                                                                 & \textbf{Mine Segmentation}                                          \\ \hline
\textbf{Landsat images} & 84k images (2014-6)                                                                                            & 28k$\times$2 images (2014-6, 2017-9)                       \\ \hline
\textbf{Area}           & \begin{tabular}[c]{@{}l@{}} within a 40km radius of mines\\ (Africa) \end{tabular} 
                         & \begin{tabular}[c]{@{}l@{}} within a 40km radius of mines\\ (Global) \end{tabular}                    \\ \hline
\textbf{Classes}        & \begin{tabular}[c]{@{}l@{}}Water\\ Urban Areas\\ Cropland\\ Other\end{tabular}                              & Mine                                                       \\ \hline
\textbf{Annotations}    & \begin{tabular}[c]{@{}l@{}}ESA (Water, Urban)\\ GHSL (Urban) \\ Facebook (Urban) \\ NASA (Croplands)\end{tabular} & \cite{Maus2020} \\ \hline
\textbf{Train-Val-Test} & 80\%-5\%-15\% & 80\%-5\%-15\% \\
\hline
\end{tabular}
\end{adjustbox}
\begin{tablenotes}[flushleft]
\scriptsize \item \textit{Note:} Further details on annotation sources are listed in App.~\ref{satim:app:data_appendix}. 
\end{tablenotes}
\end{center}
\end{table}

Our model architecture has a U-Net backbone \citep{ronneberger2015u} with 16M parameters. The encoder side of the U-Net model is a ResNet-50 model \citep{He2016}, but adapted to input 7 image channels instead of 3 RGB channels at the first convolutional layer. As the thermal band is at a resolution 4 times lower than the remaining 6 bands, we use 4-pixel dilated convolutions on this channel in the first layer. For the land use model, the model weights are initialized by the ResNet-50 model pre-trained model on ImageNet \citep{Deng2009}. The model weights corresponding to the non-RGB bands in the first convolutional layer are initialized as the average weights corresponding to the RGB bands. For the mine segmentation model, the weights are initialized using the trained land use segmentation model. We consider image segmentation as a pixel classification problem and train using cross-entropy loss. \\

The evaluation metric is the $R^2$ value between the predicted shares of a class and the ground truth shares of a class within each image tile. This is because in our analyses, our observation unit is an image tile and not an image pixel; we are interested in computing the share of each land use category in each image tile and thus do not require pixel-level accuracy. We firstly train the land use model for 75 epochs (passes of the training set through the model) for 12 days on a single NVIDIA Tesla K80 GPU and choose epoch 72, as it has the best evaluation metric on the validation set. We then initialize the weights of the mine segmentation model with those of the land use model and train this model for a further 2 days or 40 epochs on a single NVIDIA Quadro RTX 3000 GPU, and choose epoch 29 based on the evaluation metric on the validation set. 

\begin{figure}[htb]
\centering
\caption{Test Set: Predicted vs. Actual Shares by Segmentation Class}\label{satim:fig:landuse_mine_predictions}
\subfloat[$R^2$ Urban: $0.96$ \label{satim:fig:a_perf}]{\includegraphics[width=0.25\textwidth,height=\textheight,keepaspectratio]{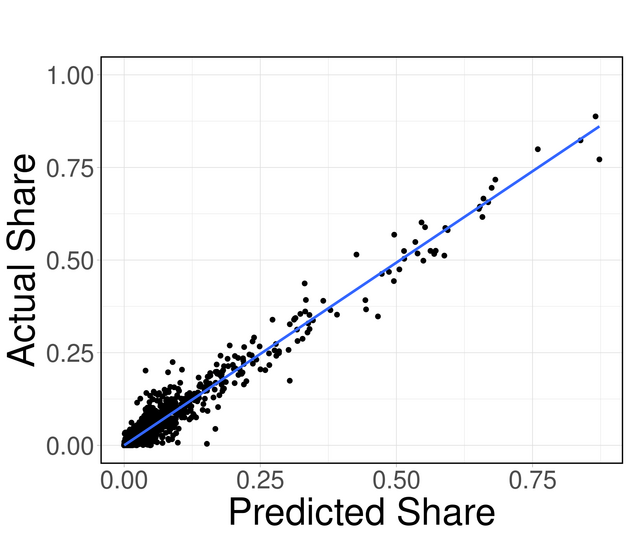}}
\subfloat[$R^2$ Cropland: $0.82$\label{satim:fig:b_perf}]{\includegraphics[width=0.25\textwidth,height=\textheight,keepaspectratio]{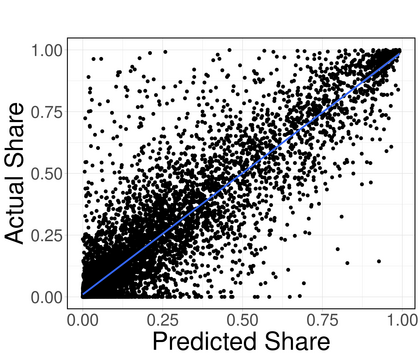}}
\subfloat[$R^2$ Water: $0.98$\label{satim:fig:c_perf}]{\includegraphics[width=0.25\textwidth,height=\textheight,keepaspectratio]{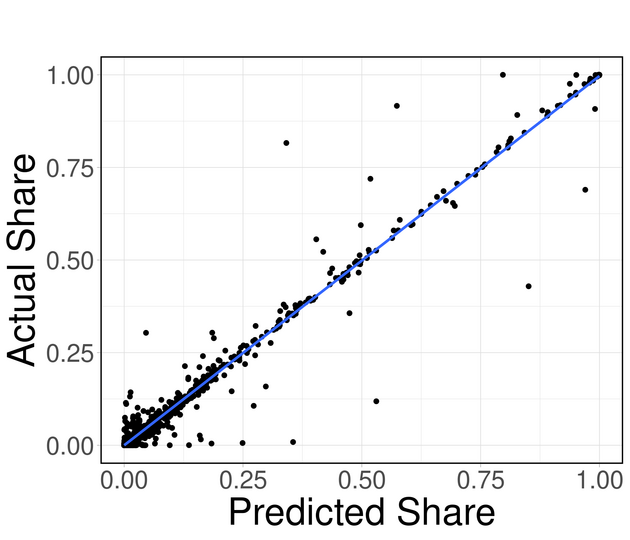}}
\subfloat[$R^2$ Mine: $0.78$\label{satim:fig:d_perf}]{\includegraphics[width=0.25\textwidth,height=\textheight,keepaspectratio]{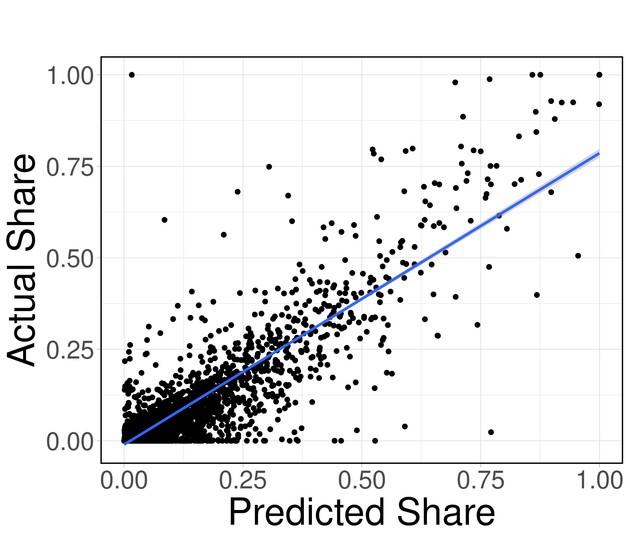}}
\begin{tablenotes}[flushleft]
\scriptsize \item \textit{Note:} The $R^2$ values show strong correlations for all classes, but particularly for the Urban and Water classes.
\end{tablenotes}
\end{figure}

\begin{figure}[!ht]
\centering
\caption{Examples of Automatic Segmentation Results (2011-3)}
\label{satim:fig:landuse_mine_visualisation}

\subfloat[Urban \label{satim:fig:a_ex}]{\includegraphics[width=0.24\textwidth,height=\textheight,keepaspectratio]{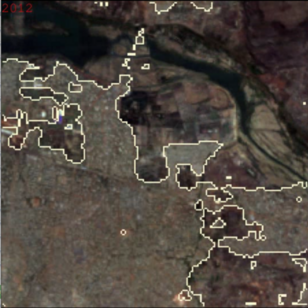}}\hspace{0.5pt}
\subfloat[Cropland \label{satim:fig:b_ex}]{\includegraphics[width=0.24\textwidth,height=\textheight,keepaspectratio]{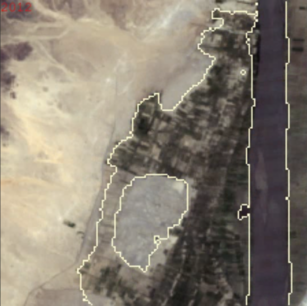}}\hspace{0.5pt}
\subfloat[Water \label{satim:fig:c_ex}]{\includegraphics[width=0.24\textwidth,height=\textheight,keepaspectratio]{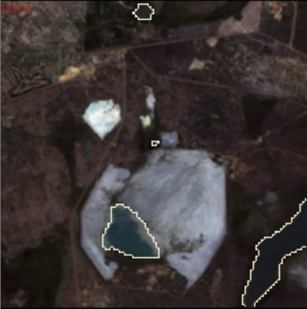}}\hspace{0.5pt}
\subfloat[Mine \label{satim:fig:d_ex}]{\includegraphics[width=0.24\textwidth,height=\textheight,keepaspectratio]{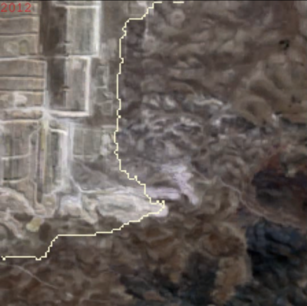}}
\begin{tablenotes}[flushleft]
\scriptsize \item \textit{Note:} These examples illustrate that our models are capable of identifying areas of interest.
\end{tablenotes}
\end{figure}

Performance of the land use model and the mine segmentation model on the test set is provided in Figure~\ref{satim:fig:landuse_mine_predictions}. The $R^2$ value is 0.96 for urban areas, 0.82 for cropland areas and 0.98 for water bodies, demonstrating that our model is a strong predictor of the share of land use categories in each image tile. The $R^2$ value for mine prediction is also high, at 0.78. Our model is thus capable of predicting image tiles with a high or low presence of mines. Qualitative examples of the land use and mine segmentation models in Figure~\ref{satim:fig:landuse_mine_visualisation} show that our model is capable of accurately segmenting urban, cropland, water and mining areas.

We apply our land use model to compute the log of 
the share of urban, cropland and water areas in each image tile, using our mine segmentation model to exclude the area of the mine. Moreover, we exclude outliers by first flagging all observations that are more than 2 interquartile distances below the first quartile or above the third quartile. We then conduct a generalized extreme studentized deviate (ESD) test to sequentially test if the flagged observations are outliers at the 90\% confidence level \citep{Rosner1983}.

\subsection{Material Wealth Index}\label{satim:subsec_materialwealthindex}

In \cite{yeh2020using}, the authors train a CNN with ResNet-18 architecture \citep{He2016} to learn a local material wealth index from multi-spectral Landsat images. The idea behind this approach is that there are material manifestations of local household wealth in satellite images such as the shape, density and roof reflectance of the urban area, or the length, size and color of road infrastructure. 
The CNN is trained using satellite images and corresponding asset wealth indices based on 43 Demographic and Health Surveys (DHS) conducted in 23 countries in Africa from 2009 to 2016. In the cross-section, \cite{yeh2020using} obtain an $R^2$ value on the relationship between the predicted and true (survey) values of the material wealth index on unseen data of around 0.65. \\

\begin{figure}[!ht]
    \caption{DHS-Based Wealth Score and Satellite Based Predictions}
    \label{satim:fig:val_material_wealth}
    \centering
    \includegraphics[width=0.32\textwidth]{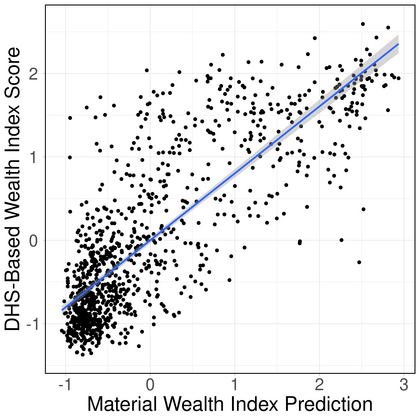}

    \begin{tablenotes}[flushleft]
    \scriptsize \item \textit{Note:} The wealth index shows a strong correlation with DHS survey results from our study area of interest in 2014-6 ($R^2$~value of 0.67).
    \end{tablenotes}
\end{figure}

As a sanity check, 
we create a DHS-based asset wealth index for all tiles with DHS respondents in 2014-6 (N=1046) and compare it to the material wealth predictions based on Landsat imagery.\footnote{To protect the privacy of DHS respondents, their geo-coordinates are generally displaced by up to 2km in urban areas and 5km in rural areas, but 1\% of tiles are further displaced to up to 10km. We assign DHS clusters to tiles when their geocoordinates lie within the central region of a tile, at least 1km from the edges.} We obtain an $R^2$ value of 0.67, in coherence with the results presented in \cite{yeh2020using}. Figure~\ref{satim:fig:val_material_wealth} plots the wealth index values from the DHS survey and the model predictions for tiles in our study area. \\

\begin{figure}[ht]
\centering
\caption{Qualitative Examples of Increases in Wealth Index over Time}
\label{satim:fig:time_poor_vs_rich}
\begin{subfigure}{0.5\hsize}\centering
    \caption{\scriptsize \centering Prior to the mine (1991, left) the wealth index is low, but post-mining (2018, right) the wealth index is high}
    \includegraphics[width=0.47\textwidth]{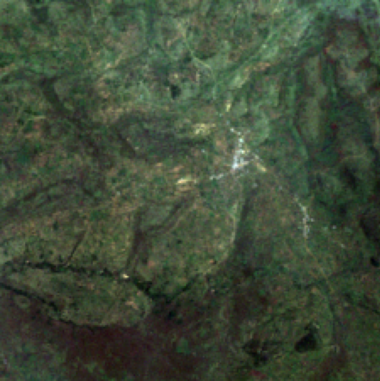}
    \includegraphics[width=0.47\textwidth]{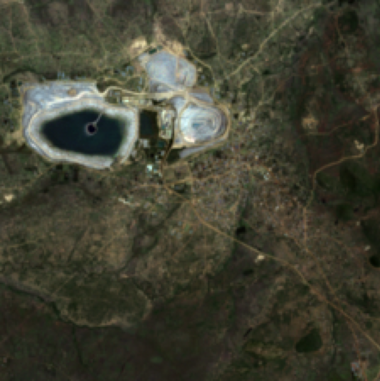}
\end{subfigure}
\begin{subfigure}{0.5\hsize}\centering
    \caption{\scriptsize \centering An area with a low wealth index (1988, left) evolves into an area with a  high wealth index (2018, right)}
    \includegraphics[width=0.47\textwidth]{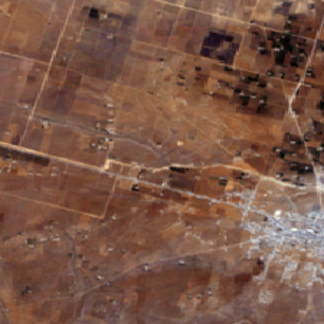}
    \includegraphics[width=0.47\textwidth]{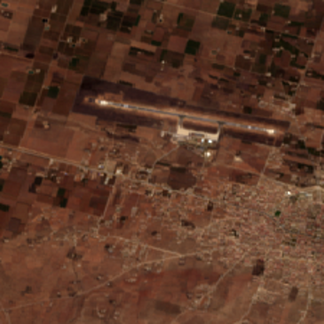}
\end{subfigure}
\begin{tablenotes}[flushleft]
\scriptsize \item \textit{Note:} The growth in material wealth index over time seems to be coherent with our intuition for numerous examples from our dataset. We see the index increasing over time when towns and cities become more developed. 
\end{tablenotes}
\end{figure}

Moreover, the wealth index can be used to track changes in local material wealth over time.
Living Standards Measurement Surveys (LSMS) surveys from the World Bank contain geolocalized household asset data for the same households at different points in time, providing an estimate of the change in wealth.\footnote{For each pair of years, we compute the mean of the households that were surveyed in both years. We then run a PCA of these asset-differences across the 5 countries for which we have panel LSMS surveys (Ethiopia, Malawi, Nigeria, Tanzania, Uganda). The value of the first principle component is the household-level index of asset differences. Within each small geographical cluster, we then compute the mean household-level index of asset differences to get the cluster-level index of asset differences.} There is a positive correlation (coefficient 0.35) between the predicted wealth change from the model and the estimated wealth change from LSMS panel data surveys, suggesting that the material wealth model is capable of detecting changes in wealth over
time (Figure~\ref{satim:fig:corr_wealth_time}). However, this correlation is relatively weak, due to large measurement errors in the LSMS survey data and the predicted wealth index from satellite image acquisitions in comparison to the true change in wealth over just a few years. %
\\

Qualitative examples of changes in the wealth index over time are coherent with our intuition. Figure~\ref{satim:fig:time_poor_vs_rich} illustrates changes in wealth index over time at two different locations. Further examples are provided in Figure~\ref{satim:fig:poor_vs_rich}. Saliency maps in \cite{yeh2020using} suggest that the model appears to weight urban areas, farmland, water bodies, and desert terrain when making predictions, in support of the view that the index actually reflects material wealth. However, the complexity of the model makes it difficult to guarantee that the image characteristics learned by the model are relevant explanatory variables for predicting material wealth. In our context, an additional complication is that the model may learn to detect the mine itself and for example learn that mines are associated with higher wealth. This would lead to spurious predictions of wealth and might lead to overestimating the impact of mining. In contrast, the land use classes described in Section~\ref{satim:subsec_landusemine} omit the area of the mine. We exclude outliers using the same method as for land use and mine classes, also described in Section~\ref{satim:subsec_landusemine}.

\FloatBarrier

\section{Empirical Strategy} \label{satim:sec_empirical_strategy}

Our empirical strategy uses an event study approach with various choices of counterfactual groups in order to estimate the local impact of mine openings and closings. We measure three outcome variables: log urban land cover, log agricultural land cover and material wealth.

In Section~\ref{satim:sect_descriptives}, we provide a descriptive analysis of spatial and temporal variations in our main outcome variable - urban land cover share - across various mine activity statuses. Section~\ref{satim:sect_counterfact} discusses our choices of counterfactual groups used to estimate the causal impact of mine openings on local development. Finally, Section~\ref{satim:sec_ident} describes our stacked event study approach and difference-in-difference (DiD) model using control groups to estimate the causal effects driving the trends observed in our descriptive analysis.

\subsection{Descriptive Trend Analysis of Urban Land Cover Share} \label{satim:sect_descriptives}

To justify our empirical strategy, we identify key spatial and temporal variations in our main outcome variable: log urban land cover. We refer to App.~\ref{satim:app:additional_figures} for additional analyses on our other outcome variables: agricultural land cover and material wealth. Our method exploits these variations to estimate the impact of mine openings and closings on local development. \\

\begin{figure}[htb]
\caption{Cross-Sectional Comparisons of Urban Land Cover Share in Periods 1 \& 12}\label{satim:descr_crosssect}
\begin{subfigure}{0.5\hsize}\centering
    \caption{Urban Land Cover (Period 1)} \label{satim:descr_a}
    \includegraphics[width=0.9\hsize]{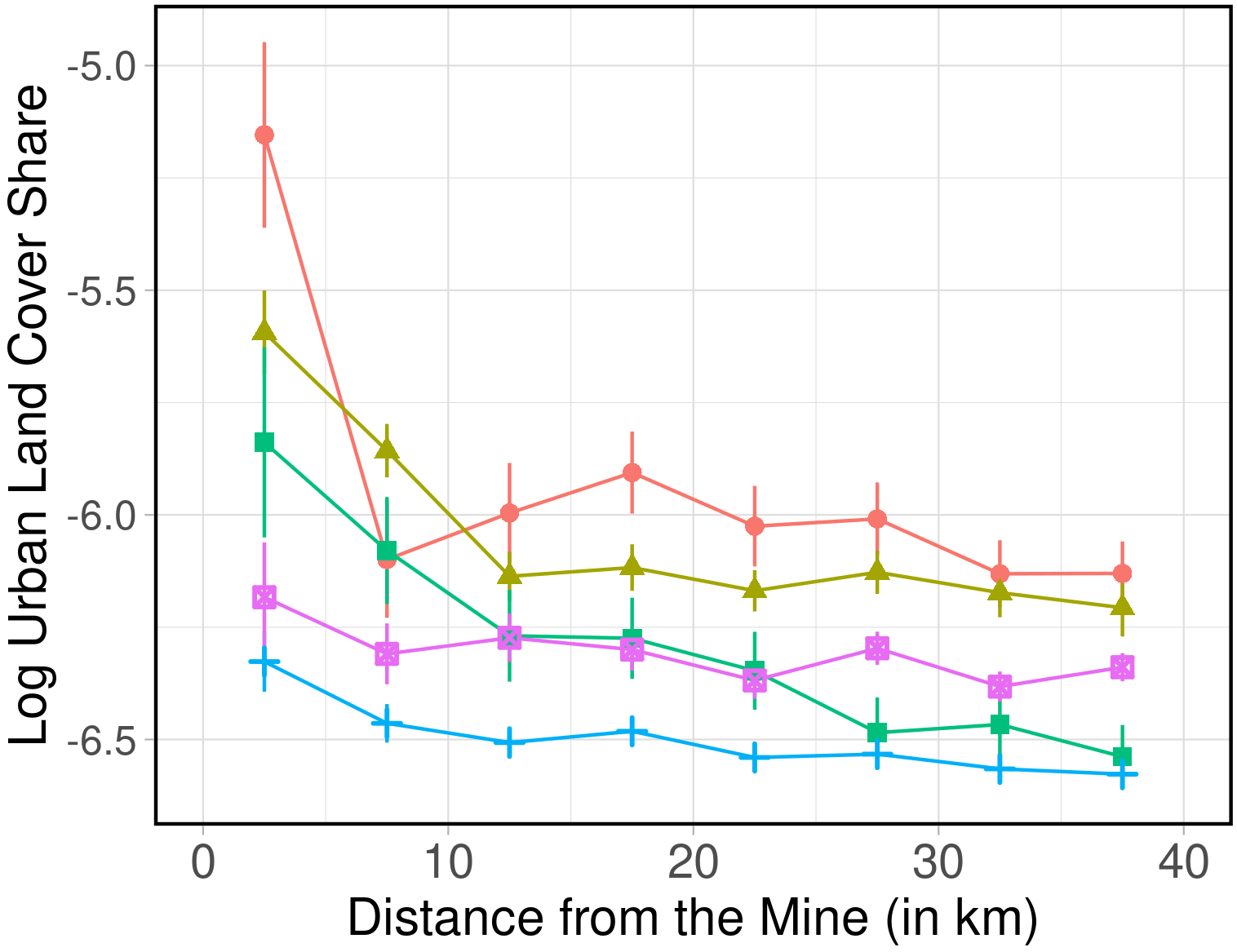}
\end{subfigure}
\begin{subfigure}{0.5\hsize}\centering
    \caption{Urban Land Cover (Period 12)} \label{satim:descr_b}
    \includegraphics[width=0.9\hsize]{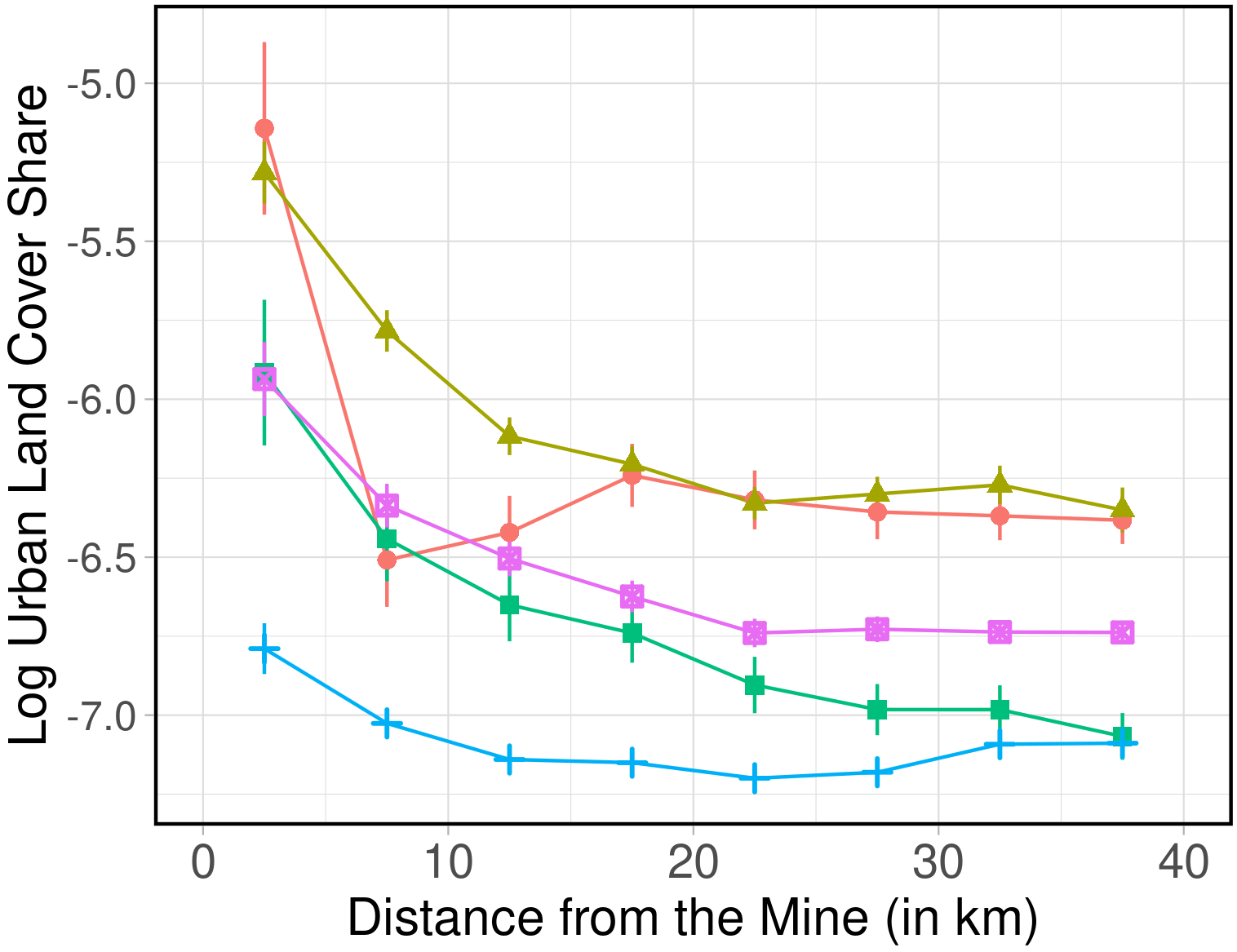}
\end{subfigure}
\begin{subfigure}{\hsize}\centering
\includegraphics[width=0.9\hsize]{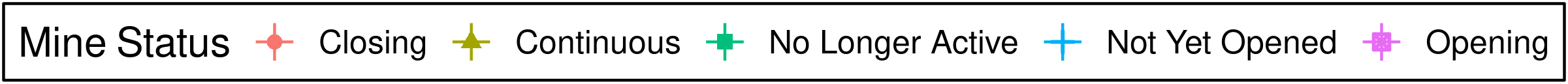}
\end{subfigure}
\begin{tablenotes}[flushleft]
\scriptsize \item \textit{Note:} 
Urban land cover tends to decrease with distance from the mine. Areas near a current or past active mine have higher urban land cover than areas which do not. Urban areas around \emph{Continuous} mines tend to grow between Period 1 and 12, relative
to mines that are \emph{Closing}.
\end{tablenotes}
\end{figure}

The graphs in Figure~\ref{satim:descr_crosssect} plot the urban land cover share in Period 1 (1984-86) and Period 12 (2017-19) at distances from the mine in 5km intervals.\footnote{The corresponding graphs for agricultural land use and material wealth index can be found in Figure~\ref{satim:descr_crosssect_append_agri} and Figure~\ref{satim:descr_crosssect_append_wealth}.} Mines with different activity statuses are displayed separately (see Figure~\ref{satim:mine_groups_overview} for an overview of different mine activity statuses). There are three striking patterns. Firstly, urban land cover is high in proximity to the mine and decreases with distance from the mine in areas where there is a currently active or previously active mine (\emph{Closing}, \emph{Continuous} and \emph{No Longer Active} mines at both Period 1 and Period 12 as well as \emph{Opening} mines at Period 12). This suggests that urban areas develop within a $\sim$20km radius surrounding mines. Secondly, areas with a mineral deposit but a mine that has not yet opened tend to have low urban land cover shares (\textit{Not Yet Opened} at both Periods 1 and 12 as well as \textit{Opening} at Period 1). Such areas can be used to estimate the level of urban development prior to mining. Thirdly, areas with active mining (\emph{Opening}, \emph{Closing}, and \emph{Continuous}) have higher urban growth between Periods 1 and 12 relative to areas without active mining (\textit{Not Yet Opened} and \textit{No Longer Active}), suggesting urban growth occurs during periods of active mining. \\

\vspace{-0.5cm}

\begin{figure}[H]
\caption{Log Urban Land Cover Share Relative to \emph{Not Yet Opened} Mines} \label{satim:demeaned_examples}
\centering
\begin{subfigure}{0.5\hsize}\centering
    \caption{\emph{Opening}} \label{satim:demeaned_open}
    \includegraphics[width=0.99\hsize]{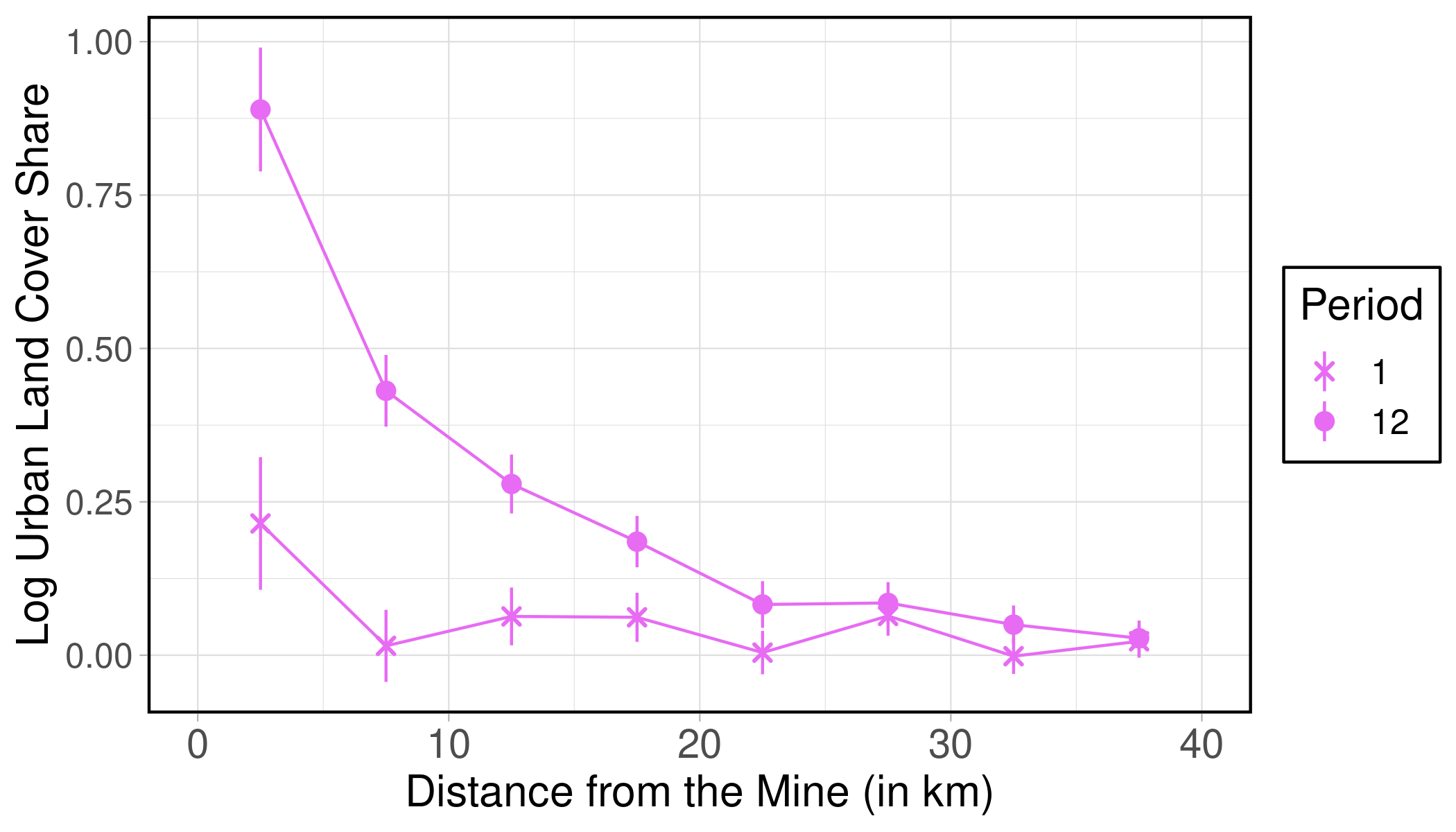}
\end{subfigure}\\
\hspace{0.3cm}
\begin{subfigure}{0.5\hsize}\centering
    \caption{\emph{Continuous}} \label{satim:demeaned_cont}
    \includegraphics[width=0.99\hsize]{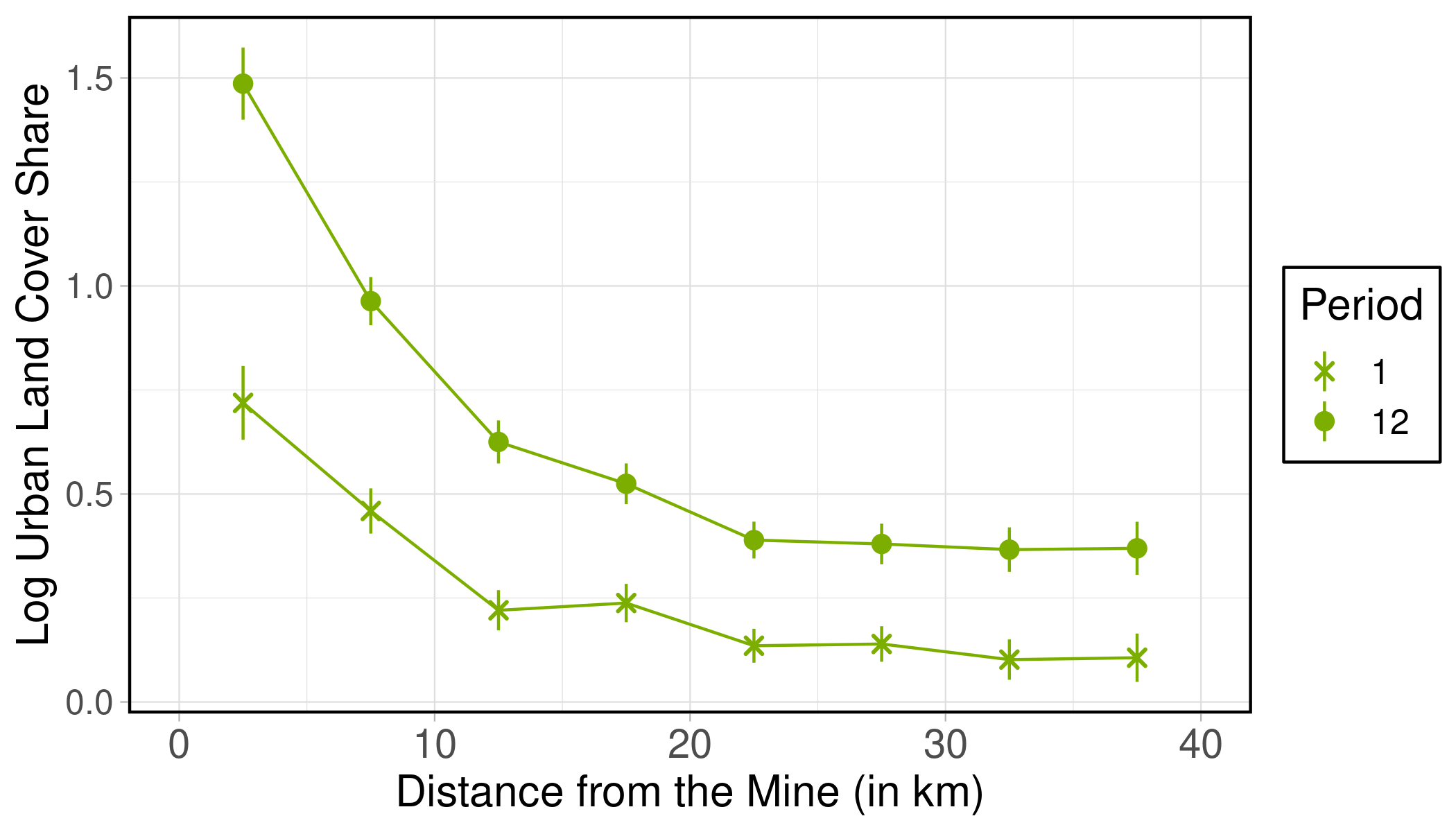}
\end{subfigure}
\hspace{0.3cm}
\begin{subfigure}{0.5\hsize}\centering
    \caption{\emph{Closing}} \label{satim:demeaned_close}
    \includegraphics[width=0.99\hsize]{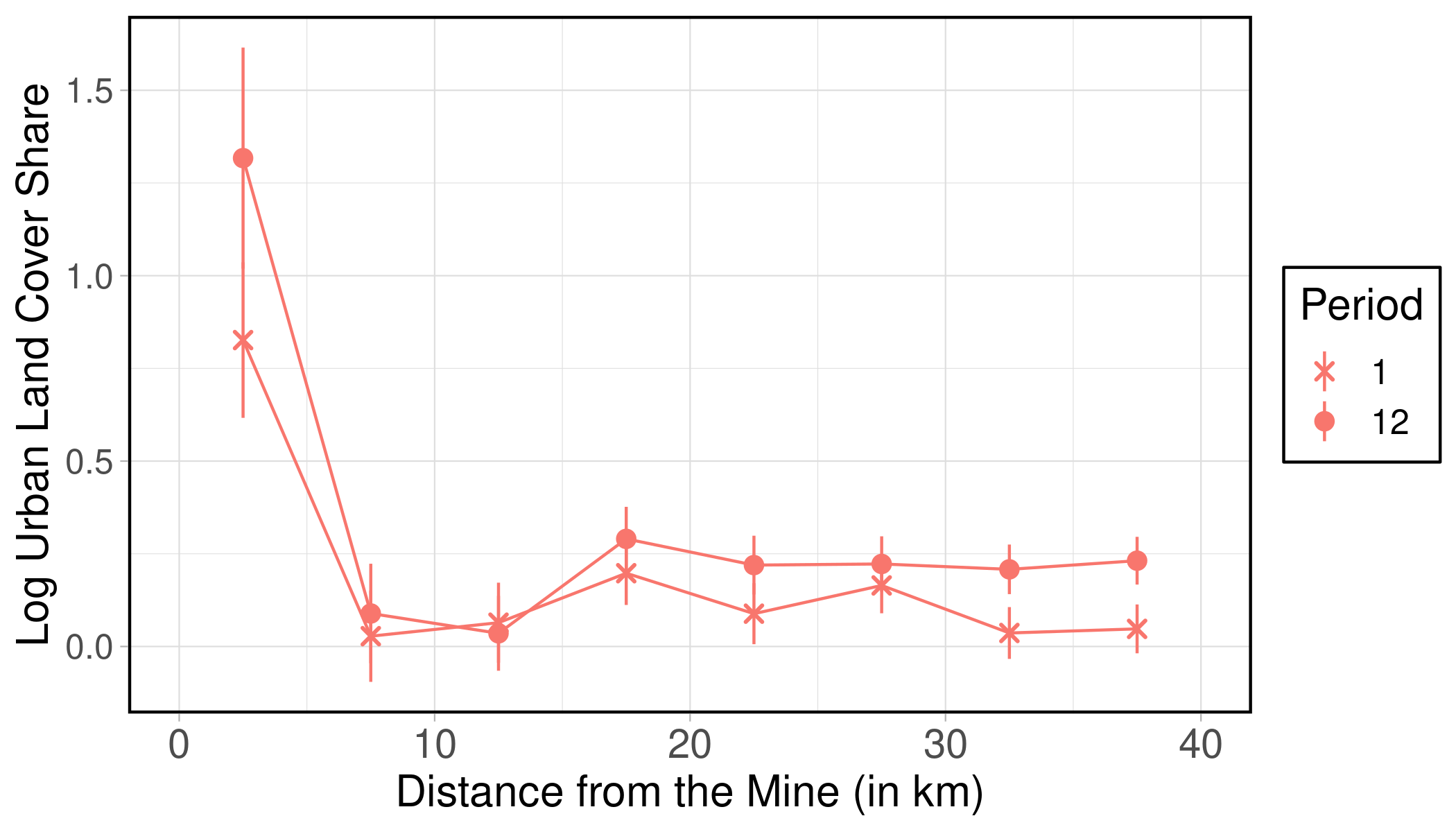}
\end{subfigure}
\begin{tablenotes}[flushleft]
\scriptsize \item \textit{Note:} The vertical axis is the log urban land cover share of the respective mine area categories, minus the country and period average urban land cover share of \emph{Not Yet Opened} mine areas. A relative log urban land cover share of~1 thus means the urban land cover share is $\textrm{exp}(1) \approx 2.7$ times the urban land cover share of \emph{Not Yet Opened} mine areas from the same country and period. \emph{Opening} mines have similar urban land cover shares as \emph{Not Yet Opened} mine areas in Period 1, but areas near the mine have comparatively large urban areas by Period 12. \emph{Closing} mines seem to stop growing between Periods 1 and 12, in comparison to \emph{Continuous} mines. 
\end{tablenotes}
\end{figure}

Assuming that areas around \emph{Not Yet Opened} mines would develop comparatively to areas around other mine categories, if not for the differences in mine activity status, we can estimate the impact of mine openings and closings on urban development by comparing \emph{Opening}, \emph{Continuous} and \emph{Closing} mines to \emph{Not Yet Opened} mines. We address the validity of this strong assumption in Section~\ref{satim:sect_counterfact}, but this simple preliminary analysis illustrates our empirical approach. In Figure~\ref{satim:demeaned_examples}, we subtract the average outcome in each country and in each period of the group of \emph{Not Yet Opened} mines to control for country-level and time fixed effects.\footnote{The equivalent graphs for our other two outcome variables - log agricultural land cover and material wealth - can be found in Figure~\ref{satim:demeaned_agri} and Figure~\ref{satim:demeaned_wealth}.} Simple calculations give a first estimate of the magnitude of the effects of mine openings and closings on urban development. \\

In Period 1, \emph{Opening} mine areas have fairly similar land cover shares to \emph{Not Yet Opened} mine areas. Figure~\ref{satim:demeaned_examples}a shows that the log urban land cover share relative to \emph{Not Yet Opened} mine areas is between 0 (equal land cover share) and 0.25 (1.3 times more urban land cover in \emph{Opening} mine areas). Between Period 1 and Period 12, the urban land cover share in areas near ($<20$km) \emph{Opening} mines increases to around 1.2-2.4 times\footnote{$\textrm{exp}(0.18) \approx 1.2$ and $\textrm{exp}(0.88) \approx 2.4$, where 0.18 and 0.88 are the minimum and maximum relative log urban land cover shares within 20km from the mine in Period 12, shown in Figure~\ref{satim:demeaned_examples}a.} the urban land cover of \emph{Not Yet Opened} mine areas. Areas within 0-5km an \emph{Opening} mine are on average 2.4 times more urban than areas within the same distance of \emph{Not Yet Opened} mines in Period 12, but were only 1.3 times more urban in Period 1, prior to the mine's opening. 

Whilst \emph{Closing} mines are similar to \emph{Continuous} mines in Period 1, \emph{Continuous} mines have higher urban growth until Period 12. In Period 1, areas near \emph{Continuous} mines have 1.3-2.1 times the urban land cover share of \emph{Not Yet Opened} mine areas, and areas near \emph{Closing} mines have 1-2.2 times the urban land cover share of \emph{Not Yet Opened} mines.\footnote{$\textrm{exp}(0.25) \approx 1.3$, $\textrm{exp}(0.75) \approx 2.1$, $\textrm{exp}(0) = 1$ and $\textrm{exp}(0.8) \approx 2.2$, where 0.5 and 1.5 are the minimum and maximum relative log urban land cover shares within 20km from the mine in Period 1, shown in Figure~\ref{satim:demeaned_examples}b, and 0 and 2.2 are the minimum and maximum relative log urban land cover shares within 20km from the mine in Period 1, shown in Figure~\ref{satim:demeaned_examples}c.} In Period 12, areas near \emph{Continuous} mines have 1.6-4.5 times the urban land cover share of \emph{Not Yet Opened} mine areas, but areas near \emph{Closing} mines have 1-3.5 times the urban land cover share of \emph{Not Yet Opened} mines.\footnote{$\textrm{exp}(0.5) \approx 1.6$, $\textrm{exp}(1.5) \approx 4.5$, $\textrm{exp}(0) = 1$ and $\textrm{exp}(1.25) \approx 3.5$, where 0.5 and 1.5 are the minimum and maximum relative log urban land cover shares within 20km from the mine in Period 12, shown in Figure~\ref{satim:demeaned_examples}b, and 0 and 1.25 are the minimum and maximum relative log urban land cover shares within 20km from the mine in Period 12, shown in Figure~\ref{satim:demeaned_examples}c.}  \\

These patterns suggest that mining has a positive economic impact on local communities in areas near the mine. During activity, mines appear to continuously induce urban growth. However, after a mine closes, the local area seems to stop growing at the same rate as during active mining. \\

Nevertheless, there are numerous reasons why \emph{Opening}, \emph{Continuous} and \emph{Closing} areas may not have evolved at the same rate \emph{Not Yet Opened} areas in the absence of mining. Mine openings might self-select into areas where in locations with more developed infrastructure due to reduced fixed costs and higher investor protection. Similarly, mine closings may be due to misgovernance or local conflict, which could explain decreased growth. For this reason, we require a more sophisticated identification strategy. %
In particular, we demonstrate parallel trends in the evolution of \textit{Opening} mine areas and \textit{Not Yet Opened} mine areas prior to active mining, and compare their divergence after the opening of the mine. We also compare the trend of \textit{Closing} mine areas prior and post closure with \textit{Not Yet Opened} or \textit{Continuous} mine areas in the same country and period.

\subsection{Choice of Counterfactual Groups} \label{satim:sect_counterfact}

In Section~\ref{satim:sect_descriptives}, we controlled for country and period fixed-effects using \emph{Not Yet Opened} mine areas, even though these different mine categories may not be valid counterfactuals for each other. Nevertheless, areas in the same country and period are similarly affected by spurious shocks, giving us an indication of how a treated group would have evolved in the absence of treatment and thereby facilitating the estimation of treatment effects. \\ 

\textit{Partially Active} mines are most suitable for understanding the impact of mine openings and closings, as we can track their evolution before and after treatment. Consequently, we use \textit{Opening} mines as the treatment group to assess the impact of mining onset, and \textit{Closing} mines as the treatment group to study what happens to mining areas after the closure of the mine.\footnote{Mines that are both \textit{Opening} and \textit{Closing} are also of interest, but we have dropped these observations as they usually change their status within short time intervals, making it difficult to separate the effects of different treatments of opening or closing.} \\

For \textit{Opening} mines, the most suitable counterfactual group are \textit{Not Yet Opened} deposits within the same country. The advantage for this comparison is that both groups are in the same state (inactive mine areas) at the beginning of the study period, but the former group undergoes treatment (a mine opens in the area). A potential pitfall of using \textit{Not Yet Opened} areas as controls is that they might not exhibit parallel trends due to being endogenously selected into their group. For example, investors might avoid certain mineral deposits due to an underdeveloped local economy. %
Yet, there are many potential reasons that determine whether or not a deposit is mined at a certain point in time, many of which are unrelated to local economic performance, such as time since discovery, support of local leadership, legal obstacles or global commodity demand. Our approach of dealing with this potential selection bias is twofold. Firstly, we make both groups more comparable by restricting our sample to recent discoveries during or after 1984. This means that deposits discovered a long time ago without an active mine - which we interpret as a signal of endogenous obstacles to mining - are dropped from our sample. Conducting balancing tests reveal that \textit{Opening} and \textit{Not Yet Opened} deposits are indeed similar regarding geographical characteristics and pre-treatment outcomes (see Figure~\ref{satim:fig:balancing_opening_not_yet}). The only significant difference between both groups is that \textit{Opening} mines exhibit around 12\% additional agricultural land use prior to treatment. Secondly, and more importantly, we will show that both groups exhibit parallel trends prior to treatment. \\

As an alternative to using \textit{Not Yet Opened} mines as a counterfactual, we could use mines that only start operating at a later point in time, and then compare the evolution of economic development in areas with active mines to that of areas that will host mines in the future - early vs. late (or future) treated \citep{Goodman-Bacon2021}. Using future treated units as controls for current treated areas is a relatively common approach in studies of the causal effect of location-based policy interventions - see for example \cite{Busso2013}. In the technical and institutional context of mining, there are various site-specific obstacles unrelated to the level of development prior to treatment, but which determine the precise timing of the mine opening. To support this argument empirically, we conduct pre-period balancing tests for \emph{Opening} mine areas, which indicate that the period of mine opening is unrelated to pre-treatment outcomes (see Figure \ref{satim:fig:balancing_opening_early_late}). Therefore, future mine sites are likely to constitute a valid counterfactual for current treated areas, and allow us to control for confounding factors when investigating the local economic impact of mining. A drawback of using future treated units as controls is that they might actually already experience gains in anticipation of treatment during the development stage of the mine. This would lead to underestimating the magnitude of the effects. For this reason, we interpret estimates using future treated units as controls as lower-bound estimates. \\

In addition to understanding how mine openings shape local communities, we aim to estimate the effect of the closure of the mine site - which is mostly triggered by random geological factors such as the exhaustion of the ore body. Since \emph{Closing} mines experience two treatments - becoming an active mine and closing - it is hard to find a valid counterfactual. Nevertheless, we can compare the evolution of local economic indicators in \emph{Closing} mine areas to \textit{Not Yet Opened} and \textit{Continuous} mine areas. The corresponding balancing tests in Figure~\ref{satim:fig:balancing_closing_not_yet} and Figure~\ref{satim:fig:balancing_closing_contin} show how \textit{Closing} mine areas compare to \textit{Not Yet Opened} and \textit{Continuous} mine areas respectively. \emph{Closing} mine areas tend to be at a higher elevation, further distance from a capital city, and with lower mean and minimum temperatures than \emph{Not Yet Opened} mine areas. Nevertheless, outcome variables on urban and agricultural land use, as well as material wealth, are not significantly different in Period 1. \emph{Closing} and \emph{Continuous} mine areas have similar characteristics across all variables except for the material wealth index, where \emph{Closing} mines have 20\% lower material wealth than \emph{Continuous} mines in Period 1. \\

\emph{Closing} mine areas are follow a higher economic growth trajectory prior to closure than \emph{Not Yet Opened} mine areas, as suggested by Figure~\ref{satim:demeaned_examples}, thus invalidating the parallel-trends assumption. Nevertheless, it is informative to compare their relative evolution over time in order to identify potential trend breaks induced by the mine closure. \emph{Closing} mine areas may follow a different growth trajectory to \emph{Continuous} mine areas due to other characteristics such as high-intensity extraction, which may contribute to both high growth and faster closure. Similarly, we can nonetheless estimate whether or not the relative growth trends of \emph{Closing} mine areas and \emph{Continuous} mine areas change after the mine's closure. \\

We compare mine areas before and after treatment to the aforementioned reference groups based on mine status, timing of treatment or distance to the mine, by using an event study approach discussed in the next section. 

\subsection{Identification Using an Event Study and DiD Estimations} \label{satim:sec_ident}

Our main method is a stacked event study approach using control groups to obtain
causal estimates. We observe three outcome variables $Y$: Log Urban Land Cover, Log Agricultural Land Cover and Material Wealth Index ($z$-score). Over a time span of $T_{neg}$ periods before until $T_{pos}$ periods after treatment, we track the development of the treatment group against a control group to filter out unrelated shocks. This approach allows us to test the underlying identifying assumption that both groups evolved similarly prior to the treatment onset (parallel pre-trends). We can then measure the divergence between the two groups to estimate the impact of treatment. \\

\vspace{-1.1cm}

\begin{alignat}{1} \label{satim:eq_event_study}
Y_{i,t}  =  \: \sum_{\substack{t=T_{neg},\\t\neq0}}^{T_{pos}} \beta_{t} * D_{t} * Treat_{i} + b_{e,t} + b_{e, i} + \varepsilon_{i,t}
\end{alignat}
\vspace{-0.5cm}

Equation~\ref{satim:eq_event_study} formulates the event study equation and we estimate the $\beta_t$ coefficients: $i$ refers to a tile, $t$ to relative 3-year periods between $T_{neg}$ and $T_{pos}$, with Period 1 corresponding to the first period after treatment (e.g. a mine opening or closing), $e$ to an event (e.g. mine openings or closings in a given country and period belong to one `event'), $D_{t}$ represents period dummies, $Treat_{i}$ is a dummy for the treatment group, $b_{e,t}$ are binary vectors of event $\times$ period fixed effects, $b_{e, i}$ are binary vectors of event $\times$ tile fixed effects, and $\varepsilon_{i,t}$ are error terms. In the case of parallel trends prior to treatment, we expect $\beta_{t} = 0$ for $t\in \{T_{neg}, \dots, -1\}$. The coefficient $\beta_{t}$ for $t>0$ estimates the effect at $t$ periods post-treatment. We keep only events for which we have observations over the entire interval $\{T_{neg}, \dots, T_{pos}\}$. This is necessary to ensure that time specific $\beta$-coefficients are not sensitive to distortions caused by varying compositions of the treatment group, due to treatment heterogeneities between mines. \\

Furthermore, we use ordinary DiD and stacked DiD estimations to complement our event study. We estimate $\beta$ in Equation~\ref{satim:eq_did}, where $i$ refers to a tile, $p$ to a period, $Treat_{i,p}$ is a dummy for the treatment group, $b_{e,p}$ are binary vectors of event $\times$ period fixed effects, $b_{e, i}$ are binary vectors of event $\times$ tile fixed effects, and $\varepsilon_{i,t}$ are error terms. This is similar to Equation~\ref{satim:eq_event_study}, but without centering time around the event or estimating multiple coefficients for relative time periods. Instead, we estimate the treatment effect based on all units and time periods $p$ after treatment using all observations, instead of only those that are observed over the entire interval $\{T_{neg}, \dots, T_{pos}\}$. Unlike in \ref{satim:eq_event_study}, the treatment dummy $Treat_{i,p}$ depends on the period $p$, and is equal to 1 for a tile in the treatment group (e.g. \emph{Opening} or \emph{Closing} mine areas) after treatment (e.g. after the neighboring mine opens or closes). \\

\vspace{-1.1cm}

\begin{alignat}{1} \label{satim:eq_did}
Y_{i,p}  =  \: \beta * Treat_{i,p} + b_{e,p} + b_{e, i} + \varepsilon_{i,p}
\end{alignat}
\vspace{-0.5cm}

Recent research discusses the biases that might arise in ordinary DiD models, in particular in the presence of treatment heterogeneity and when treatment has a staggered onset over time \citep{Chaisemartin2020, SantAnna2020, Goodman-Bacon2021}. For this reason, we use stacked DiD estimation to ensure that earlier-treated units are not used as counterfactuals for late-treated units after treatment \citep{Cengiz2019}. To do this, we define \textit{events} as mine openings or closings within one period within one country. Each event is associated to control units within the same country and all events are stacked together. As a result, each event has its own control group, which prevents implicit comparisons of late vs. early-treated due to overlapping fixed effects. Since this also means that observations in the control group can be duplicated, in addition to clustering standard errors at the mine level, we also cluster standard errors at the tile level. \\

To deepen our understanding of the impact of mine openings and closings, we look for heterogeneities depending on the size of the mine, and the institutional context. In order to do this, we replace the term $\beta Treat_{i,p}$ in Equation~\ref{satim:eq_did} by $\sum_c \beta_{c} * D_{i,c} * Treat_{i,p}$, where $D_{i,c}$ is a dummy variable equal to 1 if tile $i$ is of category $c$, where $c$ is the size category of the mine, whether the area is near or far from the mine, or whether or not the country is a democracy or an autocracy. To better understand the mechanism behind the influence of institutional context on mine-related growth, we also look at the probability of conflict in areas with \emph{Opening} mines under democratic and autocratic governments.

\section{Impact of Mine Openings and Closings} \label{satim:sect_results}

\subsection{Mine Openings} \label{satim:sect_event_study}

In order to evaluate the impact of mine openings, we implement the identification strategy described in Section~\ref{satim:sec_empirical_strategy}. %
Our results show that a mine opening sets areas near the mine on a high economic growth trajectory, compared to areas further from the mine, as well as control areas with a mineral discovery but no active mine. Using later mine openings as a control group for earlier mine openings gives similar results. \\

\begin{figure}[H]
\caption{Event Study: \textit{Opening} vs. \textit{Not Yet Opened} mines} \label{satim:event_study_graph}
\begin{subfigure}{0.5\hsize}\centering
    \caption{Near: Log Urban Area Share}
    \includegraphics[width=0.99\hsize]{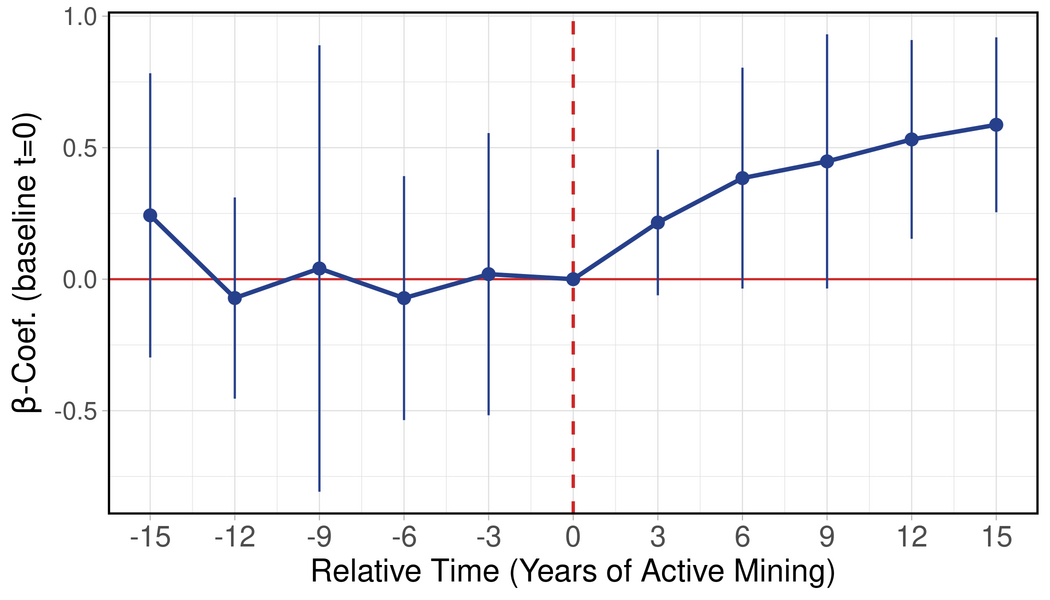}
    \label{satim:event_study_graph:near}
\end{subfigure}
\begin{subfigure}{0.5\hsize}\centering
    \caption{Far: Log Urban Area Share}
    \includegraphics[width=0.99\hsize]{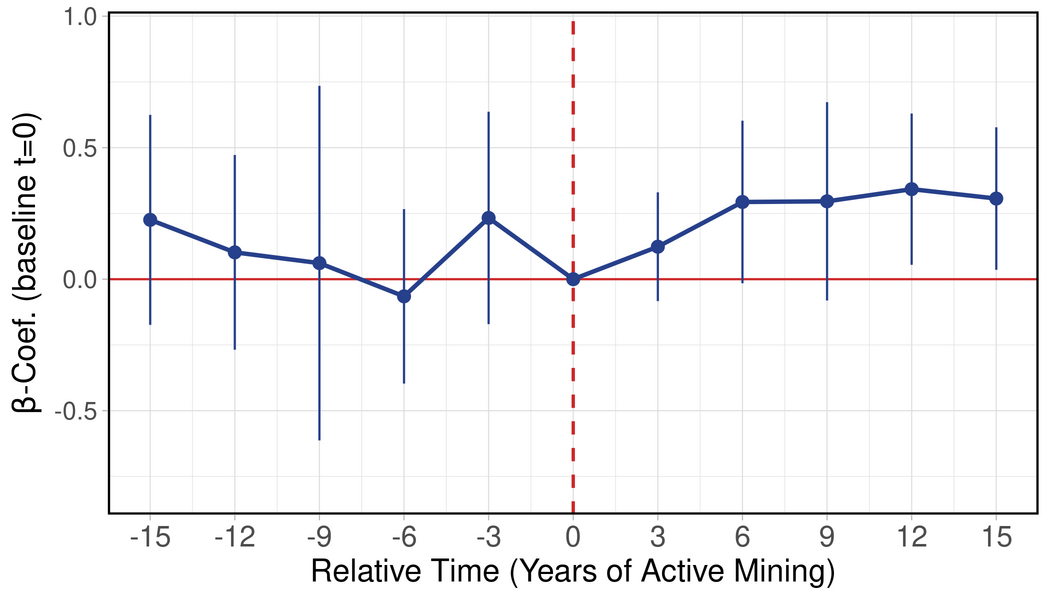}
    \label{satim:event_study_graph:far}
\end{subfigure}
\begin{subfigure}{0.5\hsize}\centering
    \caption{Near: Log Agricultural Area Share}
    \includegraphics[width=0.99\hsize]{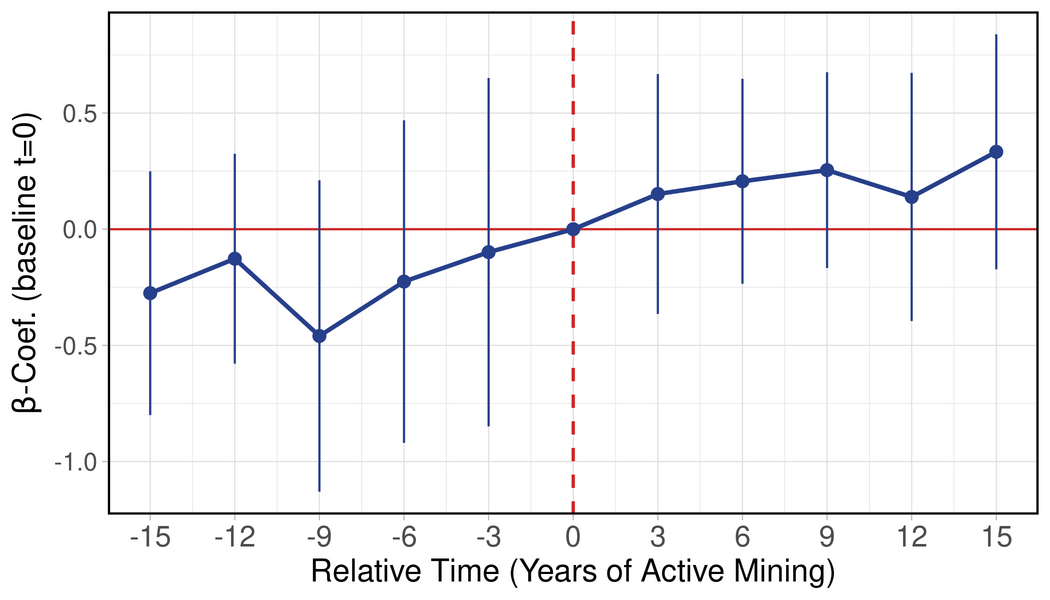}
    \label{satim:event_study_graph:near:agri}
\end{subfigure}
\begin{subfigure}{0.5\hsize}\centering
    \caption{Far: Log Agricultural Area Share}
    \includegraphics[width=0.99\hsize]{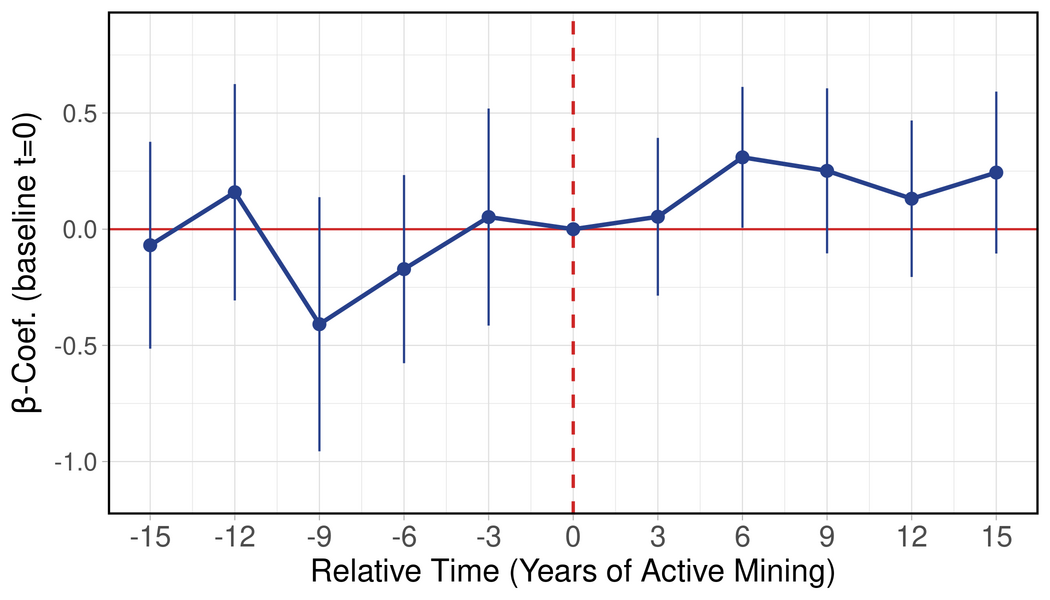}
    \label{satim:event_study_graph:far:agri}
\end{subfigure}
\begin{subfigure}{0.5\hsize}\centering
    \caption{Near: Material Wealth Index}
    \includegraphics[width=0.99\hsize]{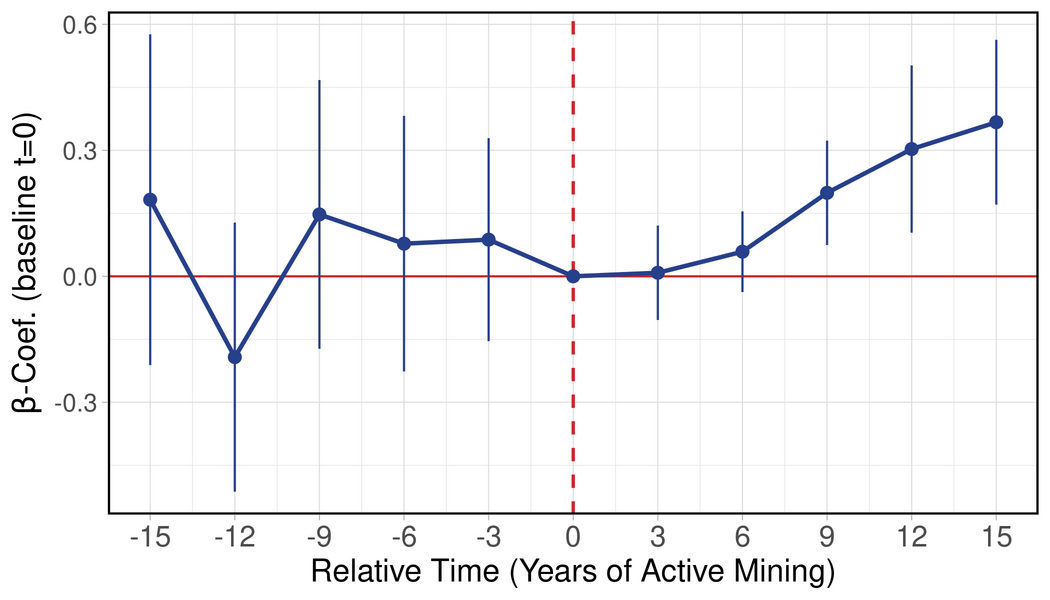}
    \label{satim:event_study_graph:near:wealth}
\end{subfigure}
\begin{subfigure}{0.5\hsize}\centering
    \caption{Far: Material Wealth Index}
    \includegraphics[width=0.99\hsize]{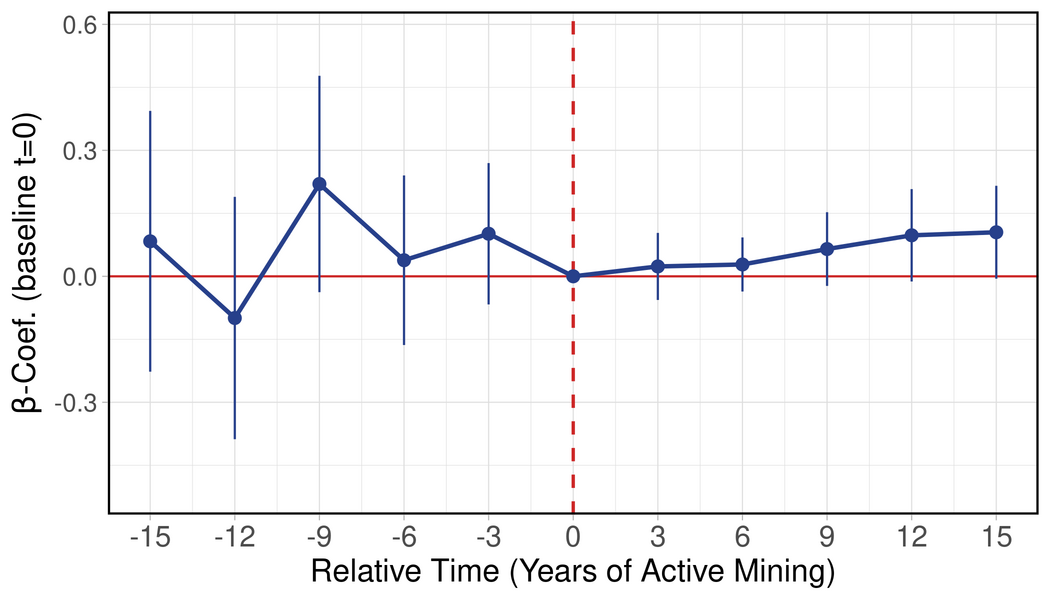}
    \label{satim:event_study_graph:far:wealth}
\end{subfigure}
\begin{tablenotes}[flushleft]
\scriptsize \item \textit{Note:} In the years prior to the mine opening, areas near the mine (\textit{Opening}) follow the similar trends to areas with a mineral discovery but no active mine (\textit{Not Yet Opened}). In the periods after the start of active mining, these two groups diverge in terms of log urban area share and material wealth index, particularly in areas near the mine ($<20$km). A mine opening sets areas near the mine on a high growth trajectory. A similar but less pronounced trend can be seen for areas further from the mine (20-40km). Although there are no significant divergence in log agricultural area share before and after $t=0$, agricultural activity does appear to grow in areas near and far to the mine after its opening. Error bars represent a 95\% confidence interval. 
\end{tablenotes}
\end{figure}

Our event study estimates the three outcome variables (urban land cover, agricultural land cover and material wealth) both near and far from \emph{Opening} mines, relative to that of \emph{Not Yet Opened} mine areas. Figure~\ref{satim:event_study_graph} plots the $\beta_{t}$-coefficients of in Equation \ref{satim:eq_event_study} over relative 3-year time periods $t$ between $T_{neg}=-5$ and $T_{pos}=5$. Relative period $t=1$ is the first period of active mining and hence $t=0$ serves as the baseline period, just before active mining. The $\beta_t$-coefficients measure the extent to which \emph{Opening} mine areas have greater ($\beta_t > 0$) or lesser (if $\beta_t < 0$) values of the outcome variable relative to \emph{Not Yet Opened} mine areas at the baseline period $t=0$. We restrict the observations to areas near the mine or areas far from the mine respectively. \\

The main concern of using \textit{Not Yet Opened} (future/never-treated) mine areas as control group for \textit{Opening} mine areas is that they may already be diverging prior to treatment, and therefore not constitute valid counterfactuals. However, the event study results in Figure~\ref{satim:event_study_graph} indicate that there are no significant differences between the evolution of the two groups prior to the onset of mining ($-5 \leq t < 0$), i.e. in the 15 years prior to the mine opening.

As soon as active mining starts at relative period $t=1$, treated units in areas near the mine significantly diverge from control areas in log urban area share and material wealth index (Figure~\ref{satim:event_study_graph}). Growth in urban areas and in material wealth is not only observed in the first period after the mine opening - the local area continues to develop relative to \emph{Not Yet Opened} mine areas at each successive period. \emph{Opening} mine areas far away from the mine also diverge from \emph{Not Yet Opened} mine areas after the onset of mining, but to a lesser extent. The fact that the divergence exactly coincides with the launch of the mine operation and that the magnitude of the divergence diminishes with distance from the mine strongly suggests that it is the mine activity that boosts the local economy, rather than other changes. \\

The magnitude of the effects is considerable. After 15 years, areas near \textit{Opening} mines gain on average around 80\%\footnote{We compute the effect based on the coefficient $\beta_5$ in Figure~\ref{satim:event_study_graph}: $\textrm{exp}(0.59) - 1 \approx 0.8$.} in urban area extent relative to \textit{Not Yet Opened} mine areas.
This result is consistent with our simple calculation in Section~\ref{satim:sec_empirical_strategy}.\ref{satim:sect_descriptives}, where we find that between Period 1 and Period 12, the urban land cover share in areas surrounding \emph{Opening} mines grows to up to 140\% more than in \emph{Not Yet Opened} mine areas. The material wealth index increases by around 0.4 standard deviations in areas near the mine after 15 years. \\ 

On the other hand, we do not observe significant divergence in agriculture land cover between \emph{Opening} mine areas and \emph{Not Yet Opened} mine areas in Figure~\ref{satim:event_study_graph}. Nevertheless, the share of agricultural areas does appear to be greater after the mine opening than prior to the mine opening, although this difference is not significant. \\

Table~\ref{satim:table_panels} shows similar results to our event study by using ordinary and stacked DiD models, as well as using \emph{Not Yet Opened} deposit areas as a control group for \emph{Opening} mine areas or \emph{Late Treated} units as a control group for \emph{Early Treated} mine areas (i.e. areas that are treated during our study period serve as controls prior to their treatment onset). These results support our event study results and indicate that mining does have a considerable positive impact on urban agglomeration (Panel A) and material wealth (Panel C) and that these effects are primarily relevant for areas near the mine. Comparing \emph{Opening} to \emph{Not Yet Opened} mine ares, we find that urban land cover increases by around 22-24\% in areas near the mine due to mining, where this figure represents the average increase of all such areas irrespective of the number of years since the mine opening. In comparison, our event study shows that urban area gains from mining increase with time and that 15 years after the mine opening, areas near mines gain on average around 80\%. Using \emph{Late Treated} units as a control group gives an estimate of 17\%, which is a lower bound estimate as future treated control units might already experience some of the gains of mining during their development stage. \\

\begin{table}[H] \centering 
  \caption{DiD Regressions: Mine Openings} 
  \label{satim:table_panels}

\resizebox{\textwidth}{!}{\begin{tabular}{@{\extracolsep{19pt}}lcc|cc|cc} 
\\[-1.8ex]\hline 
 
 \addlinespace[0.15cm]                    
                        &\multicolumn{2}{c}{Ordinary DiD}&\multicolumn{4}{c}{Stacked DiD}\\[-.5ex] 
                        \cmidrule(r{5pt}){2-3}  \cmidrule(l{5pt}){4-7} \\[-1.2em]
                        
                        &\multicolumn{4}{c}{Opening vs.}        &   \multicolumn{2}{c}{Early vs. Late}\\[-.5ex] 
                        &\multicolumn{4}{c}{Not Yet Opened}     &   \multicolumn{2}{c}{Treated}\\
                        \cmidrule(r{5pt}){2-5}  \cmidrule(l{5pt}){6-7} \\[-1.2em]
			&\multicolumn{1}{c}{(1)}&\multicolumn{1}{c}{(2)}&\multicolumn{1}{c}{(3)}&\multicolumn{1}{c}{(4)}&\multicolumn{1}{c}{(5)}&\multicolumn{1}{c}{(6)}\\[.4em]
\hline
\hline \\[-2.2ex] 
\\[-2.0ex] \multicolumn{7}{@{} l}{\underline{Panel A: Log Urban Area Share}}
 \\
 \\[-1.5ex]
Treatment Dummy  & 0.093$^{*}$ & - & 0.111$^{*}$ & - & 0.068   & - \\ 
                 & (0.051)     &   & (0.063)     &   & (0.064) &  \\ 
  & & & & & & \\[-1.8ex]  
Treatment $\times$ Near & - & 0.223$^{***}$ & - & 0.243$^{***}$ & - & 0.173$^{**}$ \\ 
                         &   & (0.061)       &   & (0.072)       &   & (0.073) \\ 
  & & & & & & \\[-1.8ex]  
Treatment $\times$ Far & - & 0.048   & - & 0.066   & - & 0.031 \\ 
                       &   & (0.050) &   & (0.062) &   & (0.063) \\ 
\\[-1.83ex] 
 \hline \\[-1.83ex]
\\[-2.0ex] \multicolumn{7}{@{} l}{\underline{Panel B: Log Agriculture Area Share}}
 \\
 \\[-1.5ex]
Treatment Dummy  & 0.127$^{*}$ & - & 0.121   & - & $-$0.004 &  - \\ 
                 & (0.067)     &   & (0.084) &   & (0.083) &  \\ 
  & & & & & & \\[-1.8ex]  
Treatment $\times$ Near & - & 0.182$^{**}$ & - & 0.173$^{*}$ & - & 0.011 \\ 
                         &   & (0.078)      &   & (0.094)     &   & (0.091) \\ 
  & & & & & & \\[-1.8ex]  
Treatment $\times$ Far & - & 0.108   & - & 0.103   & - & $-$0.009 \\ 
                       &   & (0.066) &   & (0.082) &   & (0.083) \\ 
\\[-1.83ex] 
 \hline \\[-1.83ex]
\\[-2.0ex] \multicolumn{7}{@{} l}{\underline{Panel C: Material Wealth Index (z-score)}}
 \\
 \\[-1.5ex]
Treatment Dummy  & 0.028   & - & 0.028   & - & 0.026 &  - \\ 
                 & (0.031) &   & (0.036) &   & (0.048) &  \\ 
  & & & & & & \\[-1.8ex]  
Treatment $\times$ Near  & - & 0.105$^{***}$ & - & 0.109$^{***}$ & - & 0.097$^{*}$ \\ 
                          &   & (0.035)       &   & (0.041)       &   & (0.051) \\ 
  & & & & & & \\[-1.8ex]  
Treatment $\times$ Far & - & 0.001   & - & 0.001   & - & 0.001 \\ 
                       &   & (0.030) &   & (0.035) &   & (0.048) \\ 
\hline 
\hline \\[-1.8ex] 
Country $\times$ Period FE &      Yes     &    Yes     &    -      &    -      &    -      &    -     \\
Tile FE                    &      Yes     &    Yes     &    -      &    -      &    -      &    -     \\
Event $\times$ Period FE   &      -       &    -       &    Yes    &    Yes    &    Yes    &    Yes     \\
Tile  $\times$ Event FE    &      -       &    -       &    Yes    &    Yes    &    Yes    &    Yes     \\
\hline \\[-1.8ex] 
Observations & 400,239 & 400,239 & 1,297,985 & 1,297,985 & 626,035 & 626,035 \\ 
\hline 
\hline \\[-1.8ex]
\end{tabular}}
\begin{tablenotes}[flushleft]
\scriptsize \item \textit{Note:} This table reports stacked DiD estimations based on Equation \ref{satim:eq_did}. The `Treatment Dummy' (or `Treatment') indicates if a tile's corresponding mine has started operating, it is always 0 for tiles in the control group. `Near' and `Far' in the interaction terms correspond to dummies indicating if a tile is within 20km from the mine or between 20km and 40km from the mine. The coefficients are interpreted as the mean growth of the outcome variable. Standard errors in parenthesis are clustered by mine in columns (1)-(2), and double-clustered by mine and tile in columns (3)-(6).  \hspace{4.1cm} $^{*}$p$<$0.1; $^{**}$p$<$0.05; $^{***}$p$<$0.01
\end{tablenotes}
\end{table}

We estimate that mining increases the material wealth index by around 0.1 standard deviations in areas near the mine, as shown in Table~\ref{satim:table_panels} Panel C. However, in contrast to the urban and agricultural land cover outcomes, we do not exclude the area of the mine when computing the material wealth index, which potentially confounds estimation of the development of the surrounding area with development of the mine itself. \\ %

The results in Table~\ref{satim:table_panels} Panel B indicate that mining also increases the proportion of agricultural fields in areas near the mine by around 17-18\% when using \emph{Not Yet Opened} mines areas to control for \emph{Opening} mine areas. This result is significant at the 95\%-level when using ordinary DiD estimation and at the 90\% level when using a stacked DiD estimation. This finding is in line with previous studies showing that mining stimulates the local economy through backward linkages \citep{Aragon2013}.\footnote{Mining investments might also have a positive impact on other local firms via knowledge spillovers (see for example \cite{Ghebrihiwet2019} for recent evidence on foreign direct investment (FDI) in South Africa's mining sector, or \cite{Abebe2022forth} for recent evidence on FDI in Ethiopia's manufacturing sector.) However, our dataset is not suitable to investigate such spillovers.} However, these results are of a lower magnitude and less robust when using early vs. late treated units. This is likely related to the fact that these estimates represent lower bounds. Our event study also does not show a significant increase in agriculture at the 95\%-level, however, there does seem to be an upward trend after the mine opening.

\subsection{Heterogeneities} \label{satim:subsect_hetero}

The results in the previous section demonstrate that mine openings lead to economic development around the mine. In this section, we show that the magnitude of this growth varies with mine size and institutional context. Previous research in \cite{Mamo2019}, \cite{Pokorny2019} and \citet{Bazillier2020} find that local economic outcomes depend on mine size, and hence we investigate heterogeneities between small and large mines.\footnote{See Section~\ref{satim:data} for the definition of small and large mines in our sample.} 
Moreover, research at the macro level finds evidence for the `political resource curse' i.e. natural resources are only advantageous in countries with good institutions \citep{Mehlum2006, Bhattacharyya2010}, thus we test if this relationship also applies at the micro level to mining areas. Based on the average Polity2 score\footnote{See \url{www.systemicpeace.org/polityproject.html} for more information about the Polity project by the Center for Systemic Peace.} during our study period, we categorize countries as `democratic' if they have a score greater than 0, or `autocratic' if they have a score less than 0. \\

Table~\ref{satim:table_2} presents the regression results predicting urban area share using Equation~\ref{satim:eq_did} with interaction terms between the treatment variable, mine size, and whether or not the country is a democracy or autocracy. The corresponding results for the agriculture land use and the material wealth index are similar and can be found in Table~\ref{satim:table_hetero_size_append}. The split between large vs. small and democratic vs. autocratic observations in the treatment group is relatively even: treatment group tiles around large mines represent 39\% of the tiles and 43\% of all treated tiles fall in democratic countries. In autocratic countries, there is no significant change in urban development in areas near small mines after production starts. There is even a negative urbanization trend in areas between 20-40km to the mine, relative to similar \emph{Not Yet Opened} mine areas. On the other hand, large mines in autocratic countries boost urban development both in areas near and and far to opening mines. Growth due to mining activity is significantly stronger in democracies than in autocracies. Small mines in democratic countries lead to a 50\% greater increase in urban area share near the mine than small mines in autocratic countries. Large mines in democratic countries result in a 7\% greater increase in urban area share near the mine than in large mines in autocratic countries. \\

\vspace{-0.5cm}
\begin{table}[H]
\centering 
  \caption{Stacked DiD Regressions in Areas Near the Mine - Heterogeneities} 
  \label{satim:table_2} 
\begin{adjustbox}{width=.9\textwidth,center}
\begin{tabular} {@{\extracolsep{15pt}}lcc|cc} 
\\[-1.8ex]\hline 
\hline \\[-1.8ex] 
 & \multicolumn{4}{c}{Stacked DiD Regressions: \textit{Log Urban Area Share}} \\
\cline{2-5}\\
&\multicolumn{2}{c}{Near}&\multicolumn{2}{c}{Far}\\[-.5ex] 
\cmidrule(r{5pt}){2-3}  \cmidrule(l{5pt}){4-5} \\[-1.2em] 
\\[-1.8ex] & (1) & (2) & (3) & (4) \\ 
\hline \\[-1.8ex] 
 \\[-1.5ex]
Treatment Dummy                         & $0.23^{***}$ & $-0.03$      & $0.07$    & $-0.21^{**}$ \\
                                        & $(0.08)$     & $(0.12)$     & $(0.06)$  & $(0.09)$     \\
& & & & \\[-1.8ex] 
Treat $\times$ Large Mine               &      -       & $0.29^{*}$   &     -     & $0.26^{**}$  \\
                                        &              & $(0.17)$     &           & $(0.12)$     \\
& & & & \\[-1.8ex] 
Treat $\times$ Democracy                &      -       & $0.53^{***}$ &     -     & $0.60^{***}$ \\
                                        &              & $(0.16)$     &           & $(0.13)$     \\
& & & & \\[-1.8ex] 
Treat $\times$ Large $\times$ Democracy &      -       & $-0.46^{*}$  &     -     & $-0.40^{**}$ \\
                                        &              & $(0.27)$     &           & $(0.19)$     \\   
& & & & \\[-1.8ex]                                
\hline
\addlinespace[0.1cm] 
Event x Tile FE                    & Yes          & Yes          & Yes         & Yes          \\
Event x Country x Period FE        & Yes          & Yes          & Yes         & Yes          \\
\hline
\addlinespace[0.1cm] 
Observations                       & $1,172,561$  & $1,172,561$  & $1,254,498$ & $1,254,498$    \\
Adj. R$^2$                        & $0.76$       & $0.76$       & $0.76$      & $0.77$       \\
\hline 
\hline \\[-1.8ex]
\end{tabular}
\end{adjustbox}
\begin{tablenotes}[flushleft]
\scriptsize \item \textit{Note:} This table reports stacked DiD heterogeneity tests based on Equation \ref{satim:eq_did}. The `Treatment Dummy' (or `Treat') indicates if a tile's corresponding mine has started operating, it is always 0 for tiles in the control group. In Columns (1) and (2), we restrict the treatment group to tiles within 20km from the mine, and in Columns (3) and (4) to tiles between 20km and 40km from the mine. Standard errors in parenthesis are double-clustered by mine and tile. \hfill  $^{*}$p$<$0.1; $^{**}$p$<$0.05; $^{***}$p$<$0.01
\end{tablenotes}
\end{table}

In Table~\ref{satim:tabel_regs_institutions} we compare the characteristic of being a democracy to other measures of institutions as well as having a relatively high GDP per capita. We find that democracy has a stronger effect than other institutional measures, and that GDP does not appear to be relevant to the impact of mining. 

\begin{table}[H]
\begin{center}
\caption{Heterogeneities: Country Characteristics} \label{satim:tabel_regs_institutions}
\resizebox{\textwidth}{!}{\begin{tabular}{l c c c c c}
\hline
\\[-0.9em]
&\multicolumn{5}{c}{\textit{Dependent Variable:} Log Urban Landcover Share}\\
\hline
\\[-0.9em]
 & Democracy & Voice \& Account. & Rule of Law & Decentralization & GDP \\
\hline
\\[-0.9em]
Treat                              & $0.08$      & $0.04$      & $0.09$     & $0.20$    & $0.24^{**}$ \\
                                   & $(0.10)$    & $(0.13)$    & $(0.11)$   & $(0.13)$  & $(0.09)$    \\
& & & & & \\[-1.8ex] 
Treat $\times$ Democracy           & $0.36^{**}$ &     -       &    -       &    -      &     -       \\
                                   & $(0.15)$    &             &            &           &             \\
& & & & & \\[-1.8ex] 
Treat $\times$ High Participation  &      -      & $0.33^{**}$ &    -       &    -      &     -       \\
                                   &             & $(0.16)$    &            &           &             \\
& & & & & \\[-1.8ex] 
Treat $\times$ High ROL            &      -      &     -       & $0.27^{*}$ &    -      &     -       \\
                                   &             &             & $(0.15)$   &           &             \\
& & & & & \\[-1.8ex] 
Treat $\times$ Decentralized       &      -      &     -       &     -      & $0.07$    &     -       \\
                                   &             &             &            & $(0.16)$  &             \\
& & & & & \\[-1.8ex] 
Treat $\times$ High GDP            &      -      &     -       &     -      &     -     & $-0.03$     \\
                                   &             &             &            &           & $(0.16)$    \\
& & & & & \\[-1.8ex] 
\hline
\addlinespace[0.1cm] 
Event x Tile FE             & Yes          & Yes         & Yes         & Yes         & Yes         \\
Event x Country x Period FE & Yes          & Yes         & Yes         & Yes         & Yes         \\
\hline
\addlinespace[0.1cm] 
Observations                & $1,172,561$  & $1,172,561$ & $1,172,561$ & $1,083,140$ & $1,172,561$   \\
Adj. R$^2$                  & $0.76$       & $0.76$      & $0.76$      & $0.76$      & $0.76$      \\
\hline
\end{tabular}}
\begin{tablenotes}[flushleft]
\scriptsize \item \textit{Note:} This table reports stacked DiD heterogeneity tests for the sample of treatment tiles within 20km from the mine and is based on Equation \ref{satim:eq_did}. The `Treatment Dummy' (or `Treat') indicates if a tile's corresponding mine has started operating, it is always 0 for tiles in the control group. In each column, the treatment dummy is interacted with a group dummy based on country specific institutional characteristics. `Democracy' is based on the average Polity2 score during the study period being larger than 0, `Voice \& Accountability' and `Rule of Law' are based on the being above or below the median of the World Bank's Worldwide Governance Indicators (WGI), `Decentralization' is based on the median decentralization index by Thomas Bijl and J. Vernon Henderson (LSE processed) and `GDP' based on the median of the World Bank's GDP per capita estimate (in PPP) for the year 2020. Standard errors in parenthesis are double-clustered by mine and tile. \hfill  $^{*}$p$<$0.1; $^{**}$p$<$0.05; $^{***}$p$<$0.01
\end{tablenotes}
\end{center}
\end{table}

\vspace{-0.5cm}
\subsection{Conflict as a Mechanism driving Heterogeneities} \label{satim:subsect_mech}

In order to gain a better understanding of why mine openings are particularly beneficial under democratic institutions, we investigate a potential mechanism. \cite{Armand2020} conduct a field experiment showing that increased information and community participation helps to prevent conflict, and so it is plausible that democratic institutions have a similar effect. To estimate the impact of mine openings on conflict, we combine our dataset with information about conflict events and locations from the Uppsala Conflict Data Program (UCDP) \citep{Sundberg2013}.\footnote{The main source of this database is global news reporting, but also other sources such as local news or NGO reports. The UCDP defines a conflict event as `An incident where armed force was used by an organized actor against another organized actor, or against civilians, resulting in at least 1 direct death at a specific location and a specific date'. For more information visit: \url{https://ucdp.uu.se/downloads/ged/ged211.pdf}.} A key advantage of the UCDP is that it provides a homogenous dataset of geolocalized conflicts from around the world spanning a relatively long time period from 1989 (which corresponds to Period 2 of our dataset) until today. We combine the UCDP with our dataset by creating a tile-level variable indicating the number of conflict events that occurred in that tile during the respective period. We then use our stacked DiD model and estimate the impact of mine openings and closings separately for democracies and autocracies.\footnote{We obtain similar results when using the model from Table~\ref{satim:table_2}, which also includes mine size, and we additionally find that there is no significant relationship between mine size and conflict.} \\

\vspace{-0.5cm}

\begin{table}[H]
\begin{center}
\caption{Mining \& Conflict} \label{satim:table_conflict}
\begin{adjustbox}{width=.8\textwidth,center}
\begin{tabular}{@{\extracolsep{15pt}}lcc|cc}
\hline
\\[-0.9em]
& \multicolumn{4}{c}{Prob. of Conflict in Tile (baseline: 0.3\%)}\\
\hline
\\[-0.9em]
& \multicolumn{2}{c}{{Near}} & \multicolumn{2}{c}{{Far}}\\[0.2em]
 & (1) & (2) & (3) & (4) \\
\hline
\\[-0.9em]
Treat                       & $0.003^{**}$ &      -       & $0.002^{*}$ &     -        \\
                            & $(0.001)$    &              & $(0.001)$   &              \\
\hline
& & & & \\[-1.8ex]              
Treat $\times$ Democracy    &      -       & $-0.001$     &      -      & $-0.001$     \\
                            &              & $(0.001)$    &             & $(0.001)$    \\
& & & & \\[-1.8ex]
Treat $\times$ Autocracy    &      -       & $0.006^{**}$ &      -      & $0.004^{**}$ \\
                            &              & $(0.002)$    &             & $(0.002)$    \\
& & & & \\[-1.8ex]
\hline
\addlinespace[0.1cm] 
Event x Tile FE             & Yes          & Yes           & Yes         & Yes          \\
Event x Country x Period FE & Yes          & Yes           & Yes         & Yes          \\
\hline
\addlinespace[0.1cm] 
Observations                & $1,078,351$  & $1,078,351$   & $1,153,489$ & $1,153,489$    \\
Adj. R$^2$                  & $0.12$       & $0.12$        & $0.12$      & $0.12$      \\
\hline
\hline \\[-1.8ex]
\end{tabular}
\end{adjustbox}
\begin{tablenotes}[flushleft]
\scriptsize \item \textit{Note:} This table reports stacked DiD estimates for the impact of mine openings on conflict and is based on Equation~\ref{satim:eq_did}. `Treat' indicates if a tile's corresponding mine has started operating, it is always 0 for tiles in the control group. All models are linear probability models and the dependent variable indicates whether a tile experience any conflict during a given period. In columns (1) and (2), the treatment group is restricted to areas within 20km from the mine, and to areas between 20km and 40 km in Columns (3) and (4). Please note, that period 1 (1984-1986) is omitted from the sample as the Uppsala Conflict Data Program (UCDP) only starts in period 2 (1989). The baseline probability of conflict is 0.3\% and refers to the average conflict probability per tile in the treatment group prior to treatment onset. Standard errors in parenthesis are double-clustered by mine and tile. $^{*}$p$<$0.1; $^{**}$p$<$0.05; $^{***}$p$<$0.01
\end{tablenotes}
\end{center}
\end{table}

In Table~\ref{satim:table_conflict} columns (1) and (3), we replicate the finding in the literature that mining fuels conflict \citep{Berman2017}.\footnote{All regressions in Table \ref{satim:table_conflict} are linear probability models. Using binomial regressions is computationally difficult due to the high dimensional fixed effects and large number of observations.} However, when decomposing this effect, we find that the onset of mining leads to a significant increase in conflict in autocracies, but not in democracies. The impact is also significant in areas further away from the mine, although it is of slightly smaller magnitude. The value of the estimates indicate that the onset of mining increases the probability of conflict by 0.6\% in areas near the mine and 0.4\% further away from the mine. The average probability of conflict prior to the onset of mining is 0.3\%, and hence the increase in areas near the mine corresponds to twice the baseline probability of conflict. This is a very sizable effect, and its estimation is novel in the literature. While \cite{Berman2017} undertake a similar exercise, they do not find significant effects associated with institutions. This is likely due to the fact that their dataset is much smaller, less granular and covers a much shorter time period between 1997-2010. \\

Our findings in Sections \ref{satim:sect_event_study}, \ref{satim:subsect_hetero} and \ref{satim:subsect_mech} indicate that the onset of mining has a considerable positive impact on the local economy, particularly in areas within 20km of the mine. However, there are important heterogeneities: growth due to mining is significantly greater in democracies, and although large mines lead to local development in autocracies, small mines might even have a negative impact on the local economy. One mechanism that explains this differential finding in regards to institutional context is increased conflict. Whilst mine openings lead to significant increases in conflict in autocracies, democratic institutions prevent a rise in conflict.

\subsection{Mine Closures} 

In this section, we investigate what happens to mine areas after the closure of the mine. We analyze the evolution of \emph{Closing} mine areas, i.e. areas surrounding mines which are active in Period 1 but which close by Period 12. Since closing mines experience two treatments - firstly becoming an active mine and secondly ceasing operation - it is much harder to find a suitable control group with similar pre-trends. For this reason, rather than using a control group indicating what would have happened if the mines had not closed, we use comparison groups to benchmark the performance of areas surrounding closing mines. We compare \emph{Closing} mine areas to \emph{Not Yet Opened} mines areas as well as to \emph{Continuous} mine areas, in order to identify divergent trends between these groups after the mine's closure. \\

We use an event study following Equation~\ref{satim:eq_event_study}. We center relative time around the last period of active mining ($t=0$), hence $t=1$ is the first period without any mine activity. Since we are ultimately interested in whether the mine closure triggers a trend break, we need to compare the relative evolution between $t=-5$ and $t=0$ (during the 15 years prior to closing) to the evolution between $t=0$ and $t=5$ (during the 15 years after closing). For this purpose, we add a dashed horizontal line with an intercept corresponding to the coefficient in period $t=0$. We can then test whether \textit{Closing} mine areas developed prior to the closure by comparing the confidence interval of the estimate in period $t=0$ to the horizontal line at $y=0$. In the next step, we can test if \textit{Closing} mines gained after the closure by comparing the confidence interval of the estimate at $t=5$ to the horizontal line with the intercept of the coefficient at period $t=0$.  

\begin{figure}[htb]
\caption{Event Study: \textit{Closing} vs. \textit{Not Yet Opened} Mines} \label{satim:event_study_closing_notyet}
\vspace{0.1cm}
\begin{subfigure}{0.5\hsize}\centering
    \caption{Near: Log Urban Area Share} 
    \includegraphics[width=0.99\hsize]{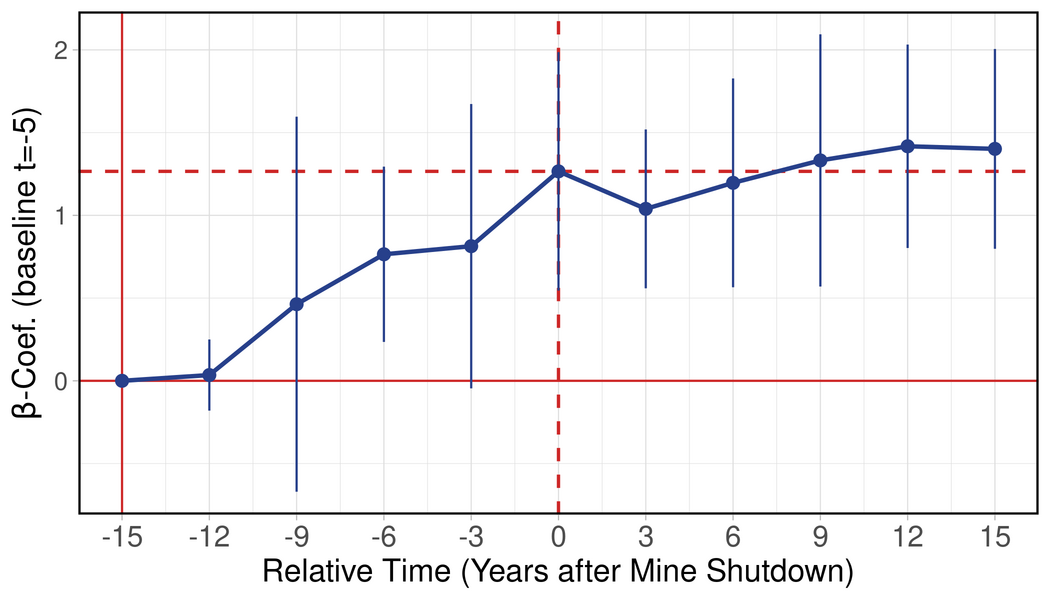}
\end{subfigure}
\begin{subfigure}{0.5\hsize}\centering
    \caption{Far: Log Urban Area Share} 
    \includegraphics[width=0.99\hsize]{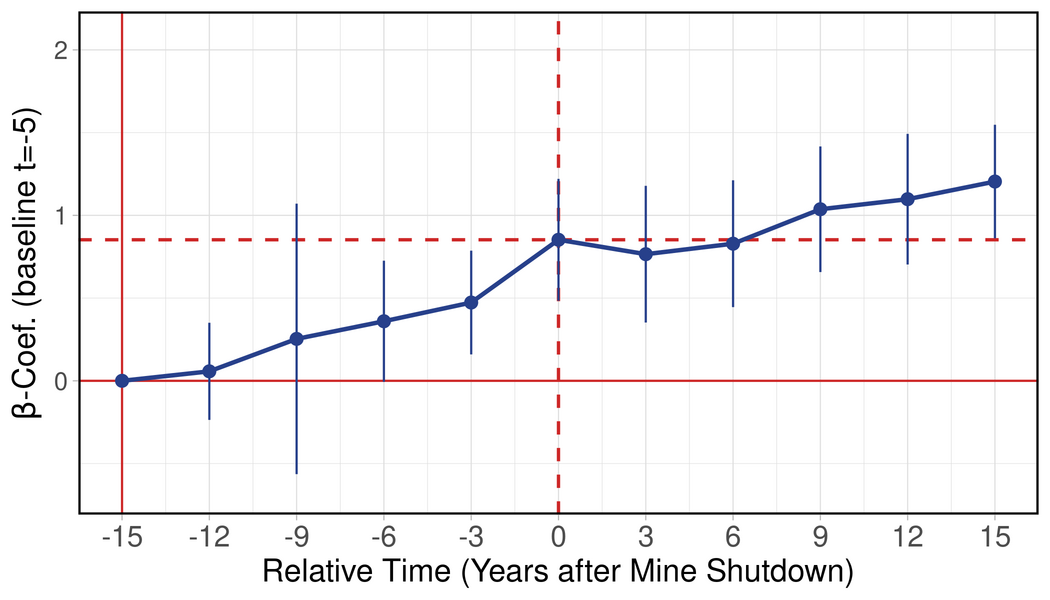}
\end{subfigure}

\begin{tablenotes}[flushleft]
\scriptsize \item \textit{Note:} Prior to closure, \emph{Closing} mine areas have higher urban growth than \emph{Not Yet Opened} mine areas. However, after closure, the urban growth in areas in proximity to \emph{Closing} mines slows in comparison to non-mine areas. Error bars represent a 95\% confidence interval. 
\end{tablenotes}
\end{figure} 

\vspace{-0.5cm}
\begin{figure}[htb]
\caption{Event Study: \textit{Closing} vs. \textit{Continuous} Mines} \label{satim:event_study_closing_cont}
\vspace{0.1cm}
\begin{subfigure}{0.5\hsize}\centering
    \caption{Near: Log Urban Area Share}
    \includegraphics[width=0.99\hsize]{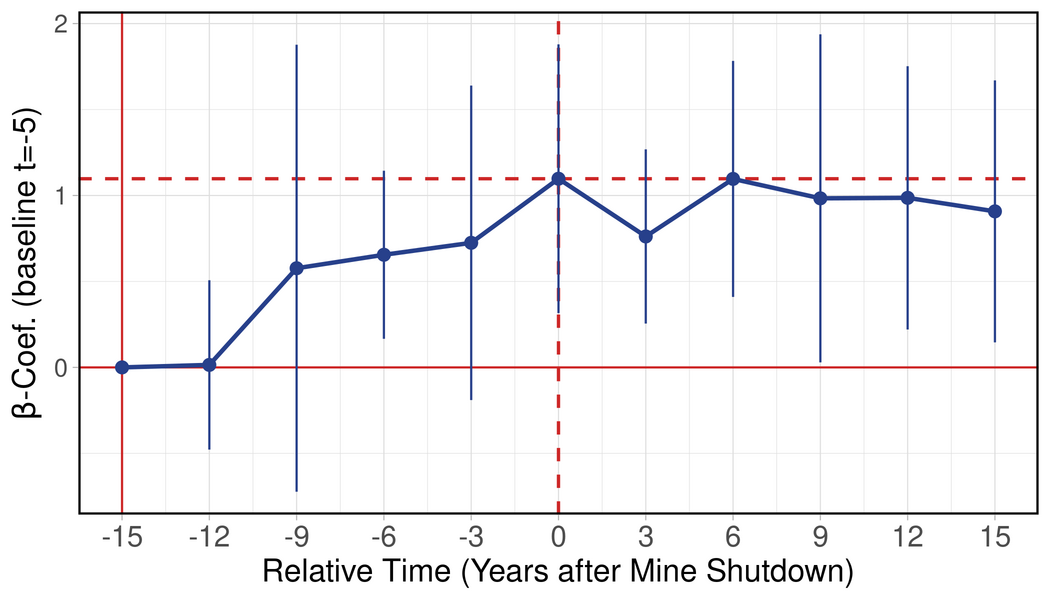}
\end{subfigure}
\begin{subfigure}{0.5\hsize}\centering
    \caption{Far: Log Urban Area Share} 
    \includegraphics[width=0.99\hsize]{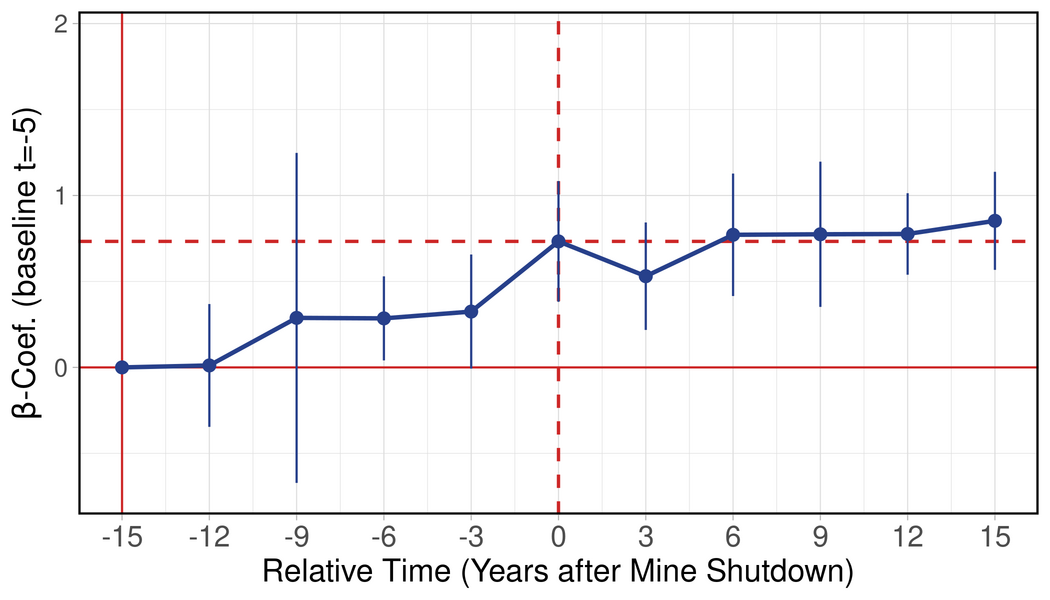}
\end{subfigure}

\begin{tablenotes}[flushleft]
\scriptsize \item \textit{Note:} Prior to closure, \emph{Closing} mine areas have higher urban growth than \emph{Continuous} mine areas. However, after closure, areas in proximity to \emph{Closing} mines evolve at the same rate as \emph{Continuous} mine areas. Error bars represent a 95\% confidence interval. 
\end{tablenotes}
\end{figure}

Figure~\ref{satim:event_study_closing_notyet} and Figure~\ref{satim:event_study_closing_cont} plot the event study results when using either \textit{Not Yet Opened} or \textit{Continuous} areas as a comparison group. Prior to the closure, \textit{Closing} mine areas outperform both \textit{Not Yet Opened} and \textit{Continuous} mine areas in terms of urban growth. However, after the closure of the mine, we observe a trend break, with diminished growth around \textit{Closing} mines relative to \textit{Not Yet Opened} and \textit{Continuous} mine areas from the same country and period. The effects are very similar for agricultural land use but are without any significant patterns for the material wealth index (see Figure~\ref{satim:closing_not_yet_evstud_append}). Overall, these findings demonstrate that mining areas are unable to maintain elevated growth rates after the closure of the mine. \\

However, the external validity of our analyses regarding \textit{Closing} mines may be undermined by the fact that we only have a total of 77 such mines across 13 countries in our sample (vs. 348 Opening mines across 40 countries, see Figure~\ref{satim:mine_groups_overview}). Our analyses may hence be influenced by the particularities of individual mines. The small sample size also prevents us from investigating heterogeneous effects of mine closures, such as across institutional contexts.

\FloatBarrier

\section{Conclusion} \label{satim:sect_conclusion}
\vspace{-0.12cm}

In this study, we create a novel dataset based on large archives of satellite imagery covering 12\% of the African continent over four decades to study the impact of mine openings on the development of local communities. We follow the trajectories of urban development, agricultural production and material wealth,
in the villages and cities surrounding mines throughout the period of active mining as well as after the mines' closure.\\ %
\vspace{-0.12cm}

Our results indicate that mineral mine operations, especially large operations, have the potential to give a considerable boost to local economic growth in areas surrounding the mine. Part of these gains are likely to be indirect gains due to backward linkages, reflected by increased agricultural activities around the mine. Nevertheless, our study also reveals important caveats with regards to the positive impact of mining. Firstly, accelerated growth rates in mining areas are only temporary and are not sustained beyond the closure of the mine. Secondly, mines have relatively little, if any, positive impact on local communities in autocratic countries. We show that one mechanism through which mining areas in democratic countries, as opposed to autocratic countries, avoid the resource curse is by managing to avoid a significant increase in conflict. \\ %

\vspace{-0.12cm}

We conclude that not all mine operations benefit local communities, and mining could even be disadvantageous for economic growth, in particular in areas with poor institutions. For this reason, our study underlines the importance of considering the institutional framework conditions when considering mine openings. 
Furthermore, in order to achieve sustained growth in areas that benefit from mining, policy makers need to develop location-specific strategies for economic transformation during operation and to prolong development after the closure of the mine. \\
\vspace{-0.12cm}

Other relevant mechanisms linking mining and institutions on the local level could be investigated in further research. One example is the fiscal channel through which the windfall gains from resource extraction fail to be redistributed to the population under extractive institutions. Another potential channel is local corruption under bad institutions with insufficient checks and balances. Furthermore, under poor institutional framework conditions, there may be reduced incentives for local public officials to negotiate and enforce regulations, such as local procurement rules, employment opportunities for local residents and other forms of resource governance that benefit local communities. \\
\vspace{-0.12cm}

\textbf{Acknowledgements:} We thank Olmo Silva and Felipe Carozzi for their guidance. We received valuable comments from Gabriel Ahlfeldt, Steve Gibbons, Vernon Henderson, Alexander Moradi, Michel Serafinelli and participants at the 15th North American meeting of the Urban Economics Association.

\textbf{Funding:} Sandro Provenzano acknowledges funding from the Economic and Social Research Council [grant number: ES/P000622/1].

\FloatBarrier

\pagebreak

\bibliography{references}

\pagebreak

\appendix

\section*{Appendix}

\pagenumbering{Roman}
\renewcommand*{\thepage}{\Roman{page}}

\setcounter{table}{0}
\renewcommand{\thetable}{A\arabic{table}}

\setcounter{figure}{0}
\renewcommand{\thefigure}{A\arabic{figure}}

\subsection{Data Sources and Description}\label{satim:app:data_appendix}

\subsection*{\underline{Mineral deposits}}

We purchase data on mining deposits from MinEx Consulting. This dataset includes the geolocalization, type, size and dates of discovery and activity of 1,658 mineral deposits in 47 African countries.

\subsection*{\underline{Land Use Label Data Sources}}

\begin{itemize}
\item[\textbf{1.}] \textbf{Agricultural Areas}
	\begin{itemize}
	\item[-] Nasa Global Food Security-support Analysis Data (GFSAD), Croplands Africa 2015, 30m resolution, available at: \url{https://lpdaac.usgs.gov/products/gfsad30afcev001/}
	\end{itemize}

\item[\textbf{2.}] \textbf{Urban Areas}
\begin{itemize}
	\item[-] Global Human Settlement Layer (GHSL) 2015 by the European Commission, available at: \url{https://ghsl.jrc.ec.europa.eu/download.php?ds=buGHS_BUILT_LDSMT _GLOBE_R2018A _3857_30_V2_0} 
	\item[-] CCI Land Cover S2 Prototype Africa 2016, resolution: 20m, available at: \url{https://2016africalandcover20m.esrin.esa.int/}) 
	\item[-] Facebook (Meta) Population Map 2015, 30m resolution, available at: \url{https://data.humdata.org/dataset/highresolutionpopulationdensitymaps}, does not cover South Sudan, Sudan, Somalia and Ethiopia
\end{itemize}

\item[\textbf{3.}] \textbf{Water Bodies}
\begin{itemize}
	\item[-] CCI Land Cover S2 Prototype Africa 2016, resolution: 20m, available at: \url{https://2016africalandcover20m.esrin.esa.int/}) 
\end{itemize}

\subsection*{\underline{Geocovariates Sources}}

\begin{itemize}

\item \textbf{Crop Caloric Index}: A measure of agricultural suitability containing the potential agricultural caloric output per year and hectar (excluding zero yields) based on \cite{Galor2016}, available at: \url{https://ozak.github.io/Caloric-Suitability-Index/}.\\

\item \textbf{Elevation}: NASA Shuttle Radar Topography Mission (SRTM) with 30m resolution, available at:\url{https://www2.jpl.nasa.gov/srtm/}. \\

\item \textbf{African Cities}: Data from OECD Sahel and West Africa Club in collaboration with e-geopolis.org, available at: \url{https://africapolis.org/data}. \\

\item \textbf{Country Borders \& Coastline}: GADM (Version 3.6), available at: \url{https://gadm.org/data.html}. \\

\item \textbf{Climate Data:} Annual precipitation (in mm), annual mean temperature, minimum temperature in the coldest month and maximum temperature in the warmest month (all in $^{\circ}$C) based on \cite{Karger2017} were obtained from \url{http://chelsa-climate.org/downloads/}.\\

\item \textbf{Ruggedness}: A measure for terrain ruggedness measured in degree of slope with an initial resolution of ($20\times20$ arcseconds) based on \cite{Nunn2012} was obtained from: \url{https://diegopuga.org/data/rugged/}.

\end{itemize}

\end{itemize}

\subsection{Additional Figures}\label{satim:app:additional_figures}

\begin{figure}[!h]
\centering
\caption{High-Resolution Images Corresponding to Figure~\ref{satim:pre_post_mine_and_zoom}B} \label{satim:pre_post_mine_zoom_google_images}

\subfloat[\centering \scriptsize{2020 high-resolution Google Earth image of city shown in Figure~\ref{satim:pre_post_mine_and_zoom}b}]{\includegraphics[width=0.35\textwidth]{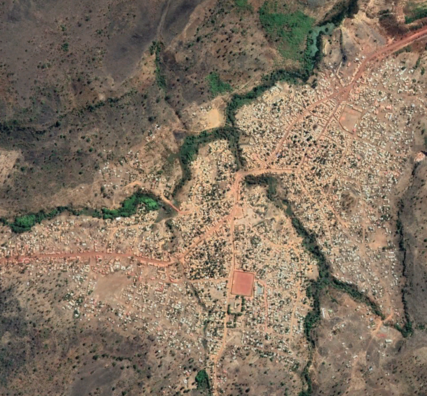}}~~
\subfloat[\centering \scriptsize{2020 high-resolution Google Earth image of mine shown in Figure~\ref{satim:pre_post_mine_and_zoom}b}]{\includegraphics[width=0.35\textwidth]{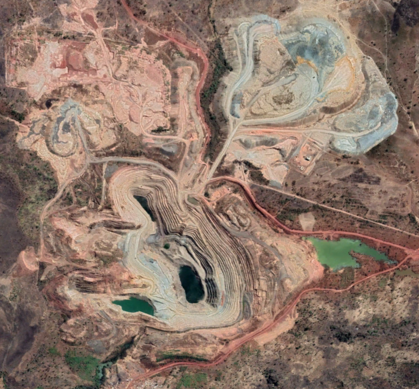}}

\vspace{0.4cm}

\begin{tablenotes}[flushleft]
\scriptsize \item \textit{Note:} Recent high-resolution satellite images can be used for better visualization of the 30m Landsat images in Figure~\ref{satim:pre_post_mine_and_zoom}b. The fact that our model excluded the area that looks like a football field in the above image (a) from the urban extent (see Figure Figure~\ref{satim:pre_post_mine_and_zoom}b) illustrates the ability of the model to identify the urban extent at a very precise and granular level.
\end{tablenotes}
\end{figure}

\begin{figure}[ht]
\centering
\caption{Landsat and Corresponding High-Resolution Images}
\label{satim:fig:poor_vs_rich}
\subfloat[\scriptsize{Landsat: Low Wealth Index}\label{satim:fig:a_poorrich}]{\includegraphics[width=0.35\textwidth]{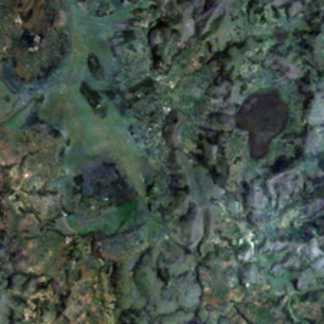}}
\quad
\subfloat[\scriptsize{Landsat: High Wealth Index}\label{satim:fig:b_poorrich}]{\includegraphics[width=0.35\textwidth]{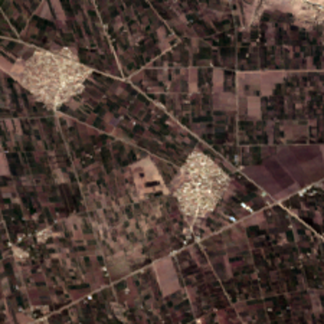}}

\vspace{0.4cm}
\centering

\subfloat[\scriptsize{Google Earth: Low Wealth Index}\label{satim:fig:c_poorrich}]{\includegraphics[width=0.35\textwidth]{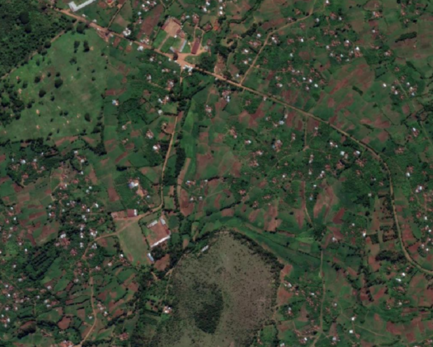}}
\quad
\subfloat[\scriptsize{Google Earth: High Wealth Index}\label{satim:fig:d_poorrich}]{\includegraphics[width=0.35\textwidth]{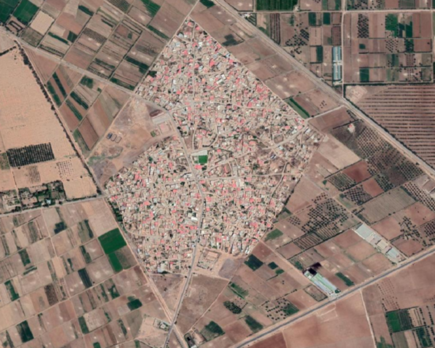}}

\begin{tablenotes}[flushleft]
\scriptsize \item \textit{Note:} High-resolution images from Google Earth provide a sanity check of the material wealth predictions using medium resolution images from Landsat. 
\end{tablenotes}
\end{figure}

\begin{figure}[htb]
    \centering
    \caption{Wealth Change: LSMS and Material Wealth Index}
    \label{satim:fig:corr_wealth_time}
    \includegraphics[width=0.4\textwidth]{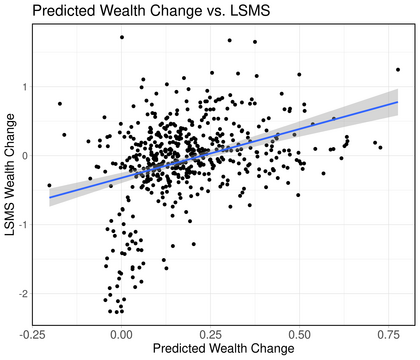}
    
 \begin{tablenotes}[flushleft]
\scriptsize \item \textit{Note:} There is a positive correlation (coefficient 0.35) between the predicted wealth change and the estimated wealth change from Living Standards Measurement Study (LSMS) panel data surveys by the World Bank, suggesting that the material wealth model is capable of detecting changes in wealth over time.
\end{tablenotes}   
\end{figure}

\begin{figure}[htb]

\caption{Cross-Sectional Comparisons of Agricultural Land Cover Share in Periods 1 \& 12}\label{satim:descr_crosssect_append_agri}

\begin{subfigure}{0.5\hsize}\centering
    \caption{Log Agricultural Land Cover (Period 1)} \label{satim:descr_c}
    \includegraphics[width=0.99\hsize]{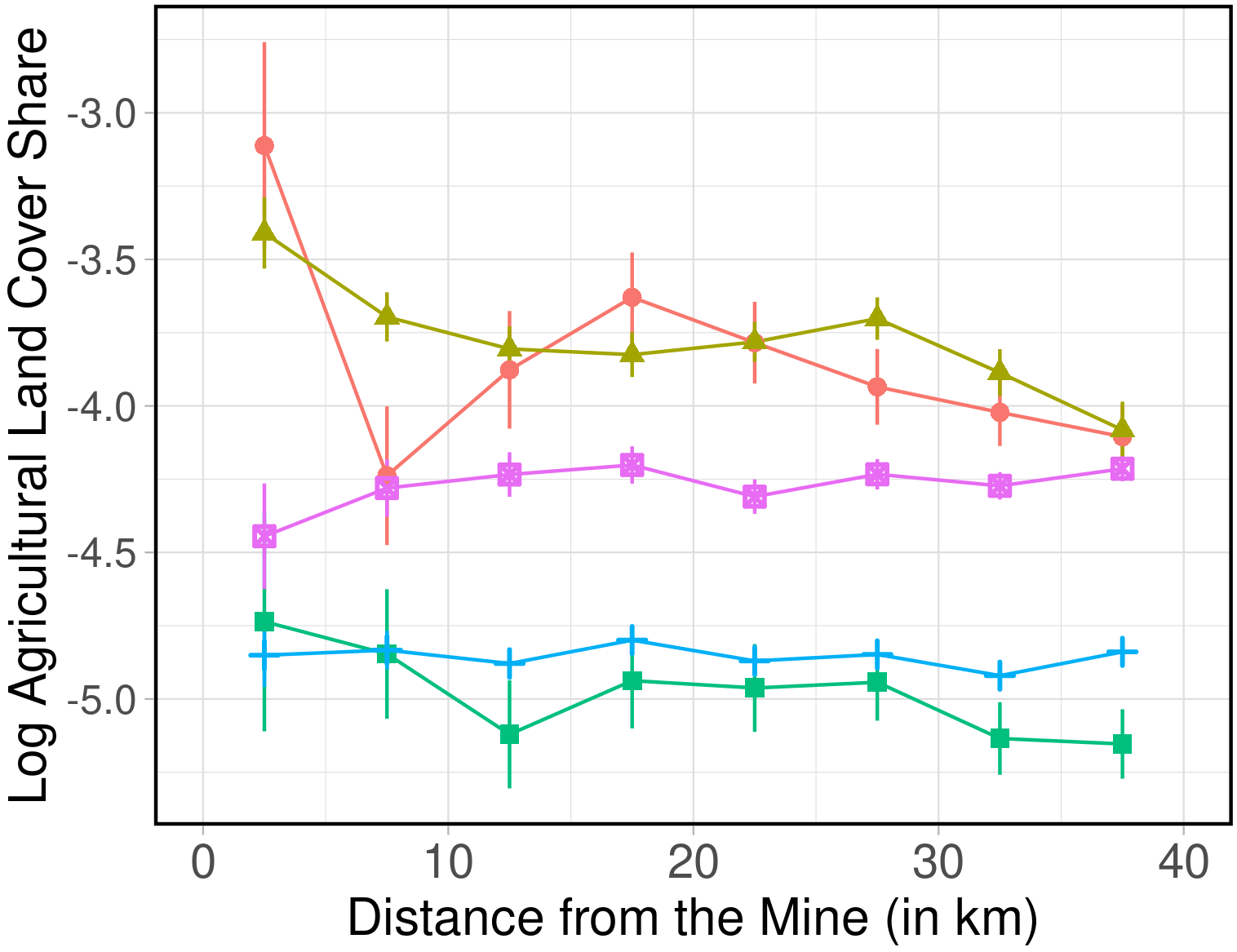}
\end{subfigure}
\begin{subfigure}{0.5\hsize}\centering
    \caption{Log Agricultural Land Cover (Period 12)} \label{satim:descr_d}
    \includegraphics[width=0.99\hsize]{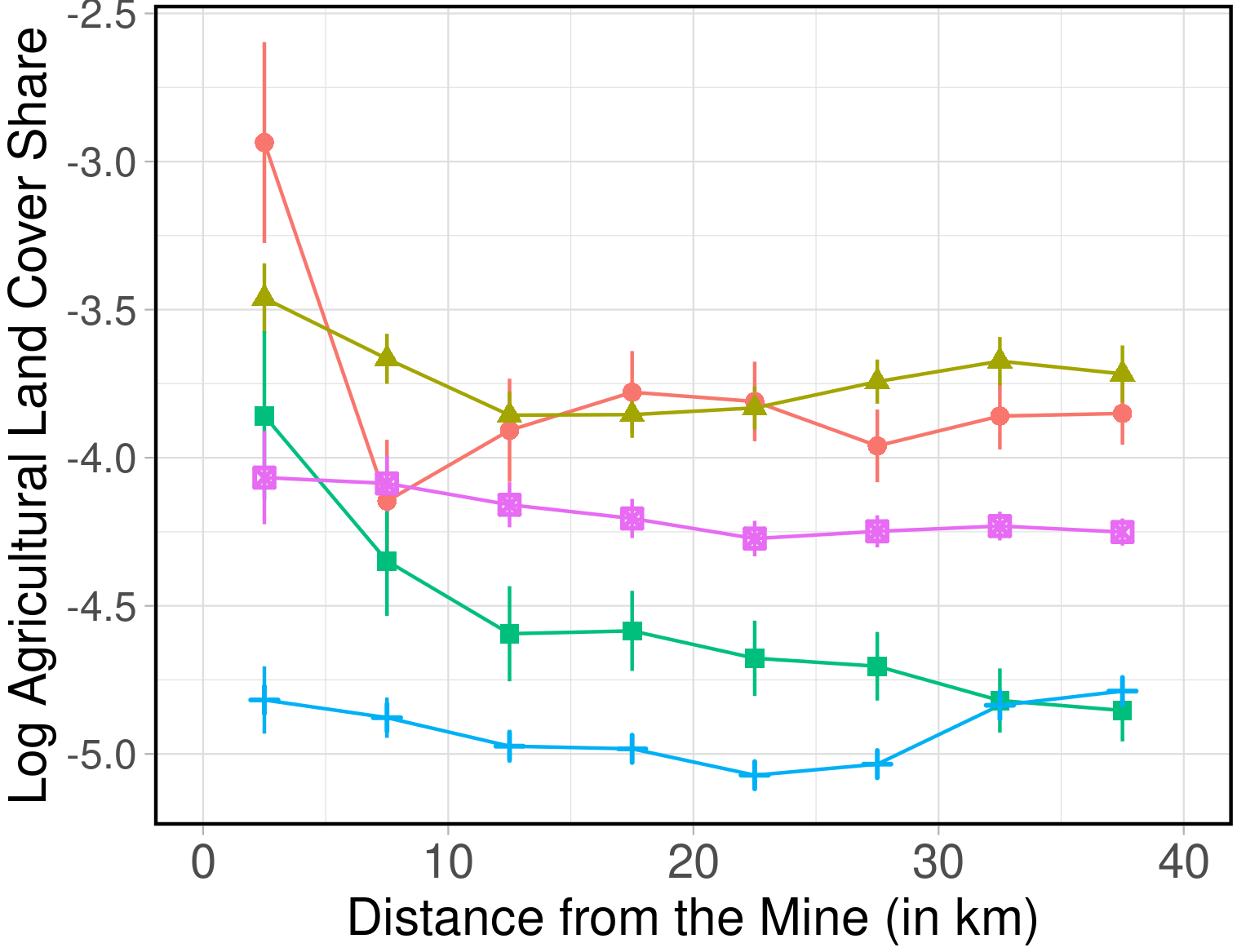}
\end{subfigure}

\begin{subfigure}{\hsize}\centering
\includegraphics[width=0.99\hsize]{figures/cross_legend.png}
\end{subfigure}

\begin{tablenotes}[flushleft]
\scriptsize \item \textit{Note:} There is a slight increase in agricultural areas surrounding \emph{Opening} mine areas between Period 1 and Period 12 relative to \emph{Not Yet Opened} mine areas, better represented in Figure~\ref{satim:demeaned_agri}. The agricultural area around \emph{No Longer Active} mines tends to grow between Period 1 and Period 12 relative to mines that are \emph{Not Yet Opened}, but we leave detailed analysis of this observation for further research. Error bars represent a 95\% confidence interval. 
\end{tablenotes}
\end{figure}

\begin{figure}[htb]

\caption{Cross-Sectional Comparisons of Wealth in Periods 1 \& 12}\label{satim:descr_crosssect_append_wealth}

\begin{subfigure}{0.5\hsize}\centering
    \caption{Wealth Index (Period 1)} \label{satim:descr_e}
    \includegraphics[width=0.99\hsize]{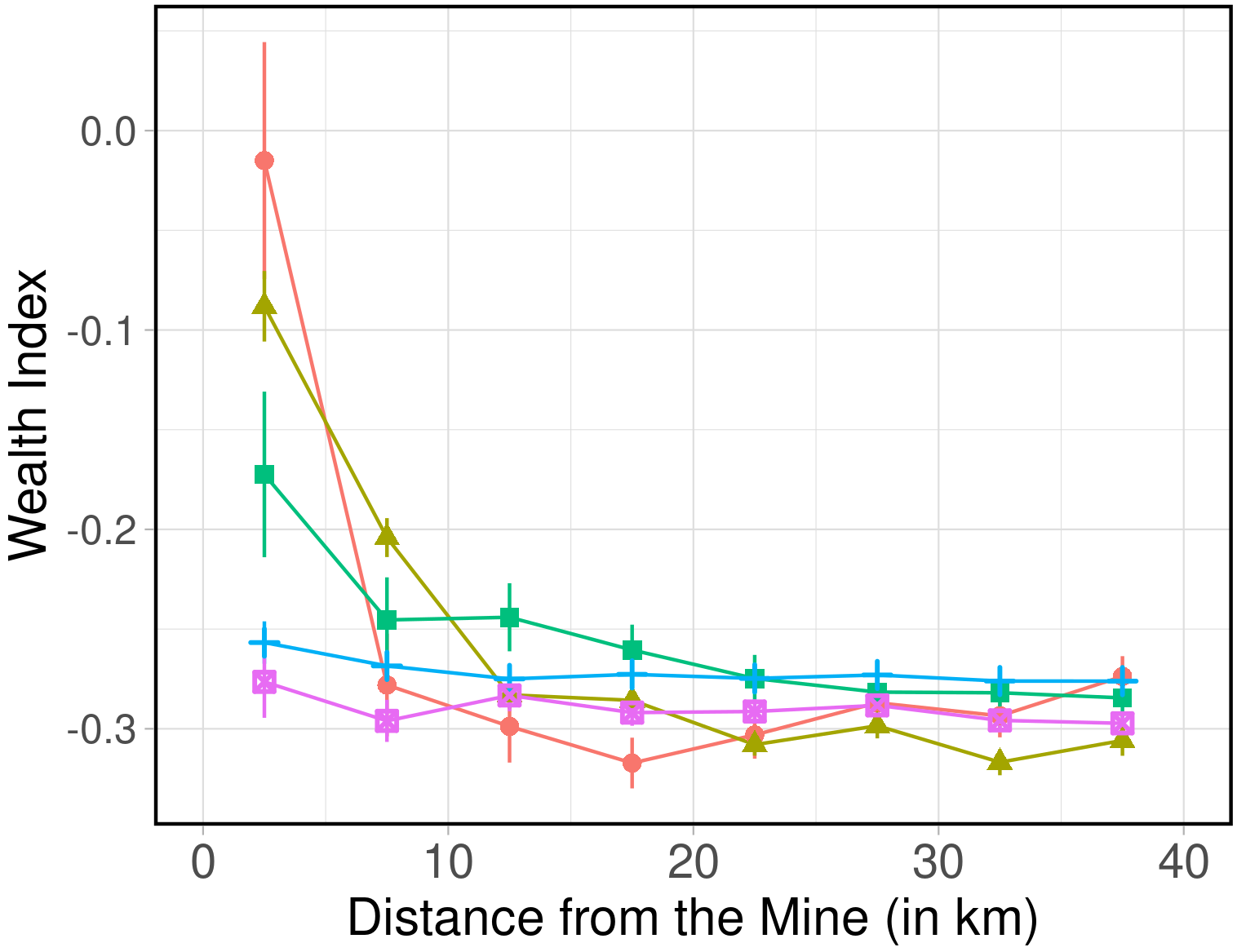}
\end{subfigure}
\begin{subfigure}{0.5\hsize}\centering
    \caption{Wealth Index (Period 12)} \label{satim:descr_f}
    \includegraphics[width=0.99\hsize]{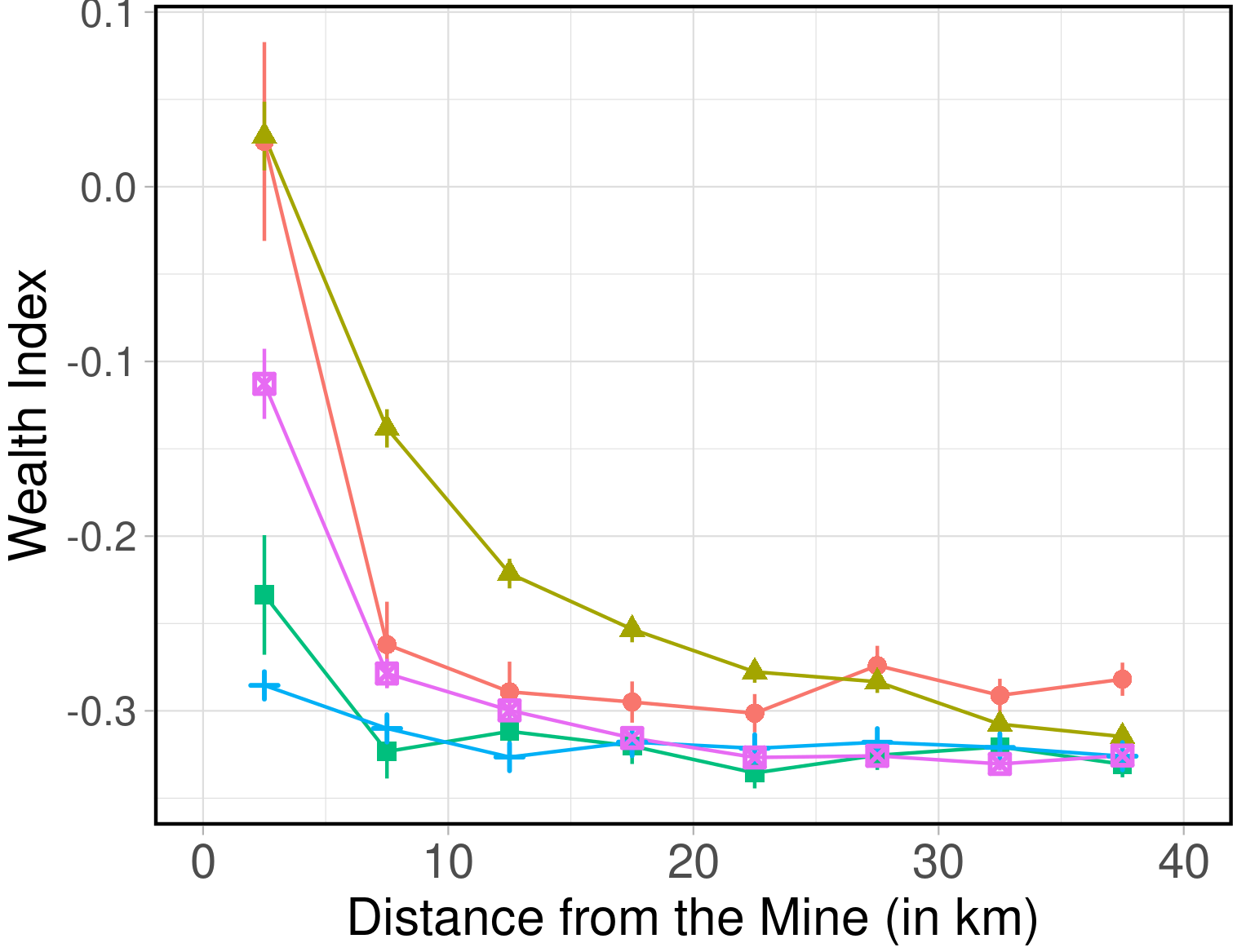}
\end{subfigure}

\begin{subfigure}{\hsize}\centering
\includegraphics[width=0.99\hsize]{figures/cross_legend.png}
\end{subfigure}

\begin{tablenotes}[flushleft]
\scriptsize \item \textit{Note:} The wealth index tends to decrease with distance from active mines. Mines that were open in Period 1 have a higher wealth index near the mines. The wealth index around \emph{Opening} mines grows between Period 1 and Period 12, relative
to mines that are \emph{No Longer Active} or \emph{Not Yet Opened}. Compared to areas in geographic proximity to \emph{Closing} mines, areas near \emph{Continuous} mines seem to have greater increases in the wealth index between the two periods. Error bars represent a 95\% confidence interval. 
\end{tablenotes}
\end{figure}

\begin{figure}[htb]
\caption{Log Agricultural Land Cover Share Relative to \emph{Not Yet Opened} Mines} \label{satim:demeaned_agri}
\centering
\begin{subfigure}{0.45\hsize}\centering
    \caption{\emph{Opening}} \label{satim:agri_demeaned_open}
    \includegraphics[width=0.99\hsize]{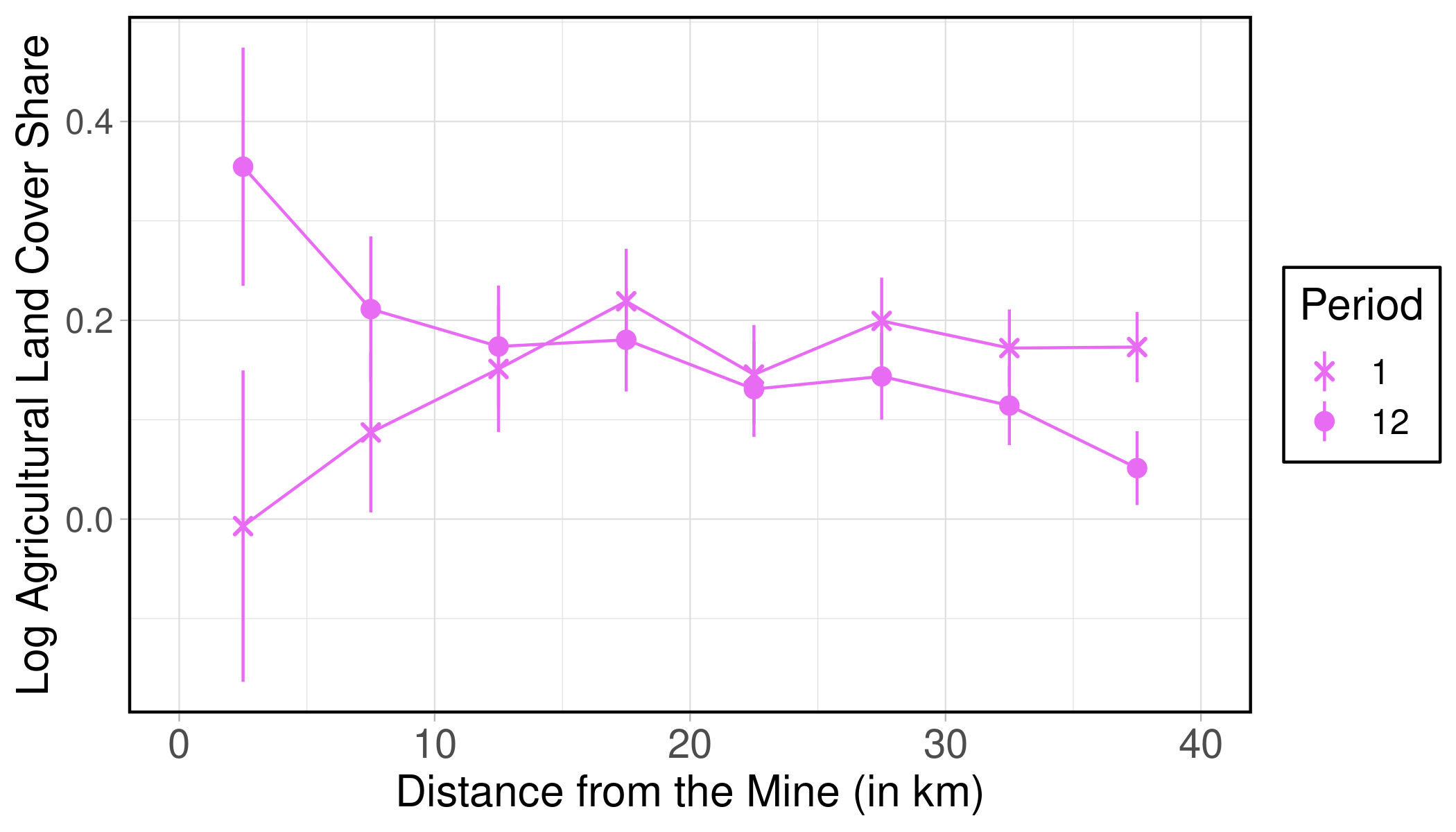}
\end{subfigure}\\
\hspace{0.3cm}
\begin{subfigure}{0.45\hsize}\centering
    \caption{\emph{Continuous}} \label{satim:agri_demeaned_cont}
    \includegraphics[width=0.99\hsize]{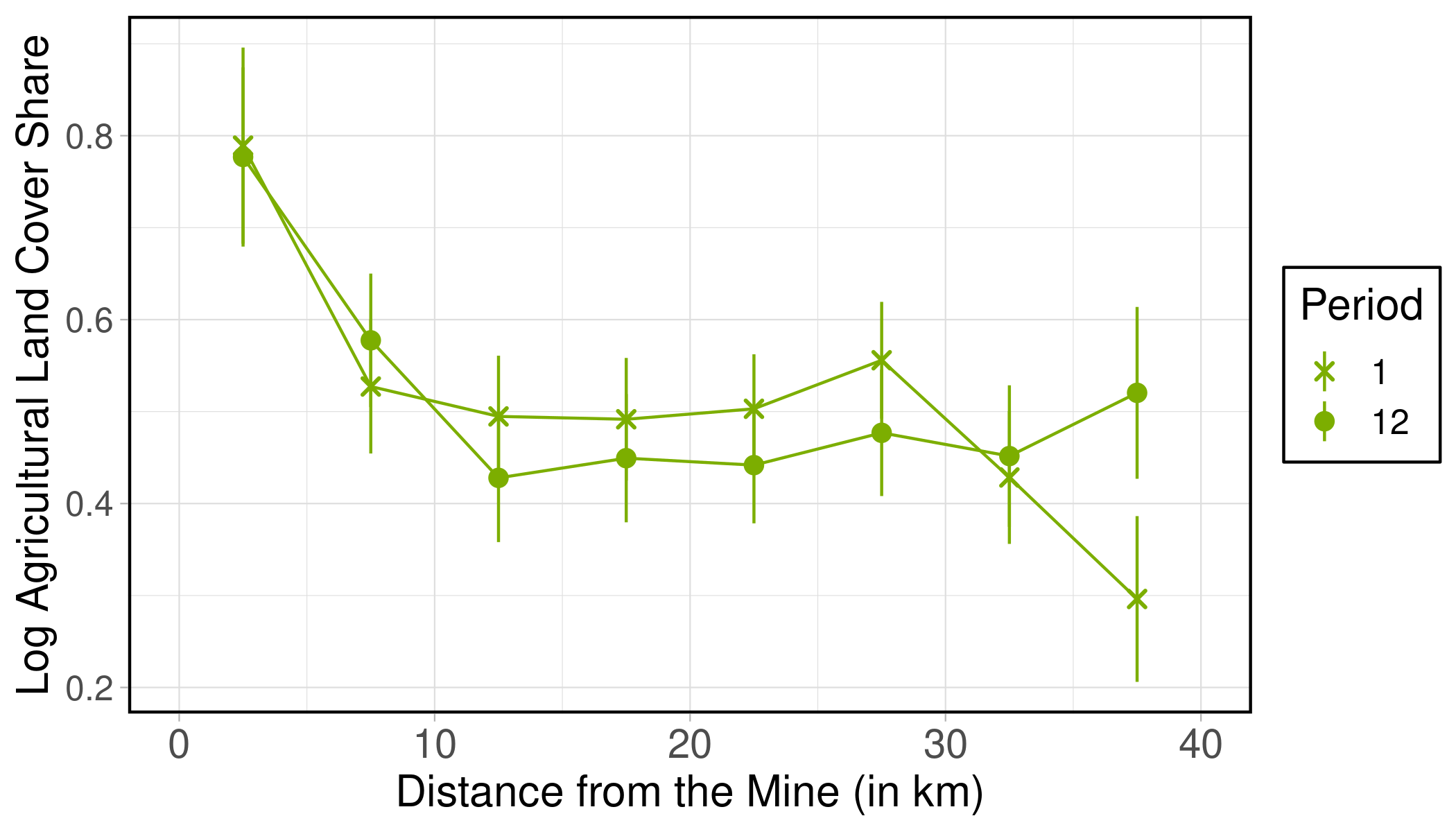}
\end{subfigure}
\hspace{0.3cm}
\begin{subfigure}{0.45\hsize}\centering
    \caption{\emph{Closing}} \label{satim:agri_demeaned_close}
    \includegraphics[width=0.99\hsize]{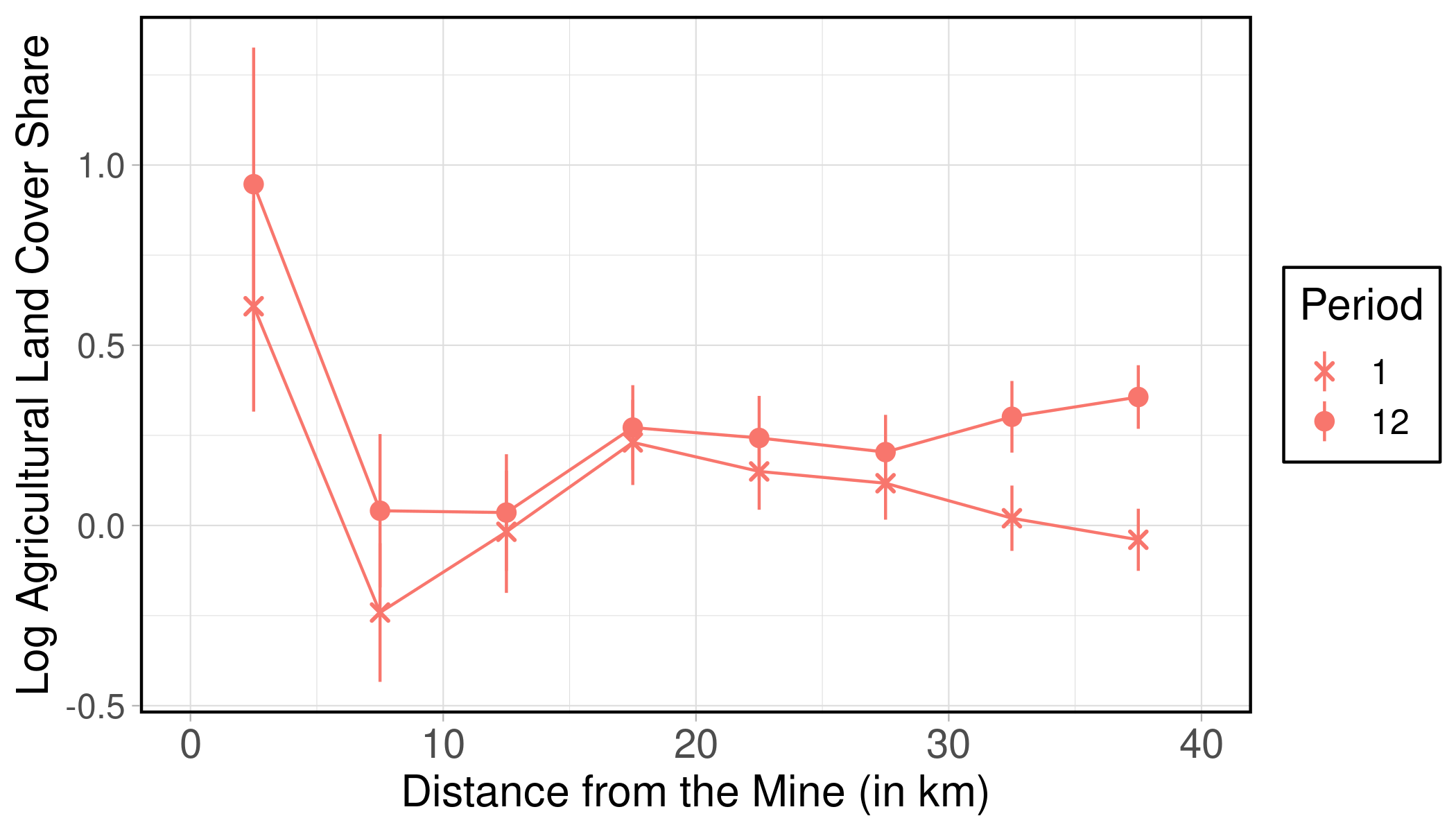}
\end{subfigure}
\begin{tablenotes}[flushleft]
\scriptsize \item \textit{Note:} The vertical axis is the log agriculture share of the respective mine area categories, minus the country and period average log agriculture land cover share of \emph{Not Yet Opened} mine areas. A relative log agriculture land cover share of 1 thus means the agriculture land cover share is $\textrm{exp}(1) \approx 2.7$ times the agriculture land cover share of \emph{Not Yet Opened} mine areas from the same country and period. \emph{Opening} mines have similar agriculture land cover shares as \emph{Not Yet Opened} mine areas in Period 1, but areas near the mine have comparatively large agricultural areas by Period 12. 
\end{tablenotes}
\end{figure}

\begin{figure}[htb]
\caption{Wealth Index Relative to \emph{Not Yet Opened} Mines} \label{satim:demeaned_wealth}
\centering
\begin{subfigure}{0.45\hsize}\centering
    \caption{\emph{Opening}} \label{satim:wealth_demeaned_open}
    \includegraphics[width=0.99\hsize]{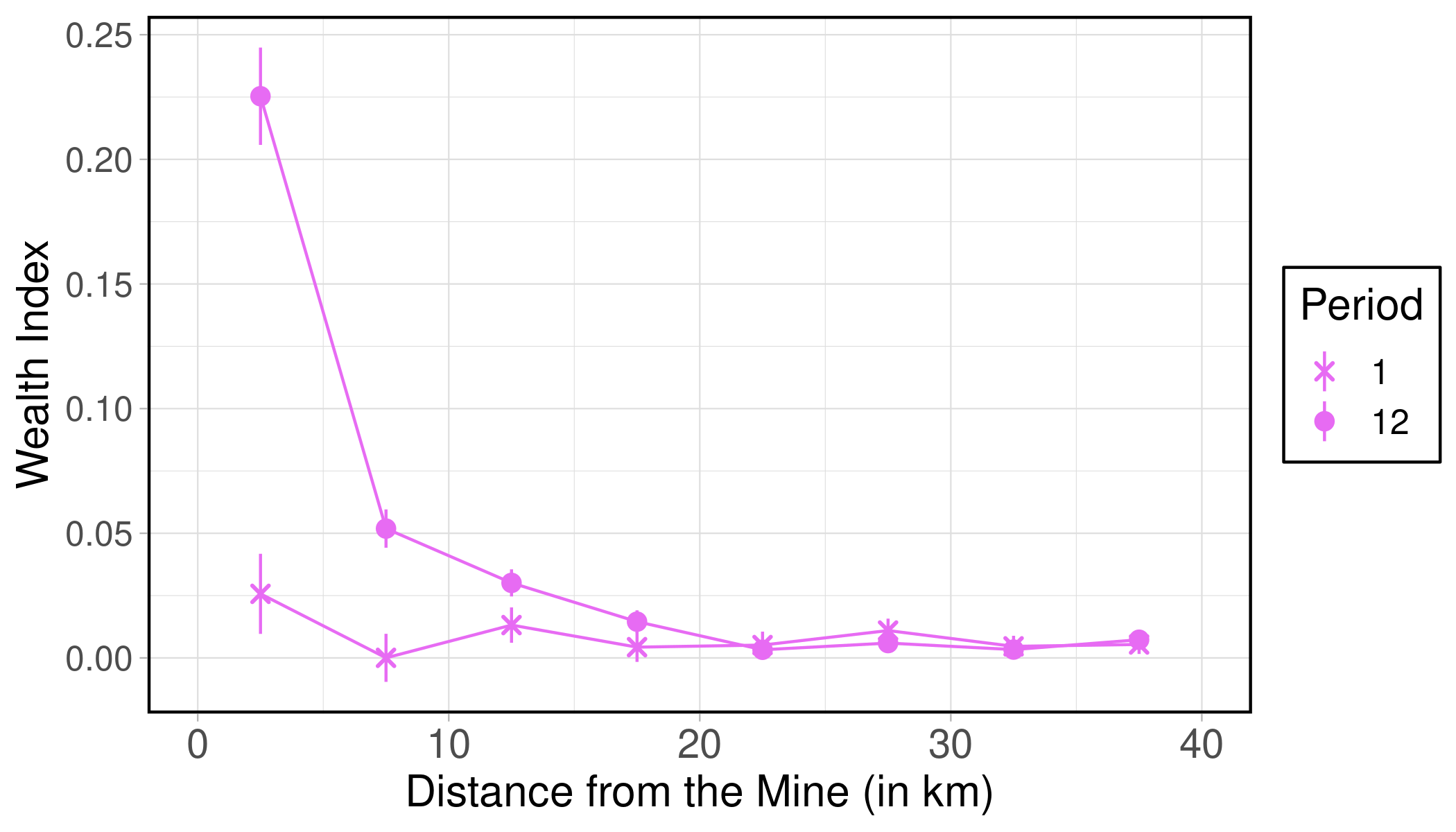}
\end{subfigure}\\
\hspace{0.3cm}
\begin{subfigure}{0.45\hsize}\centering
    \caption{\emph{Continuous}} \label{satim:wealth_demeaned_cont}
    \includegraphics[width=0.99\hsize]{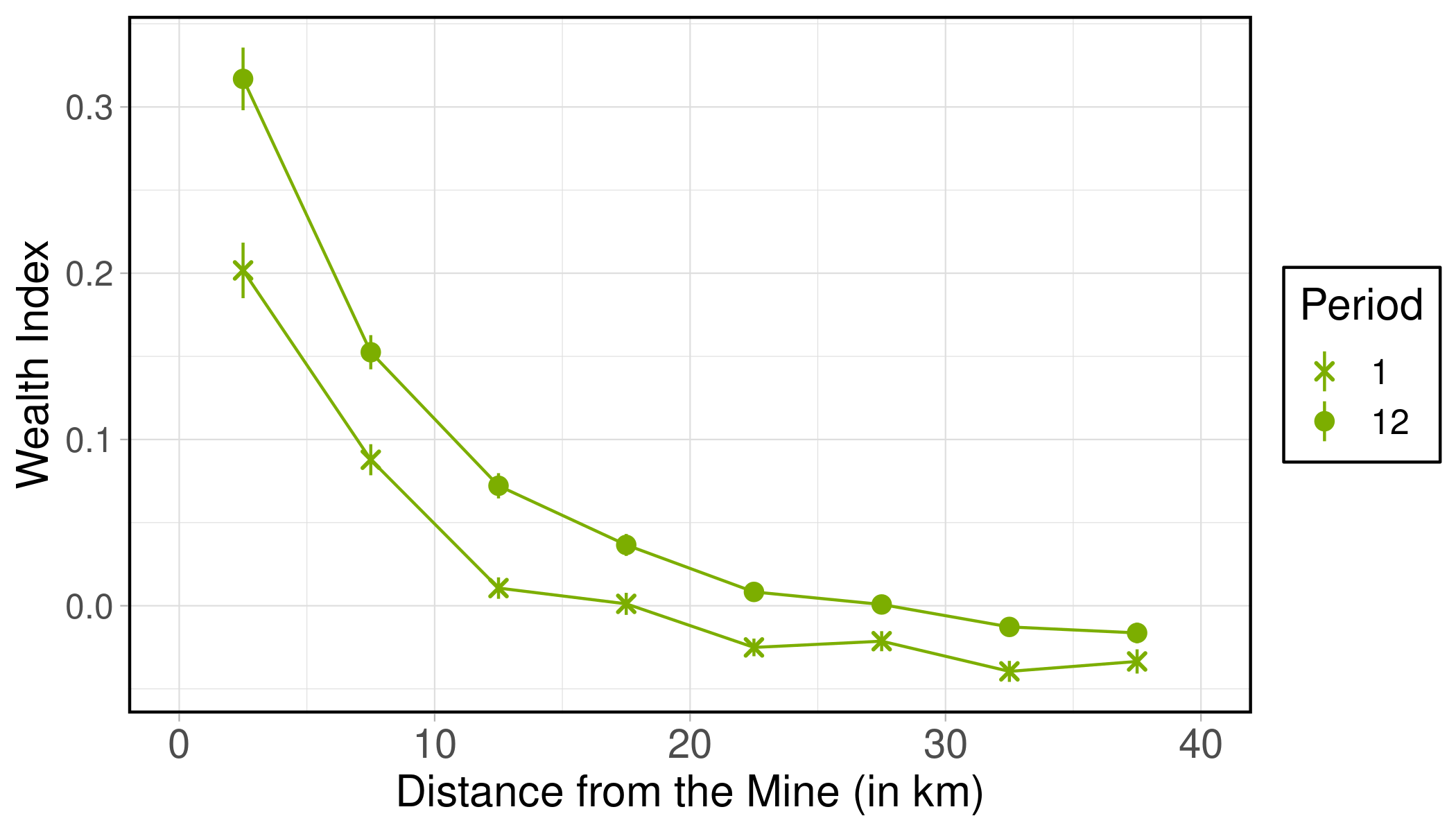}
\end{subfigure}
\hspace{0.3cm}
\begin{subfigure}{0.45\hsize}\centering
    \caption{\emph{Closing}} \label{satim:wealth_demeaned_close}
    \includegraphics[width=0.99\hsize]{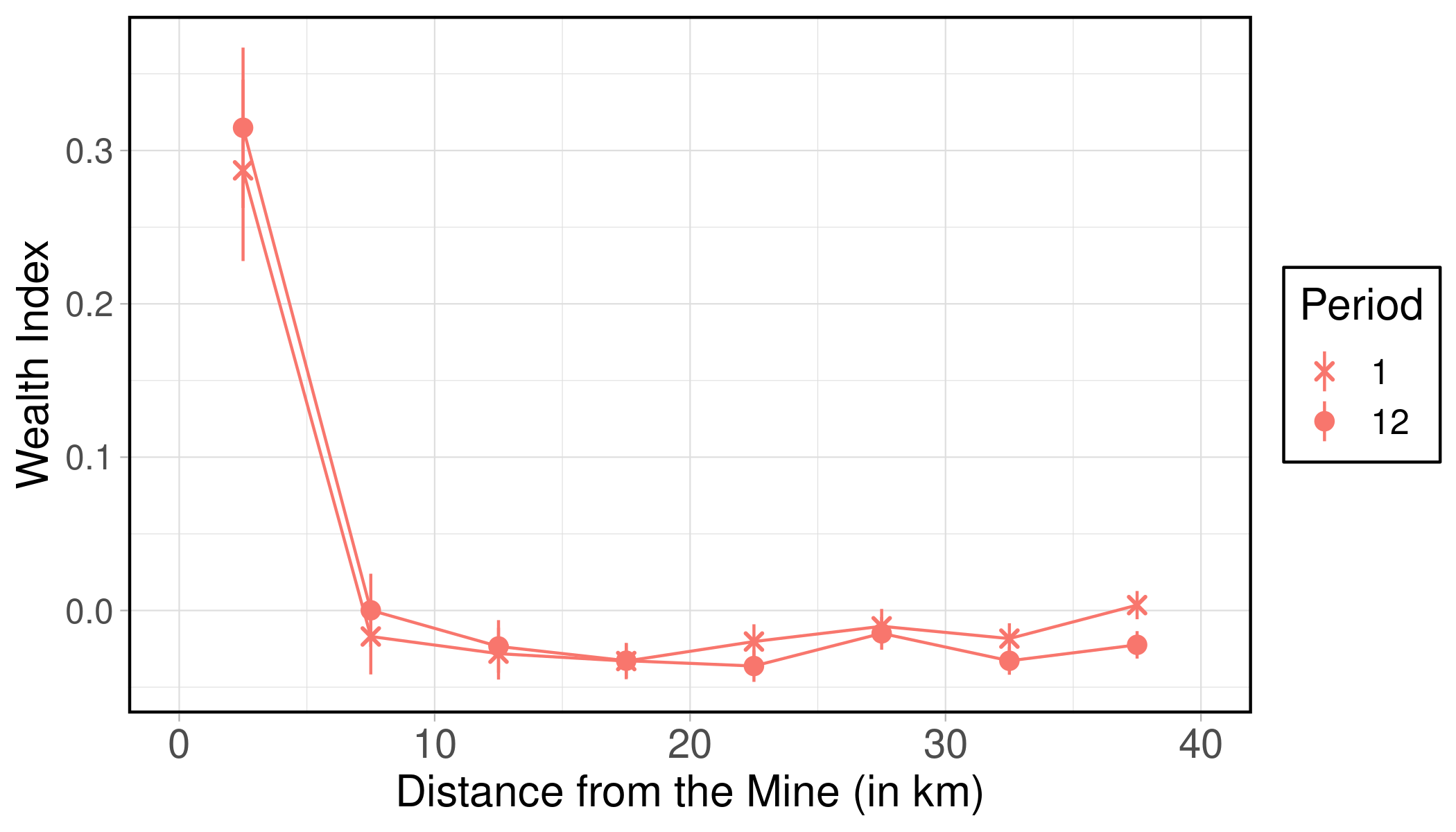}
\end{subfigure}
\begin{tablenotes}[flushleft]
\scriptsize \item \textit{Note:} The vertical axis is the wealth index of the respective mine area categories, minus the country and period average wealth index of \emph{Not Yet Opened} mine areas. \emph{Opening} mine areas have similar wealth indices as \emph{Not Yet Opened} mine areas in Period 1, but areas near the mine have comparatively greater wealth indices by Period 12. \emph{Closing} mine areas experience low wealth increases between Periods 1 and 12, in comparison to \emph{Continuous} mine areas. 
\end{tablenotes}
\end{figure}

\FloatBarrier

\subsection{Balancing Tests}

\vspace{3cm}

\begin{figure}[!h]
\caption{Balancing Graphs: Opening vs. Not Yet Opened} \label{satim:fig:balancing_opening_not_yet}

\vspace{0.4cm}

\begin{center}
\subfloat[Time Invariant Covariates]{
    \includegraphics[width=0.49\hsize,valign=t]{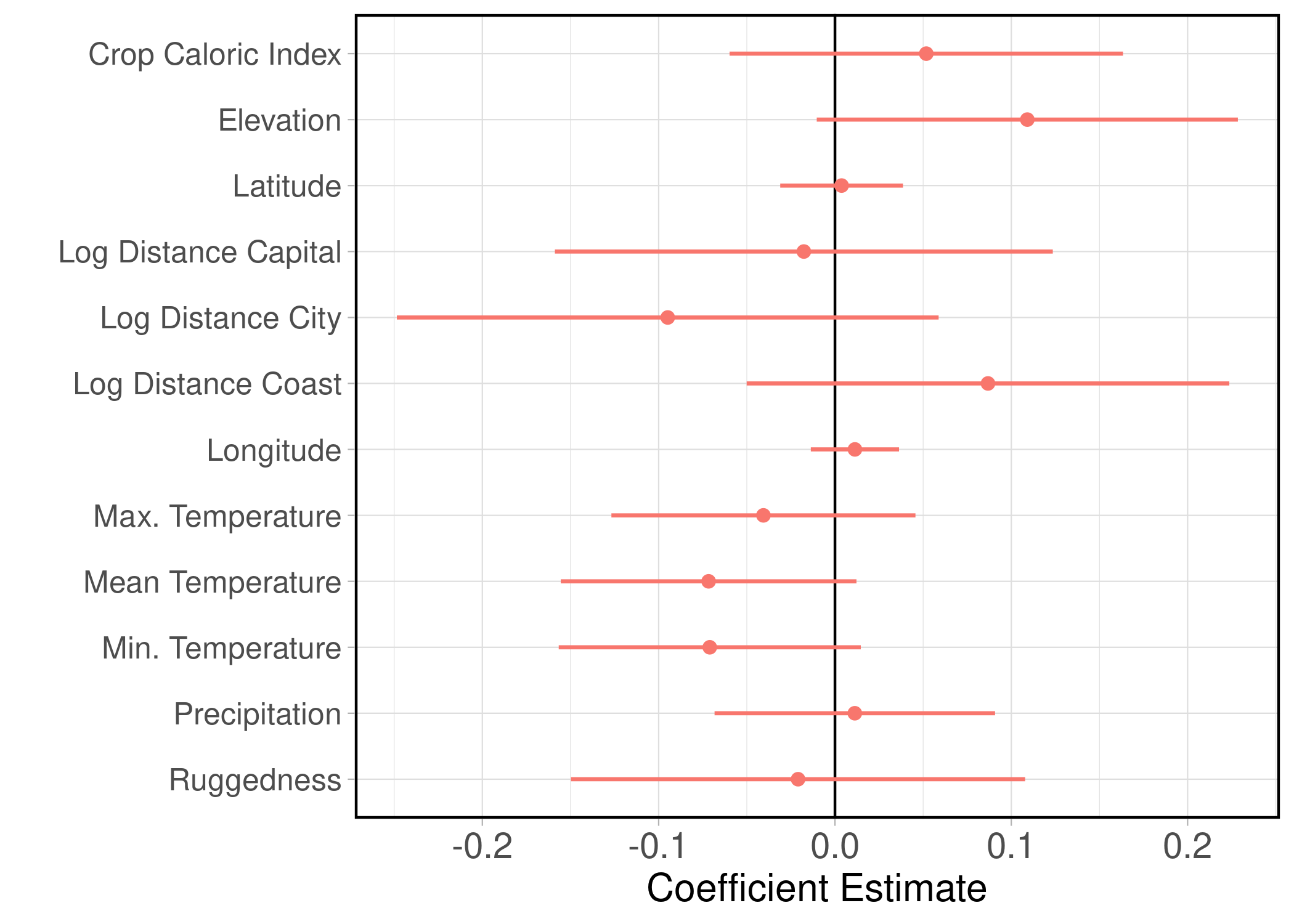}}
\subfloat[Outcomes in Period 1]{
    \includegraphics[width=0.49\hsize,valign=t]{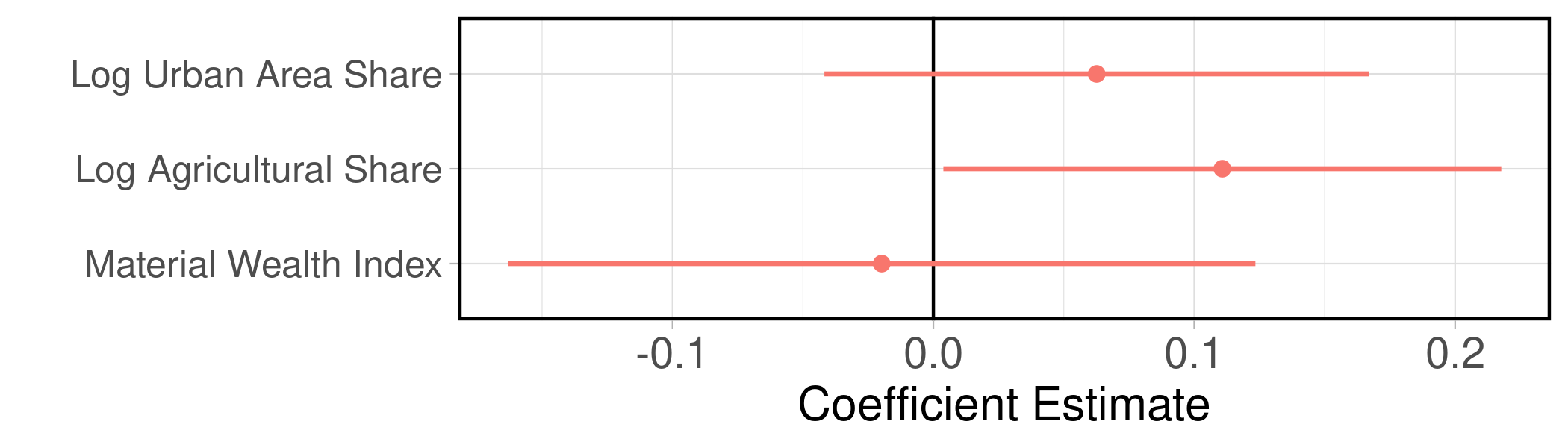}}
\end{center}

\vspace{-0.1cm}

\begin{tablenotes}
\scriptsize \item \textit{Note:} These figures report balancing test between \textit{Opening} and \textit{Not Yet Opened} deposits based on estimating: $Y_{i}  = \beta \; OpeningDummy_{i} + b_{c} + \varepsilon_{i} $, where $b_{c}$ are country fixed effects and with SEs clustered at the mine level. Error bars represent a 95\% confidence interval. 
\end{tablenotes}

\end{figure}

\vspace{-0.6cm}

\begin{figure}[!h]
\centering
\caption{Balancing Graphs: Early vs. Late Opening} \label{satim:fig:balancing_opening_early_late}
\subfloat[Time Invariant Covariates]{
    \includegraphics[width=0.49\hsize,valign=t]{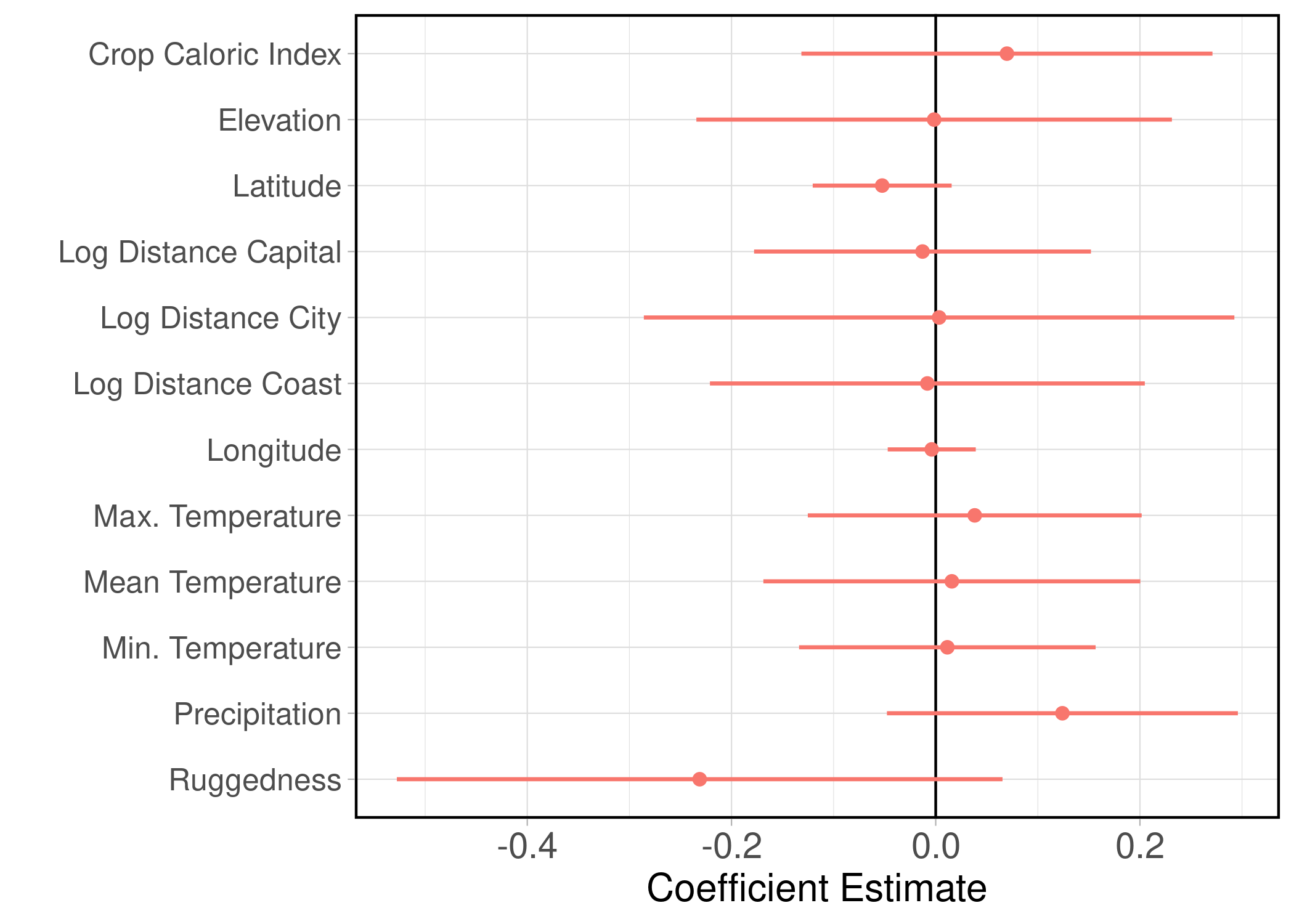}}
\subfloat[Outcomes in Period 1]{
    \includegraphics[width=0.49\hsize,valign=t]{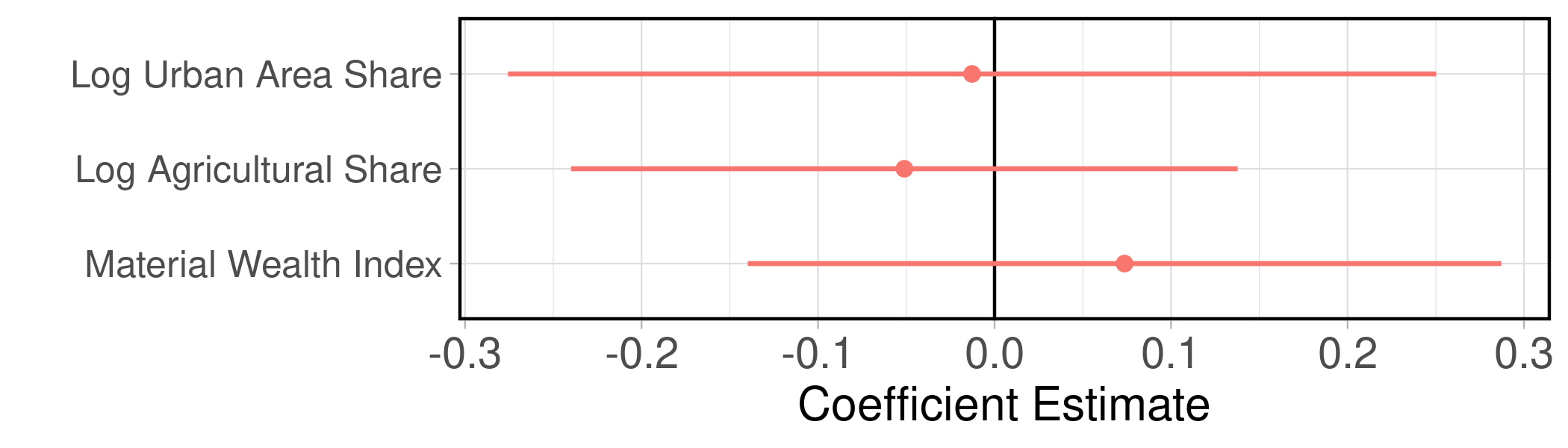}}
\vspace{-0.1cm}
\begin{tablenotes}
\scriptsize \item \textit{Note:} These figures report balancing test between early and late \textit{Opening} mines based on estimating: $Y_{i}  =  \beta \; log(OpeningYear)_{i} + b_{c} + \varepsilon_{i} $, where $b_{c}$ are country fixed effects and with SEs clustered at the mine level. Error bars represent a 95\% confidence interval. 
\end{tablenotes}  
\end{figure}

\begin{figure}[!h]
\caption{Balancing Graphs: Closing vs. Not Yet Opened} \label{satim:fig:balancing_closing_not_yet}

\vspace{0.4cm}

\begin{center}
\subfloat[Time Invariant Covariates]{
    \includegraphics[width=0.49\hsize,valign=t]{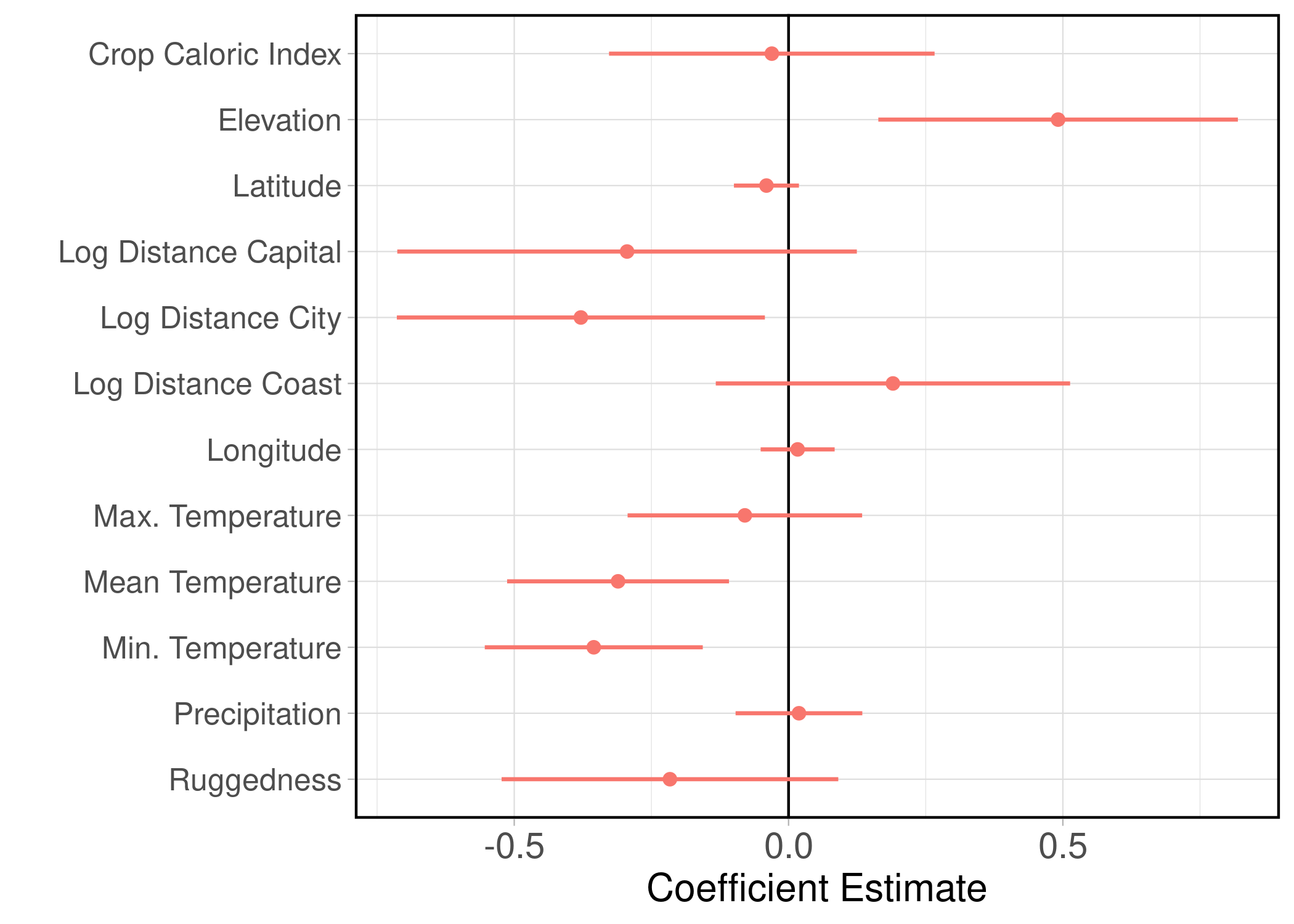}}
\subfloat[Outcomes in Period 1]{
    \includegraphics[width=0.49\hsize,valign=t]{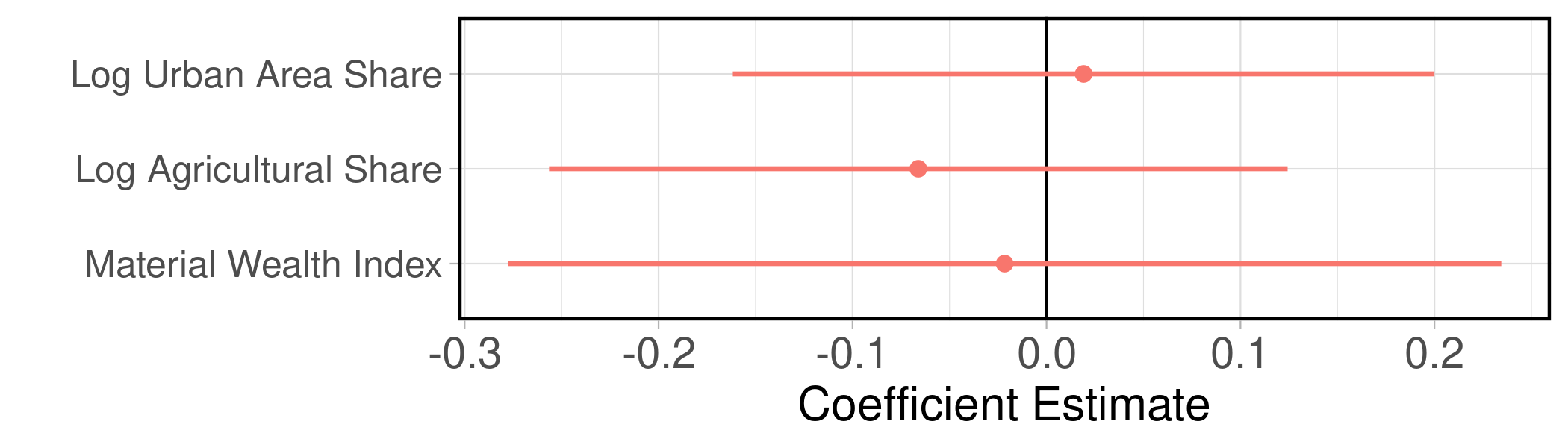}}
\end{center}

\vspace{-0.1cm}

\begin{tablenotes}
\scriptsize \item \textit{Note:} These figures report balancing test between \textit{Closing} and \textit{Not Yet Opened} deposits based on estimating: $Y_{i}  =  \beta \; ClosingDummy_{i} + b_{c} + \varepsilon_{i} $, where $b_{c}$ are country fixed effects and with SEs clustered at the mine level. Error bars represent a 95\% confidence interval. 
\end{tablenotes} 

\end{figure}

\vspace{-0.6cm}

\begin{figure}[!h]
\caption{Balancing Graphs: Closing vs. Continuous} \label{satim:fig:balancing_closing_contin}

\vspace{0.4cm}

\begin{center}
\subfloat[Time Invariant Covariates]{
    \includegraphics[width=0.49\hsize,valign=t]{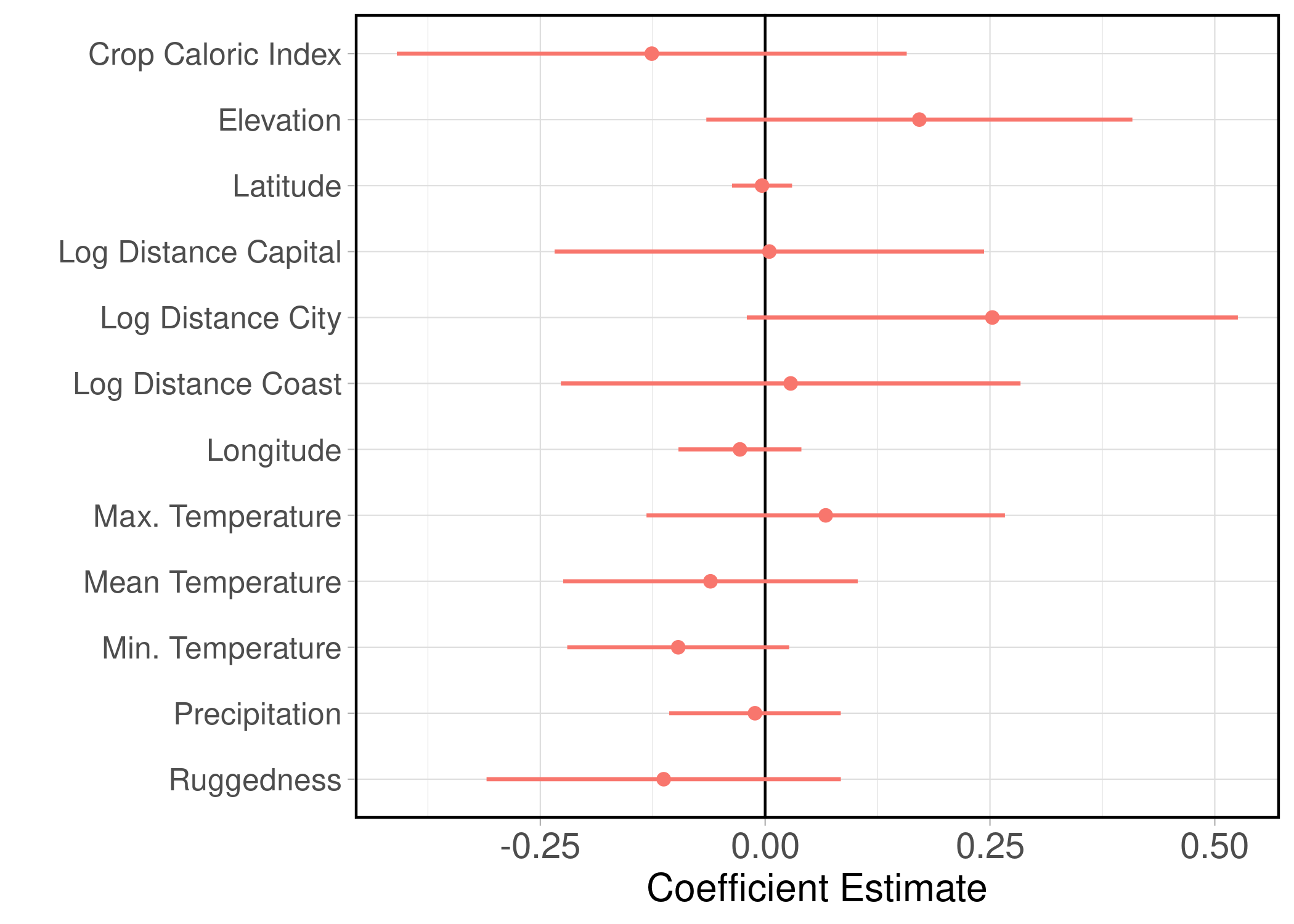}}
\subfloat[Outcomes in Period 1]{
    \includegraphics[width=0.49\hsize,valign=t]{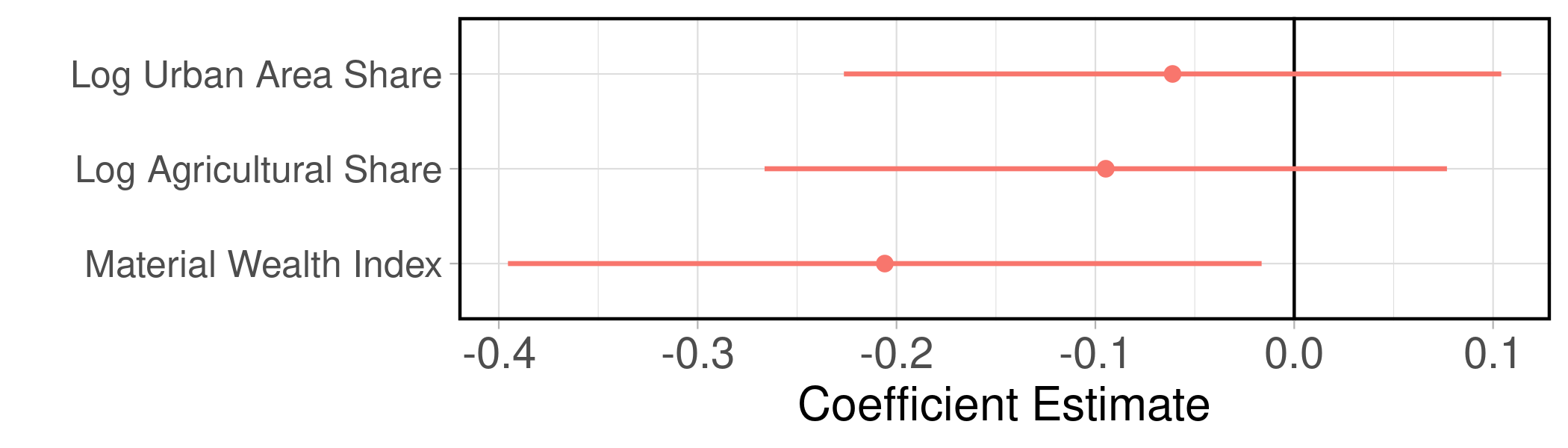}}
\end{center}

\vspace{-0.1cm}

\begin{tablenotes}
\scriptsize \item \textit{Note:} These figures report balancing test between \textit{Closing} and \textit{Continuous} deposits based on estimating: $Y_{i}  =  \beta \; ClosingDummy_{i} + b_{c} + \varepsilon_{i} $, where $b_{c}$ are country fixed effects and with SEs clustered at the mine level. Error bars represent a 95\% confidence interval. 
\end{tablenotes} 

\end{figure}

\FloatBarrier

\subsection{Additional Results}

\begin{table}[H] \centering 
  \caption{Stacked DiD Regressions in Areas Near the Mine - Heterogeneities (Cont.)} 
  \label{satim:table_hetero_size_append} 
\resizebox{\textwidth}{!}{\begin{tabular}{@{\extracolsep{5pt}}lcc:cc|cc:cc} 
\\[-1.8ex]\hline 
\hline \\[-1.8ex] 
 & \multicolumn{8}{c}{Stacked DiD Regressions} \\
\cline{2-9}\\
& \multicolumn{4}{c}{Log Agricultural Area Share} & \multicolumn{4}{c}{Material Wealth Index (z-score)} \\
\cmidrule(r{5pt}){2-5}  \cmidrule(l{5pt}){6-9} \\[-1.2em] 
&\multicolumn{2}{c}{Near}&\multicolumn{2}{c}{Far}&\multicolumn{2}{c}{Near}&\multicolumn{2}{c}{Far}\\[-.5ex] 
\cmidrule(r{5pt}){2-3}  \cmidrule(l{5pt}){4-5} \cmidrule(r{5pt}){6-7}  \cmidrule(l{5pt}){8-9}\\[-1.2em] 
\\[-1.8ex] & (1) & (2) & (3) & (4) & (5) & (6) & (7) & (8) \\ 
\hline \\[-1.8ex] 
 \\[-1.5ex]
Treatment Dummy                         & $0.15$    & $-0.16$     & $0.10$    & $-0.25^{**}$ & $0.11^{***}$ & $0.07$      & $0.00$    & $-0.01$   \\
                                        & $(0.10)$  & $(0.15)$    & $(0.08)$  & $(0.13)$     & $(0.04)$     & $(0.07)$    & $(0.04)$  & $(0.06)$  \\
& & & & & & & & \\[-1.8ex] 
Treat $\times$ Large Mine               &    -      & $0.32^{*}$  &     -     & $0.29^{**}$  &      -       & $0.20^{**}$ &     -     & $0.08$    \\
                                        &           & $(0.18)$    &           & $(0.14)$     &              & $(0.10)$    &           & $(0.07)$  \\
& & & & & & & & \\[-1.8ex] 
Treat $\times$ Democracy                &    -      & $0.60^{**}$ &     -     & $0.68^{***}$ &      -       & $-0.09$     &     -     & $-0.06$   \\
                                        &           & $(0.23)$    &           & $(0.20)$     &              & $(0.10)$    &           & $(0.08)$  \\
& & & & & & & & \\[-1.8ex] 
Treat $\times$ Large $\times$ Democracy &    -      & $-0.36$     &     -     & $-0.27$      &      -       & $0.02$      &     -     & $0.01$    \\
                                        &           & $(0.33)$    &           & $(0.25)$     &              & $(0.14)$    &           & $(0.11)$  \\
& & & & & & & & \\[-1.8ex] 
\hline
\addlinespace[0.1cm] 
Event x Tile FE                    & Yes       & Yes          & Yes       & Yes          & Yes          & Yes         & Yes       & Yes       \\
Event x Country x Period FE        & Yes       & Yes          & Yes       & Yes          & Yes          & Yes         & Yes       & Yes       \\
\hline
\addlinespace[0.1cm] 
Observations                       & $1,172,561$ & $1,172,561$   & $1,254,498$ & $1,254,498$    & $1,172,561$    & $1,172,561$   & $1,254,498$ & $1,254,498$ \\
Adj. R$^2$                         & $0.76$    & $0.76$      & $0.76$    & $0.76$       & $0.68$       & $0.68$      & $0.68$    & $0.68$    \\
\hline 
\hline \\[-1.8ex]
\end{tabular}}
\begin{tablenotes}[flushleft]
\scriptsize \item \textit{Note:} This table reports stacked DiD heterogeneity tests based on Equation \ref{satim:eq_did}. The table corresponds to Table \ref{satim:table_2} and reports the results for the log of the agricultural land cover and the z-score of the material wealth index as dependent variables. The `Treatment Dummy' (or `Treat') indicates if a tile's corresponding mine has started operating, it is always 0 for tiles in the control group. In Columns (1), (2), (5) and (6), we restrict the treatment group to tiles within 20km from the mine, and in Columns (3), (4), (7) and (8) to tiles between 20km and 40km from the mine. Standard errors in parenthesis are double-clustered by mine and tile. \hfill  $^{*}$p$<$0.1; $^{**}$p$<$0.05; $^{***}$p$<$0.01
\end{tablenotes}
\end{table}

\begin{figure}[htb]
\caption{Event Study: \textit{Closing} vs. \textit{Not Yet Opened} Mines} \label{satim:closing_not_yet_evstud_append}

\begin{subfigure}{0.5\hsize}\centering
    \caption{Close: Log Agricultural Share}
    \includegraphics[width=0.9\hsize]{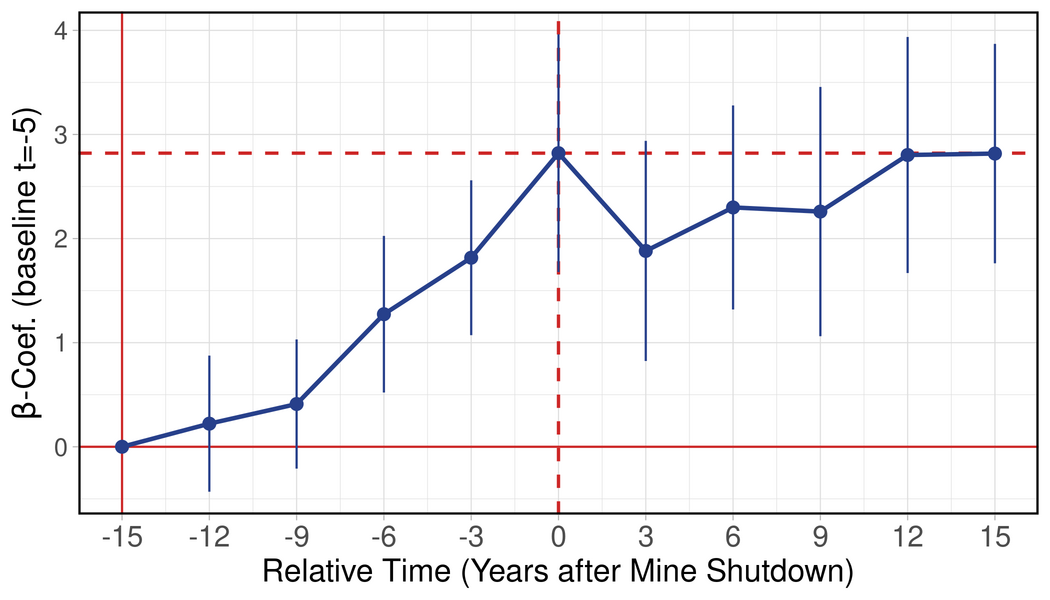}
\end{subfigure}
\begin{subfigure}{0.5\hsize}\centering
    \caption{Far: Log Agricultural Share} 
    \includegraphics[width=0.9\hsize]{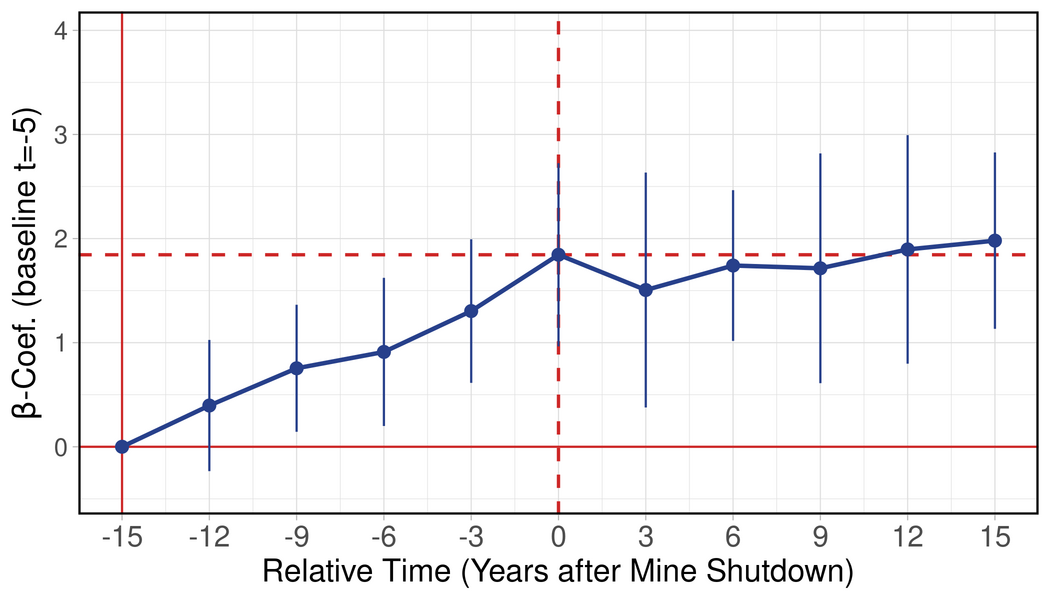}
\end{subfigure}

\vspace{0.3cm}

\begin{subfigure}{0.5\hsize}\centering
    \caption{Close: Material Wealth Index (z-score)}
    \includegraphics[width=0.9\hsize]{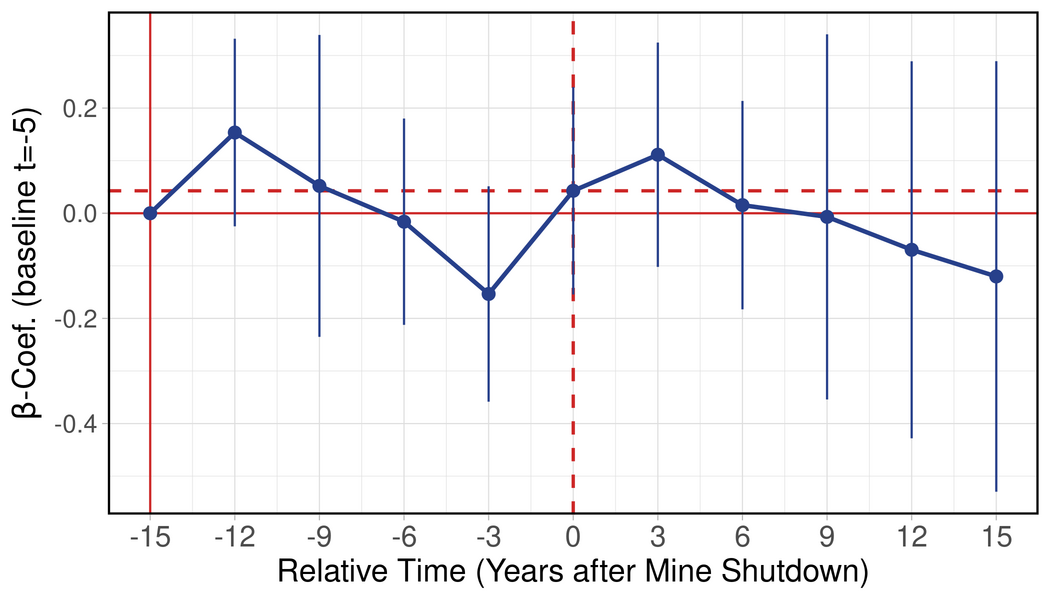}
\end{subfigure}
\begin{subfigure}{0.5\hsize}\centering
    \caption{Far: Material Wealth Index (z-score)} 
    \includegraphics[width=0.9\hsize]{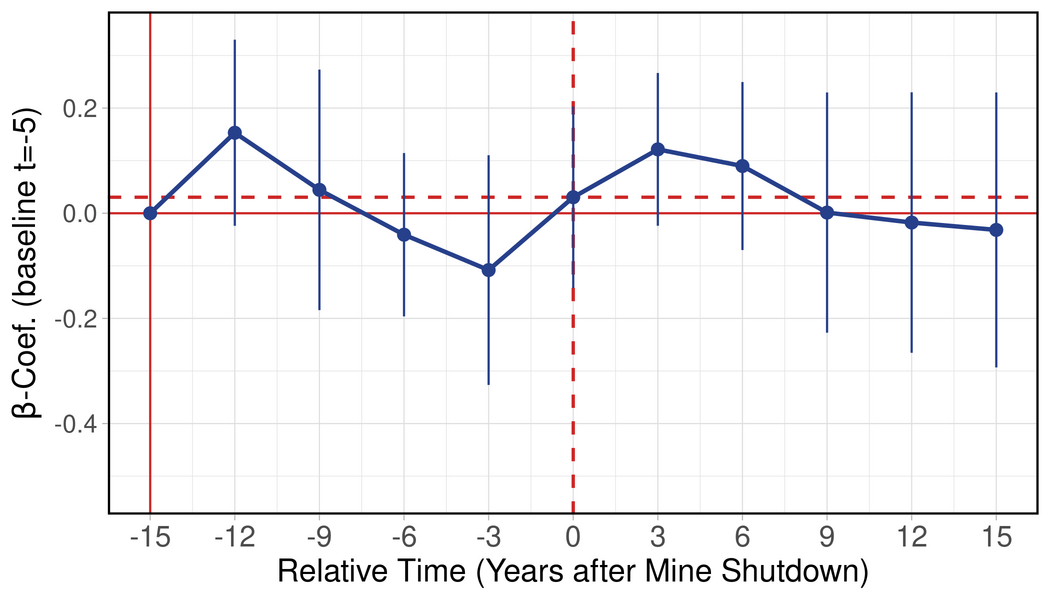}
\end{subfigure}

\begin{tablenotes}[flushleft]
\scriptsize \item \textit{Note:} Prior to closure, \emph{Closing} mine areas have higher agricultural growth compared to \emph{Not Yet Opened} mine areas. However, after closure, this growth diminishes in comparison to similar \emph{Not Yet Opened} mine areas. We do not observe any significant trend breaks for the material wealth index. Error bars represent a 95\% confidence interval. 
\end{tablenotes}
\end{figure}

\begin{figure}[htb]
\caption{Event Study: \textit{Closing} vs. \textit{Continuous} Mines}  \label{satim:closing_contin_evstud_append} 

\begin{subfigure}{0.5\hsize}\centering
    \caption{Near: Log Agricultural Share}
    \includegraphics[width=0.9\hsize]{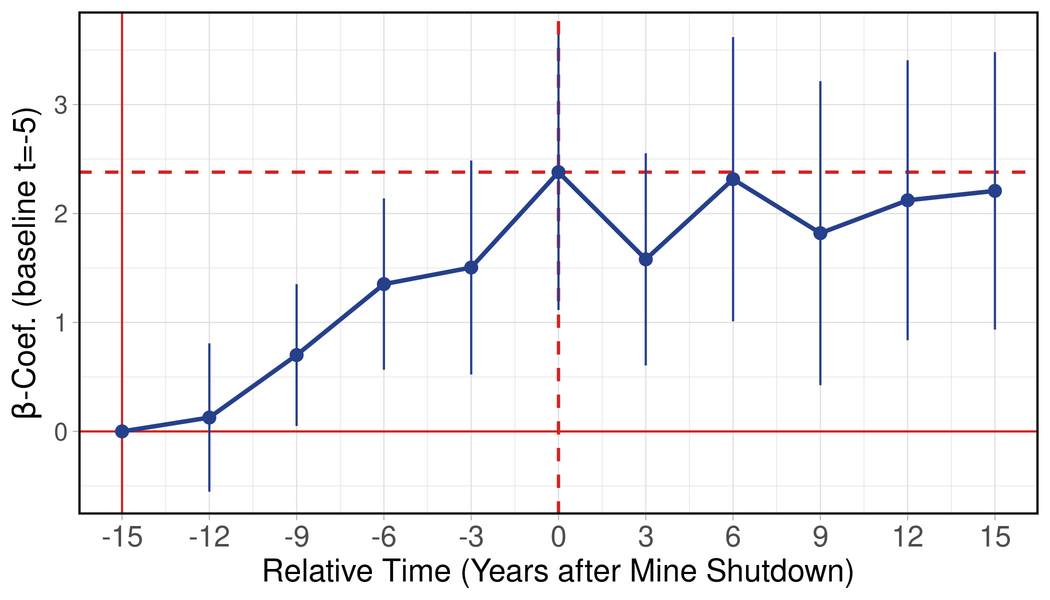}
\end{subfigure}
\begin{subfigure}{0.5\hsize}\centering
    \caption{Far: Log Agricultural Share} 
    \includegraphics[width=0.9\hsize]{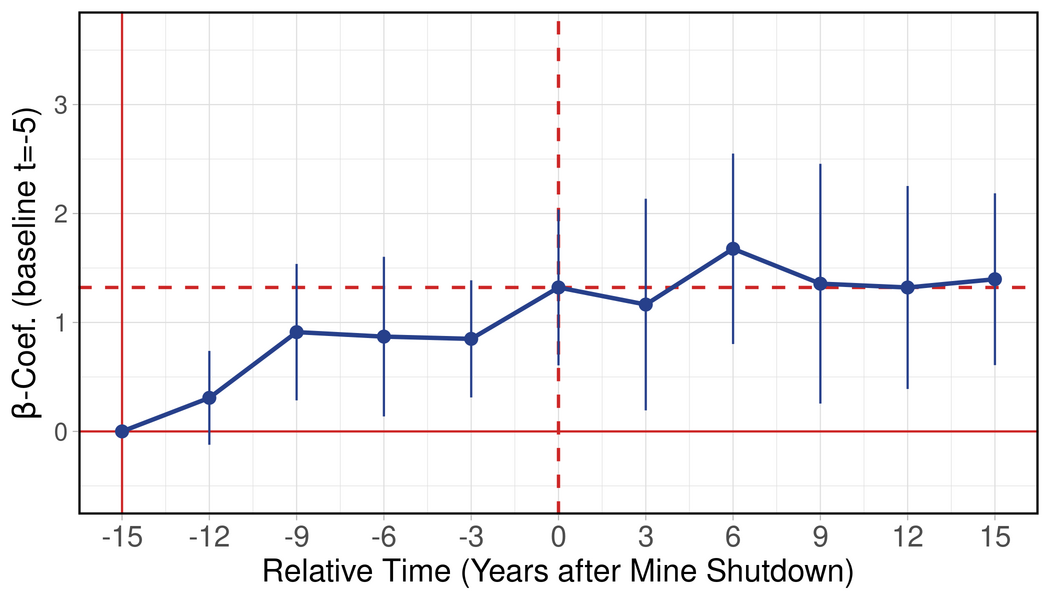}
\end{subfigure}

\vspace{0.3cm}

\begin{subfigure}{0.5\hsize}\centering
    \caption{Near: Material Wealth Index (z-score)}
    \includegraphics[width=0.9\hsize]{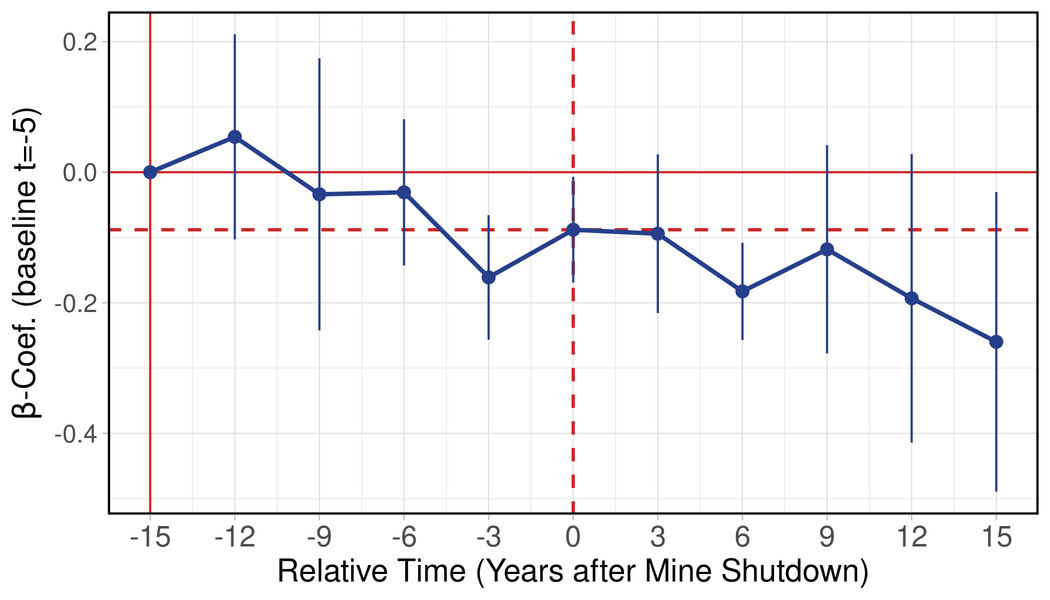}
\end{subfigure}
\begin{subfigure}{0.5\hsize}\centering
    \caption{Far: Material Wealth Index (z-score)} 
    \includegraphics[width=0.9\hsize]{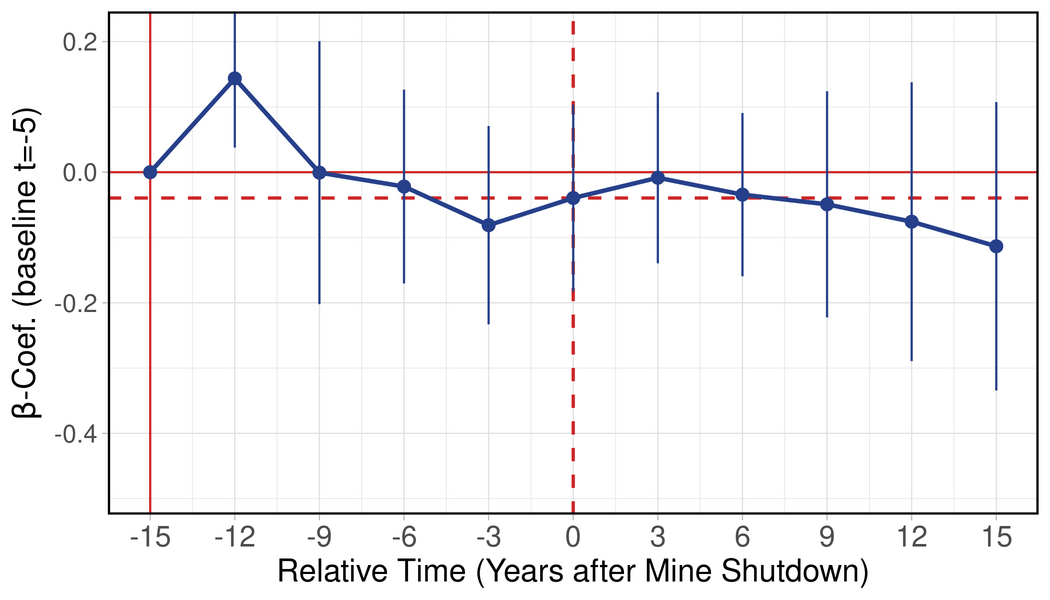}
\end{subfigure}

\begin{tablenotes}[flushleft]
\scriptsize \item \textit{Note:} Prior to closure, \emph{Closing} mine areas have higher agricultural growth compared to \emph{Continuous} mine areas. However, after closure, this growth diminishes in comparison to similar \emph{Continuous} mine areas. We do not observe any significant trend breaks for the material wealth index. Error bars represent a 95\% confidence interval. 
\end{tablenotes}
\end{figure}

\end{document}